\documentclass[10pt,prd,twocolumn,preprint,nofootinbib]{revtex4}

\usepackage{epsfig}
\usepackage{amsmath,amsfonts,amssymb}
\usepackage{t1enc}
\usepackage{verbatim}
\usepackage{float}
\usepackage{morefloats}
\usepackage{xcolor}
\usepackage[normalem]{ulem}

\providecommand{\openone}{\leavevmode\hbox{\small1\kern-3.8pt\normalsize1}}

\newcommand{\CP}{\emph{CP}\,}

\usepackage{booktabs}
\usepackage[Q=yes,pverb-linebreak=no]{examplep}
\begin{document}

\title{\boldmath The Role of the $t{\bar t}h$ Rest Frame in Direct Top-Quark Yukawa Coupling Measurements \unboldmath}

\author{
Andrea Ferroglia$^1$,
Miguel C. N. Fiolhais$^{2,3}$,
Emanuel Gouveia$^4$, 
Ant\'onio Onofre$^{5}$
\\[3mm]
{\footnotesize {\it 
$^1$ Physics Department, New York City College of Technology, The City University of New York, 300 Jay Street, Brooklyn, NY 11201 USA\\
$^2$ Science Department, Borough of Manhattan Community College, City University of New York, \\ 
     199 Chambers St, New York, NY 10007, USA \\
$^3$ LIP, Departamento de F\'{\i}sica, Universidade de Coimbra, 3004-516 Coimbra, Portugal\\
$^4$ LIP, Departamento de F\'{\i}sica, Universidade do Minho, 4710-057 Braga, Portugal\\
$^5$ Departamento de F\'{\i}sica, Universidade do Minho, 4710-057 Braga, Portugal\\
}}
}

\begin{abstract}
This paper studies new possibilities to directly measure a hypothetical \CP-odd (pseudoscalar) component in the top-quark Yukawa coupling.
In particular, the role of the $t{\bar t} h$ center-of-mass rest frame in the associated production of a top pair and a $h$ boson at the LHC is explored. The $h$ boson is assumed to have both a \CP-even (scalar) and a \CP-odd coupling to the top quark. The relative strength of the scalar and pseudoscalar components is regulated by an angle $\alpha$. Observables sensitive to the nature of the top-quark Yukawa coupling are proposed. These observables are  defined in terms of the transverse and longitudinal projections of $t$, $\bar{t}$ and $h$ momenta  with respect to the beam axis in the $t{\bar t} h$ rest frame. Distributions differential with respect to those observables are evaluated up to NLO in QCD. These distributions are found to be sensitive to the \emph{CP} nature of the coupling.
Dileptonic final states of the $t{\bar t} h$ system (with $h\rightarrow b\bar b$) are used, after fast DELPHES detector simulation and full event reconstruction through a kinematic fit, as a case study to test the observables' sensitivity to the \emph{CP} nature of the coupling. {Confidence levels are presented as a function of the total integrated LHC luminosity for the case of exclusion of a pure \emph{CP}-odd coupling against the Standard Model \emph{CP}-even hypothesis.}
By using observables evaluated in the $t{\bar t} h$  system, the luminosity needed to directly probe the \emph{CP} properties of the top-quark Yukawa coupling at the High-Luminosity run of the LHC can be decreased by a few hundred inverse femtobarns, when compared to analyses that use observables in the laboratory rest frame. In addition, transverse momentum distributions of the $h$ boson and top quarks are found to provide no more discriminant power than a counting experiment.

\end{abstract}

\maketitle

\section{Introduction}

Following the Higgs boson discovery~\cite{Aad:2012tfa,Chatrchyan:2012xdj} and the observation of the associated production of a top-quark pair and a Higgs boson at both ATLAS and CMS experiments~\cite{Aaboud:2018urx, Sirunyan:2018hoz}, the study of the Higgs-boson properties, such as Yukawa couplings, at the Large Hadron Collider (LHC) is now of utmost importance. To date, no significant deviations have been observed in the measured production and coupling properties of the Higgs boson~\cite{Aad:2013wqa,Aad:2013xqa,Aad:2015vsa,Khachatryan:2014jba,Chatrchyan:2014vua,Khachatryan:2014kca,Khachatryan:2016tnr,Aad:2015mxa,Sirunyan:2018koj}
 with respect to Standard Model (SM) predictions. However, it is by now clear that the SM cannot explain all of the observed physical phenomena. For example, the SM fails to explain the matter/anti-matter asymmetry of the Universe, for which new sources of \CP violation 
beyond the SM (BSM) are required. One possibile additional source of \CP violation could come from  the Higgs sector.
This is predicted by several BSM models, such as 2-Higgs doublets models (2HDM), where the Higgs boson(s) may not have a definite \CP quantum number, resulting in a Yukawa coupling with two components, one \CP-even and one \CP-odd \cite{Fontes:2015mea}. In order to accommodate a possible \CP-odd contribution to the top-quark Yukawa coupling, in this work the following Lagrangian was considered:
\begin{equation}
\mathcal{L} = - \, y_t \, \bar t \, (\cos \alpha + i \gamma_5 \sin \alpha) \, t \, h\, ,
\label{eq:lagrangian}
\end{equation}
where $y_t$ is the SM Higgs Yukawa coupling and $\alpha$ is the angle that regulates the relative strength of the scalar and pseudoscalar components of the coupling~\cite{Demartin:2014fia}. Note that with this Lagrangian the Higgs field $h$ has no definite \CP quantum number. The SM interaction is recovered for $\cos\alpha=\pm1$; in that case $h$ is the SM Higgs boson, indicated by $H$. The pure pseudoscalar coupling
is obtained by setting $\cos\alpha=0$; in that case the field $h$ corresponds to a purely pseudoscalar field, indicated by $A$ in this work.

Strong bounds on the electric dipole moment (EDM) of the electron indirectly constrain the magnitude of a possible \emph{CP}-violating component of the top-quark Yukawa coupling. The constraint assumes the SM values for the \CP-even part of the coupling and no cancellation among other contributions to the EDM~\cite{deBlas:2019rxi}. Although indirect limits are important to predict {the size of a possible pseudoscalar component in the top-quark Yukawa coupling}, they are complementary, rather than alternative, to direct \CP measurements. Eventual discrepancies between direct  measurements and indirect predictions could signal the presence of new physics.

The associated production of a top-quark pair and a Higgs boson is a process that provides a direct measurement of the top-quark Yukawa coupling~\cite{Ng:1983jm,Kunszt:1984ri,Marciano:1991qq,Gunion:1991kg,Goldstein:2000bp,Beenakker:2001rj,Beenakker:2002nc,Reina:2001sf,Dawson:2002tg,Dawson:2003zu,Dittmaier:2003ej,Frederix:2011zi,Garzelli:2011vp,Hartanto:2015uka,Frixione:2014qaa,Yu:2014cka,Frixione:2015zaa,Maltoni:2015ena,Broggio:2015lya,Broggio:2016lfj,Kulesza:2017ukk,Ju:2019lwp,Broggio:2019ewu}. Recent predictions at next-to-leading order (NLO) in fixed order perturbation theory, including resummation of soft emission corrections to next-to-leading-logarithmic accuracy (NLL)~\cite{Broggio:2017oyu}, have shown that several differential distributions can provide useful information on the possible presence of a pseudoscalar component in the top-quark Yukawa coupling. 
This paper considers the possibility of measuring some of the observables defined in~\cite{Gunion:1996xu} in the $t\bar{t}h$ rest frame rather than in the laboratory frame. It is found that this choice improves the sensitivity of these observables to the possible presence of a pseudoscalar component in the top-quark Yukawa coupling.
Other observables that could probe the \CP nature of other heavy-fermion couplings have been considered in the literature; for example, the case of the  $\tau$ lepton  in $\tau^+ \tau^- h$ production process is presented in~\cite{Boudjema:2015nda,Berge:2014sra,Berge:2012wm,Berge:2011ij,Berge:2008wi,Khatibi:2014bsa,Brooijmans:2014eja}.

As this work focuses on observables boosted to the $t \bar{t} h$ center-of-mass frame, the full four-momenta reconstruction of the top quarks and $h$ boson is required. The dileptonic final state of top-pair {production}, with the $h$ boson decaying to bottom quarks, is considered to study the sensitivity of the proposed observables. The two charged leptons in the final state provide a clean experimental signature, preserving as well useful information on the spin of their parent top quarks. It is interesting to observe that the specific nature of the coupling changes the angular distributions between the momenta of the $h$ boson and the top quarks in $t \bar{t} h$ events.  Clear differences are seen, in particular, when the angles are evaluated in the $t\bar t h$ center-of-mass system. This observation motivates the search for differential distributions in the $t\bar t h$ center-of-mass frame that are sensitive to the \CP nature of the top-quark Yukawa coupling.

The paper is organized as follows. In Section~\ref{sec:partonleveldist}, the parton-level distributions are introduced together with the \CP observables used in this study. Distributions of event samples after parton shower are also shown in Section~\ref{sec:showering}. Section~\ref{sec:dileptonic} describes a case study in the dileptonic final state of $t\bar t h$, presenting the main features of event generation, detector simulation and analysis. In Section~\ref{sec:CL}, expected Confidence Levels (CL) for the exclusion of a pure pseudoscalar are presented for different distributions at reconstruction level, \emph{i.e.} after the dileptonic case-study analysis. Finally, conclusions are drawn in Section~\ref{sec:conclusions}.  

\section{Parton-Level Distributions \label{sec:partonleveldist}}

This section is devoted to the study of the effect of the Lagrangian in Eq.~(\ref{eq:lagrangian}) on the total cross section and on the differential distributions that depend on the momenta of the massive particles in the final state. The total cross section and differential distributions discussed here were evaluated by means of
 {\scshape MadGraph5\Q{_}aMC@NLO}~\cite{Alwall:2014hca}
 up to NLO in QCD. The Lagrangian in Eq.~(\ref{eq:lagrangian}) is implemented in the  \texttt{HC\Q{_}NLO\Q{_}X0} model~\cite{Artoisenet:2013puc}. This calculation assumed on-shell top quark, antitop quark and $h$ boson. All of the calculations presented in this section were carried out by employing {\tt{MMHT 2014}} NLO PDFs \cite{Harland-Lang:2014zoa}. The top quark and Higgs boson masses were {set to}
 \begin{equation}
     m_t = 173 \, \mbox{GeV} \, \qquad m_h = 125 \, \mbox{GeV}\, ,
 \end{equation}
respectively. The calculations were carried out by using a default dynamic factorization and renormalization scale 
\begin{equation}
    \mu_{0,r} = \mu_{0,f} = \frac{M}{2} \, ,
\end{equation}
where $M$ indicates the invariant mass of the top-antitop-Higgs final state. The theoretical uncertainty associated to the missing beyond-NLO corrections was estimated by varying the factorization  scale in the range $\{\mu_{0,f}/2, 2 \mu_{0,f}\}$ and by considering the envelope of the values for the cross section found through this variation.

The total cross section in the case of a purely scalar coupling, including the residual scale uncertainty derived from scale variation, is
\begin{equation}
    \sigma^{t \bar{t} H} = 474.8^{+47.4 (10 \%)}_{-51.6 (11 \%)}        \, \mbox{fb} \, ,
\end{equation}
while the total cross section for the purely pseudoscalar case is     
\begin{equation}
\sigma^{t \bar{t} A} = 192.4^{+23.3 (12 \%)}_{-24.3 (13 \%)}        \, \mbox{fb} \, ,
\end{equation}
in agreement with what was found in \cite{Broggio:2017oyu}.
The total cross section for an arbitrary value of the angle $\alpha$ can be found starting from the two results above, since 
\begin{equation}
    \sigma^{t \bar{t} h} = \sigma^{t \bar{t} H} \cos^2 \alpha  +\sigma^{t \bar{t} A} \sin^2 \alpha \, .
    \label{eq:totcsalpha}
\end{equation}

\begin{figure*}
\begin{center}
\begin{tabular}{cc}
\hspace*{-1mm} \epsfig{file=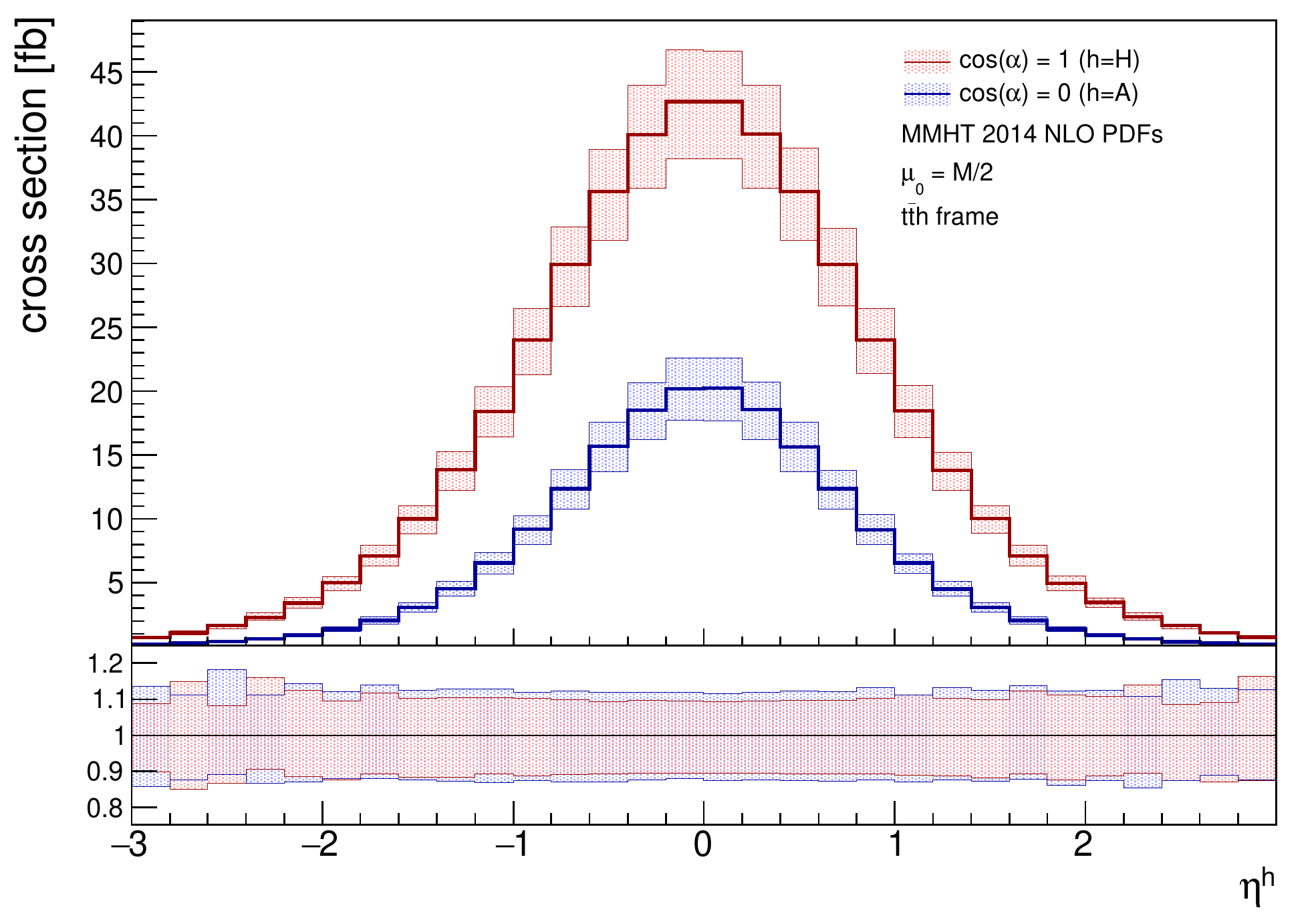,height=4.5cm,clip=} & 
                           \epsfig{file=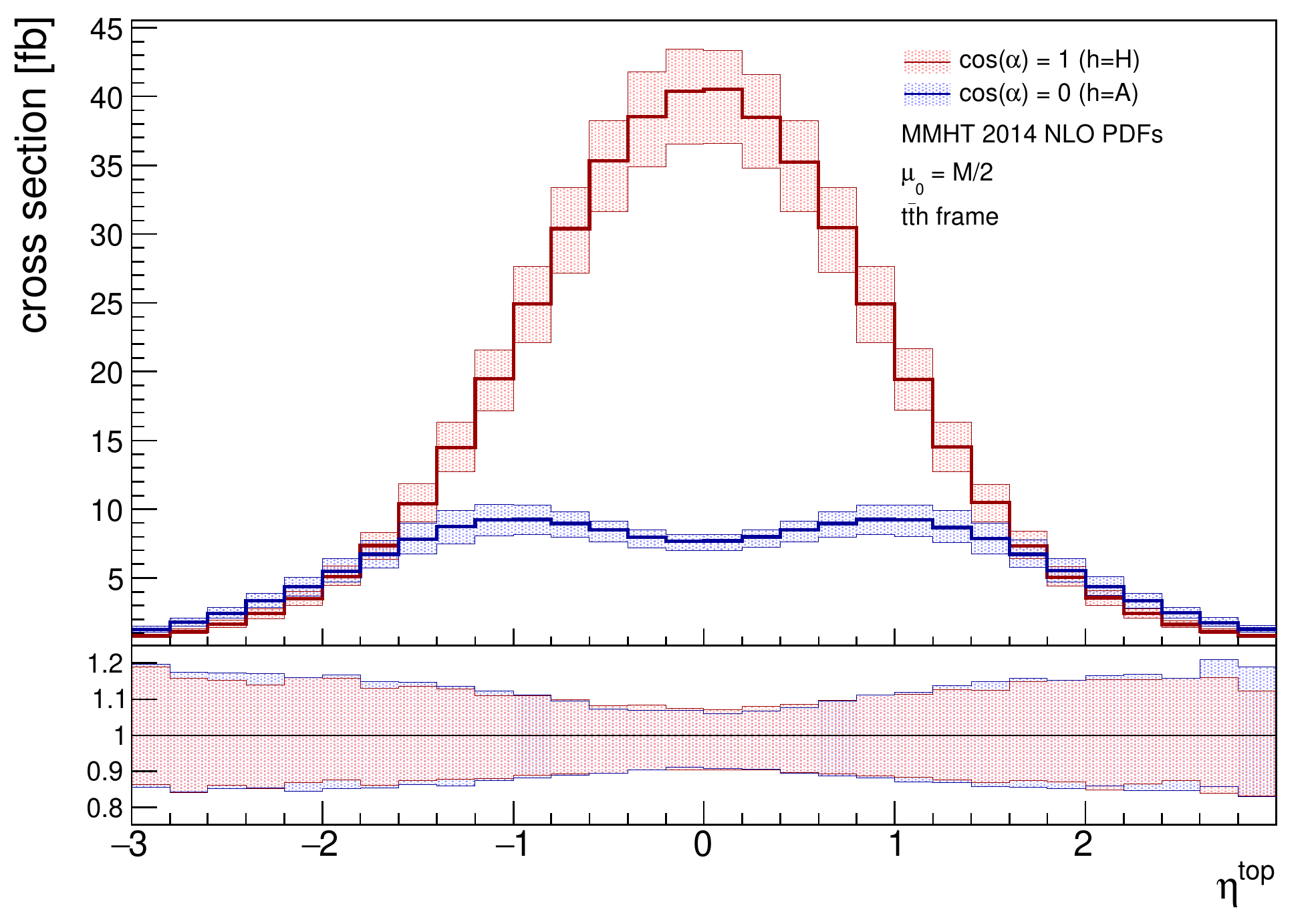,height=4.5cm,clip=} \\
\hspace*{-1mm} \epsfig{file=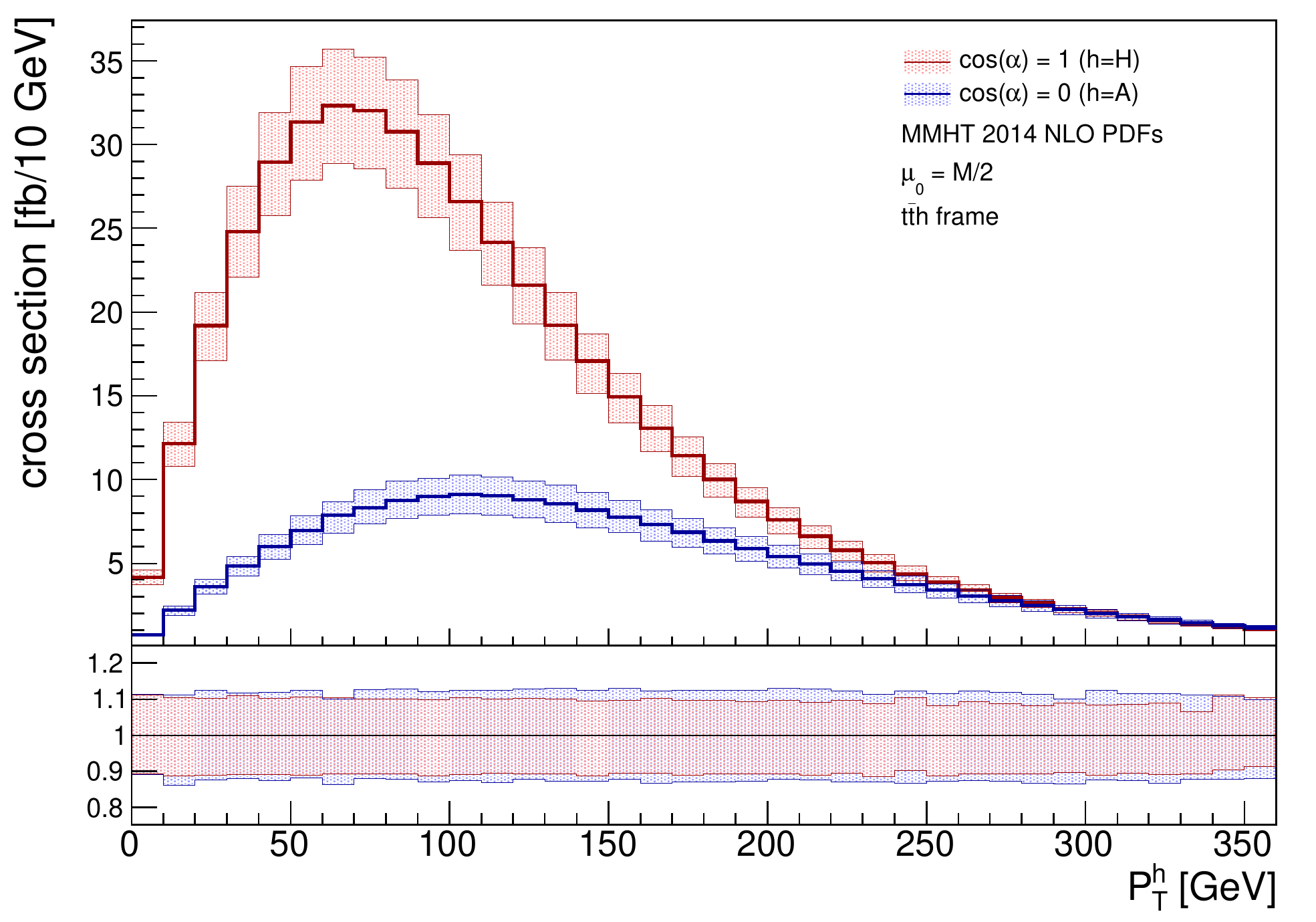,height=4.5cm,clip=} &
                           \epsfig{file=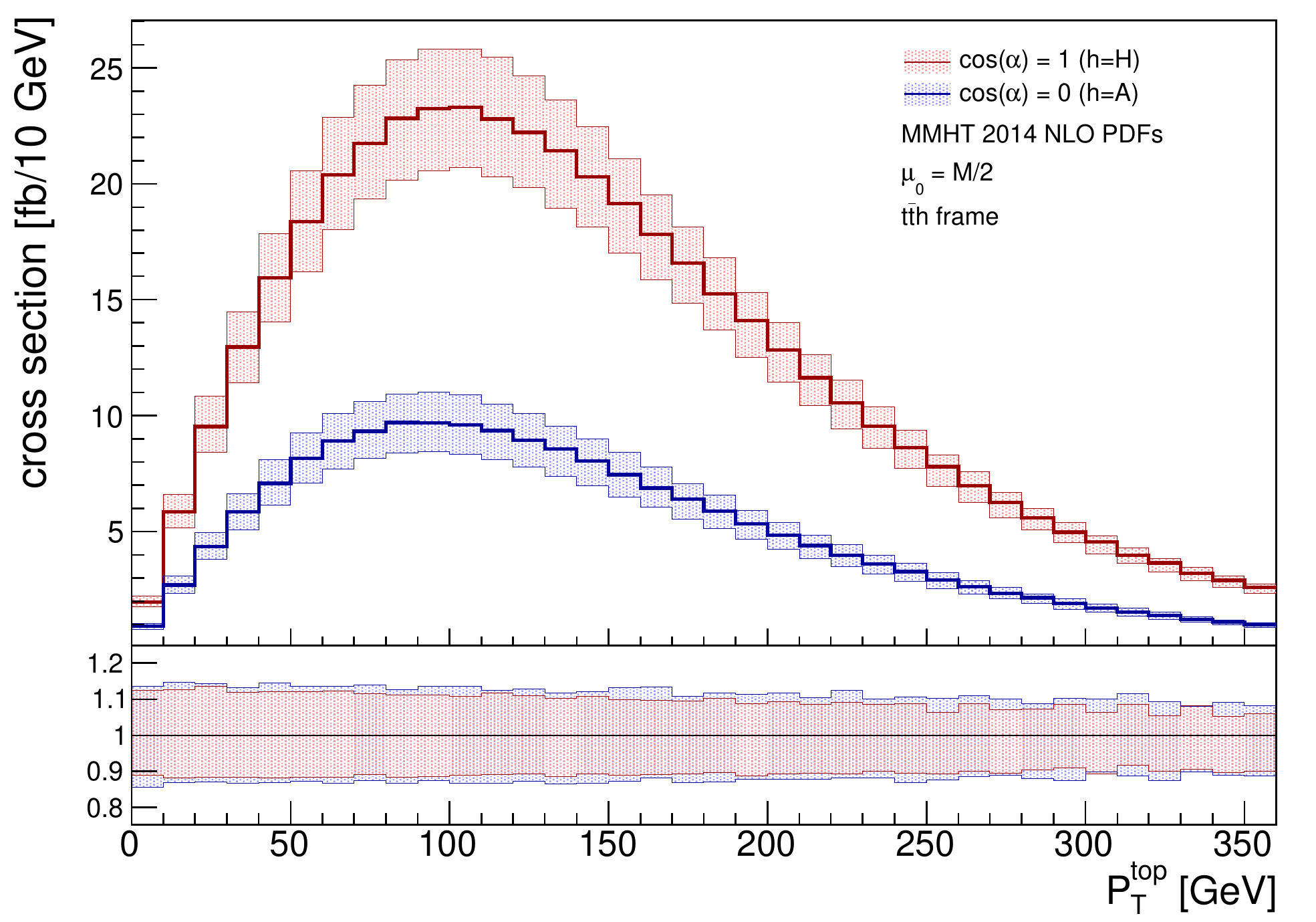,height=4.5cm,clip=} \\
\end{tabular}
\caption{Parton-level kinematic distributions in the $t\bar{t}h$ center-of-mass system. The $h$ boson and top-quark pseudorapidities are shown in the upper left and right panels, respectively. The $h$ boson and top-quark transverse momenta are shown in the bottom left and right panels, respectively.}
\label{fig:partonlevel}
\end{center}
\end{figure*}

\begin{figure*}
\begin{center}
\begin{tabular}{cc}
\hspace*{-1mm} \epsfig{file=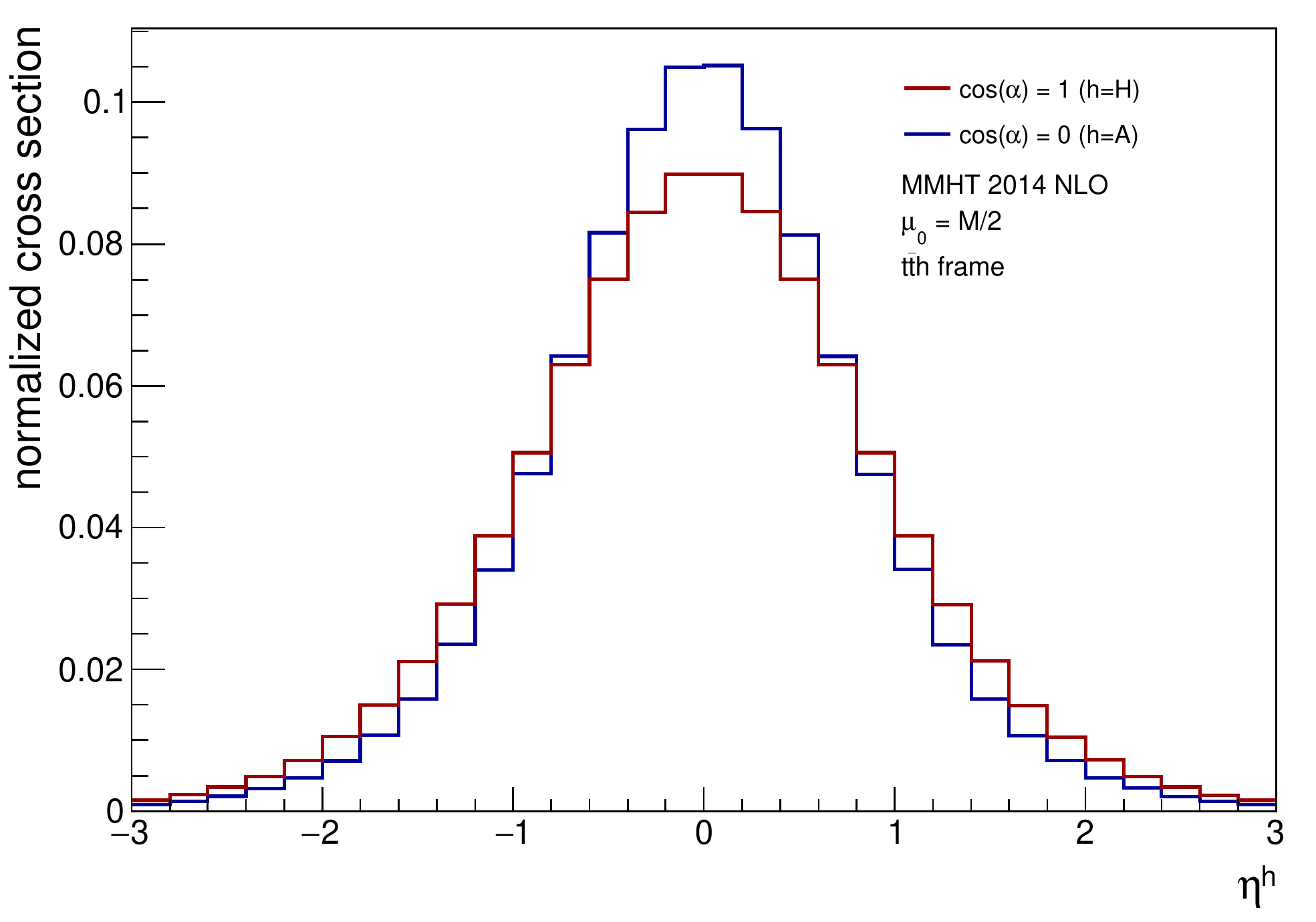,height=4.5cm,clip=} & 
                           \epsfig{file=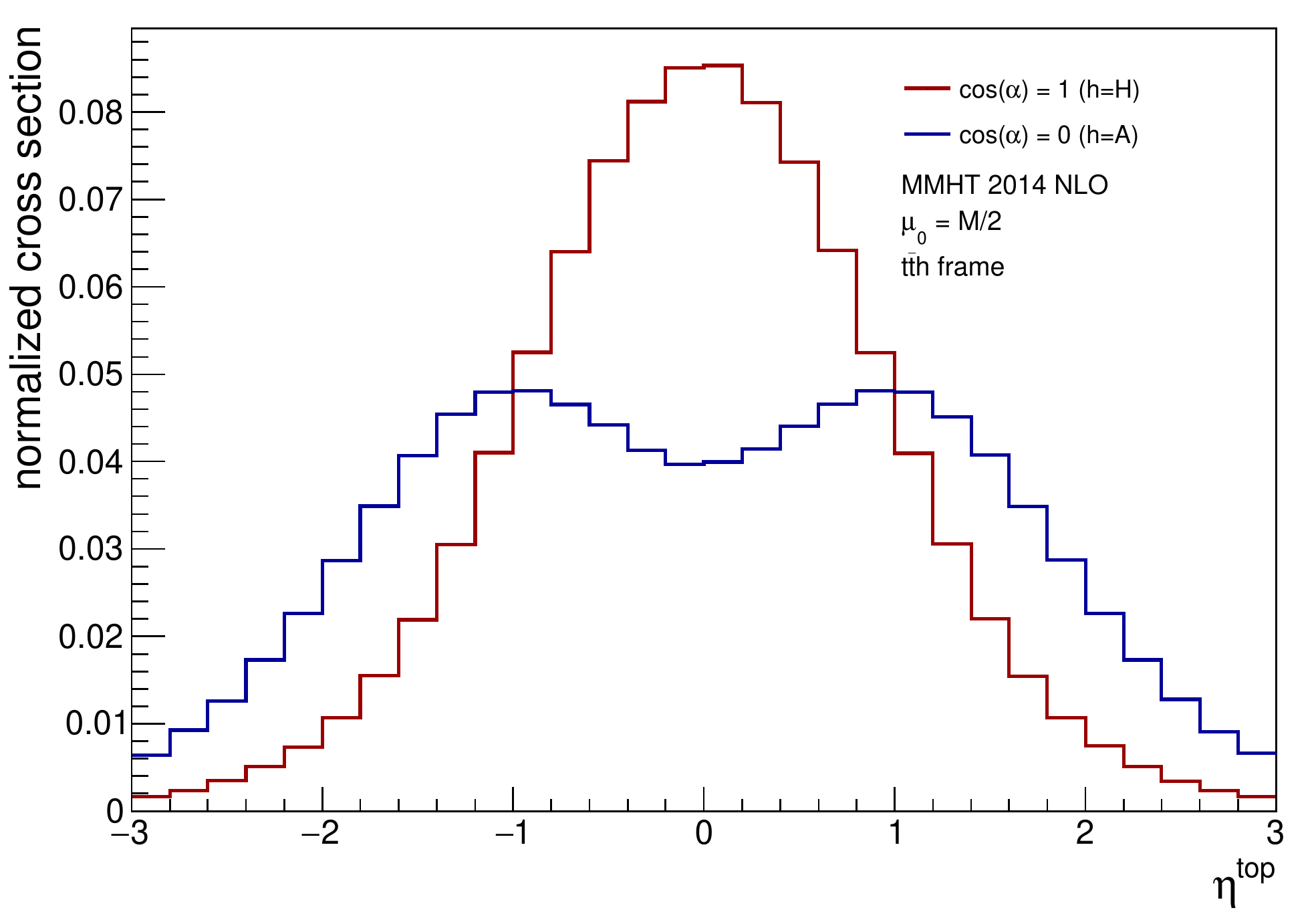,height=4.5cm,clip=} \\
\hspace*{-1mm} \epsfig{file=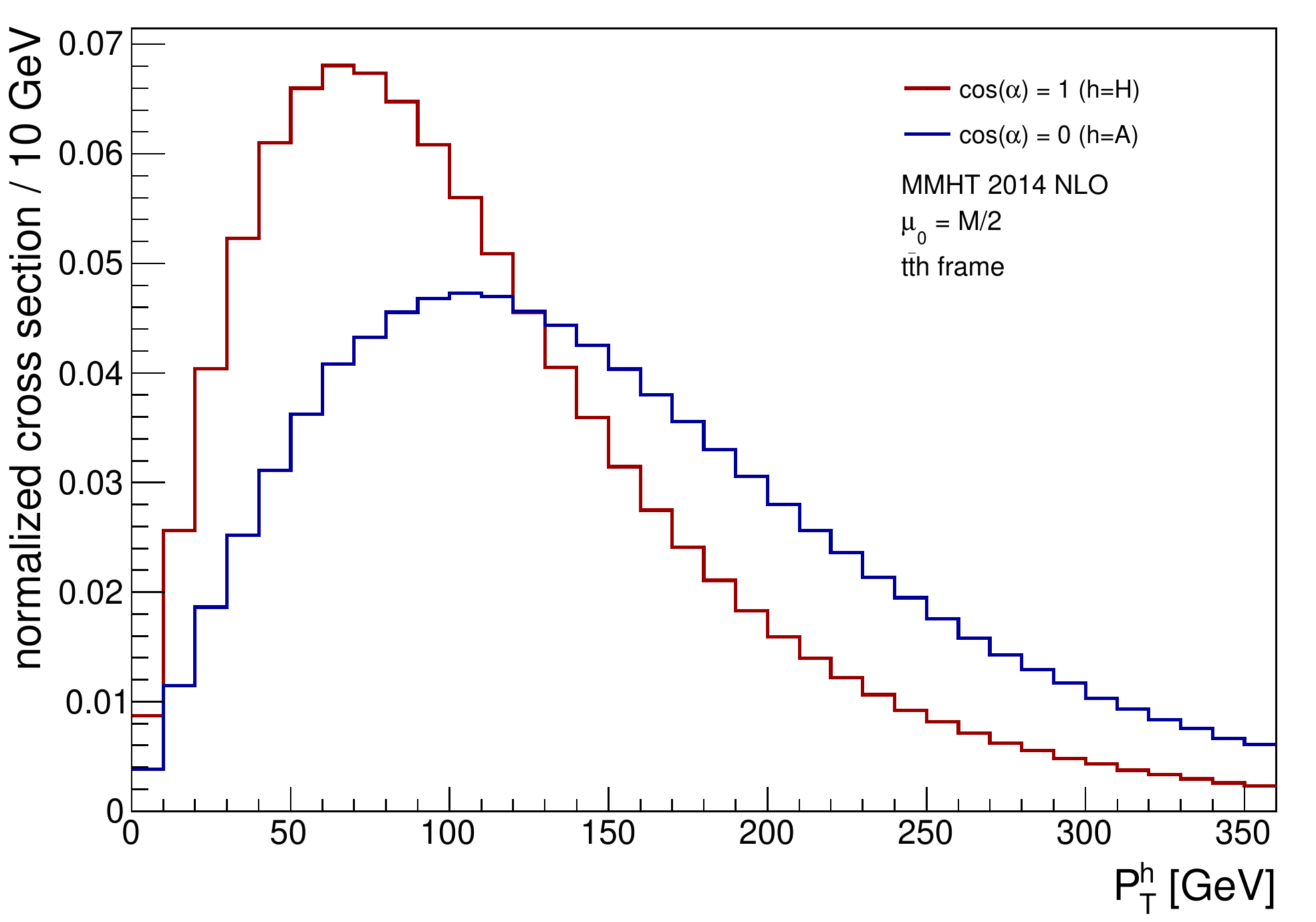,height=4.5cm,clip=} &
                           \epsfig{file=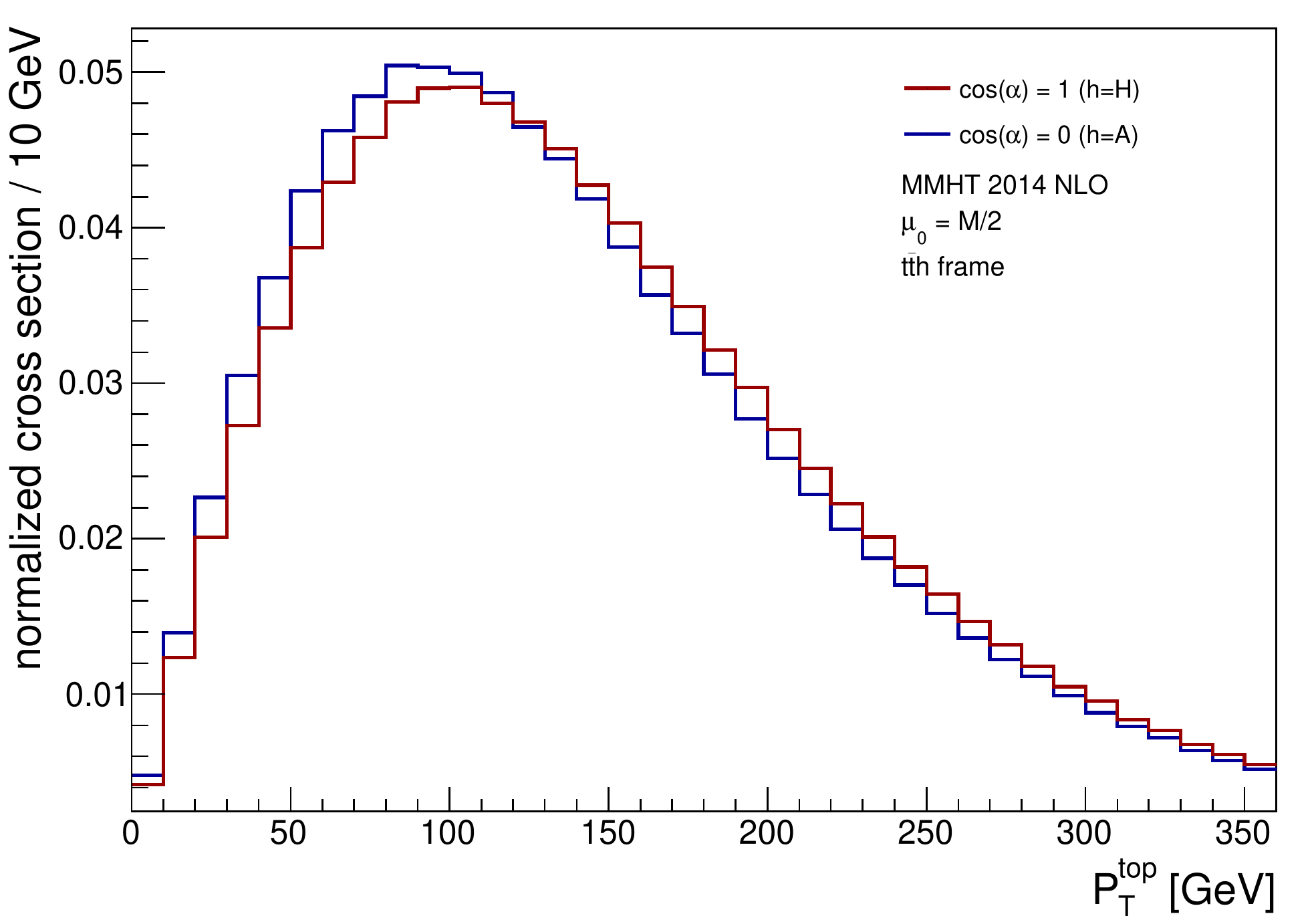,height=4.5cm,clip=} \\
\end{tabular}
\caption{Normalized parton-level kinematic distributions in the $t\bar{t}h$ center-of-mass system. The $h$ boson and top-quark pseudorapidities are shown in the upper left and right panels, respectively. The $h$ boson and top-quark transverse momenta are shown in the bottom left and right panels, respectively.}
\label{fig:normpartonlevel}
\end{center}
\end{figure*}

The distributions which are differential with respect to the top quark or
$h$-boson transverse momentum and pseudorapidity in the $t \bar{t} h$ center-of-mass system are shown in Figure~\ref{fig:partonlevel}. 
The red bands represent the distributions in the purely scalar case, while the blue bands represent the distribution in the purely pseudoscalar case.
The width of each of the bands represents the scale uncertainty obtained by varying the factorization scale as described above. The inset below each plot shows the relative scale uncertainty in each bin. By looking at these insets one can see that the scale uncertainty in percentage is very similar in the scalar and pseudoscalar cases. The shape of the differential distributions considered in Figure~\ref{fig:partonlevel} is more easily seen in Figure~\ref{fig:normpartonlevel}, where the distributions are normalized to the total cross section, i.e. the value of the distribution in each bin is divided by the total cross section, so that the heights of all of the columns in the histogram (including the ones that fall out of the range shown in the figure) add up to one. Since for normalized distributions the scale uncertainty bands become very thin, only the distributions calculated at $\mu_f = M/2$ are shown in Figure~\ref{fig:normpartonlevel}. From the figure, one can see that the scalar case (in red) and the pseudoscalar case (in blue) have similar shapes for the distributions differential with respect to the top-quark transverse momentum $p_T^t$ and boson pseudorapidity $\eta_h$. In contrast, the shape of the distributions differential with respect to the boson transverse momentum $p_T^h$ and especially top pseudorapidity $\eta_t$ are quite different in the scalar and pseudoscalar cases. The $\eta_t$ distribution   shows one single maximum at $\eta_t = 0$ in the scalar case while it shows two distinct symmetric maxima for positive and negative rapidity in the pseudoscalar case. For what concerns the $p_T$ distributions, these results are, as expected, similar to the ones found in the laboratory frame and discussed in~\cite{Broggio:2017oyu}. Conversely, the top-quark pseudorapidity  distributions in the scalar and pseudoscalar cases calculated in the $t \bar{t} h$ rest frame  (shown in the top right corner of Figure~\ref{fig:normpartonlevel}) show a more marked difference than in the laboratory frame. This indicates that other differential distributions evaluated in the $t \bar{t} h$ rest frame might help to discriminate between the case of a scalar and pseudoscalar component in the top-quark Yukawa coupling.

Indeed, in addition to the distributions described above, several other observables were introduced to isolate the pseudoscalar component of the top-quark Yukawa coupling in~\cite{Gunion:1996xu}. However, those differential distributions were designed to be measured in the laboratory frame. For the purposes of this paper, it is useful to extend the definition of two of them, $b^f_2$ and $b^f_4$, to the $t \bar{t} h$ rest frame. In particular, one can define
\begin{eqnarray}
b^{f}_2(i,j) &=& \frac{( \vec{p}^{\; f}_{i} \times \hat{k}_z ).( \vec{p}^{\; f}_{j} \times \hat{k}_z )}{ |\vec{p}^{\; f}_{i}|  |\vec{p}^{\; f}_{j}| } \, , \label{eq:b2} \\
b^{f}_4(i,j) &=& {\frac{ p^{\; f}_{i,z} \; \;p^{\; f}_{j,z} }{ |\vec{p}^{\; f}_{i}|  |\vec{p}^{\; f}_{j}| }, }
\label{eq:b4}
\end{eqnarray}
where $i,j$= $\{t, \bar{t}, h\}$ (without repetition), $\vec{p}^{\; f}_{i}$ ({$p^{\; f}_{i,z}$}) is the $i^{th}$ particle total ($z$-component) momentum measured in the $f$ frame (with  $f = t \bar{t} h$  if the observable is evaluated in the $t \bar{t} h$ rest frame while $f = \mbox{LAB}$ if the observable is evaluated in the laboratory frame). Finally,  $\hat{k}_z$ corresponds to the beam line, which defines the $z$-direction.
In the context of this section, the variables in Eqs.~(\ref{eq:b2},\ref{eq:b4}) are considered exclusively as measured in the $t \bar{t} h$ frame. 

\begin{figure*}
\begin{center}
\begin{tabular}{ccc}
\hspace*{-1mm}\epsfig{file=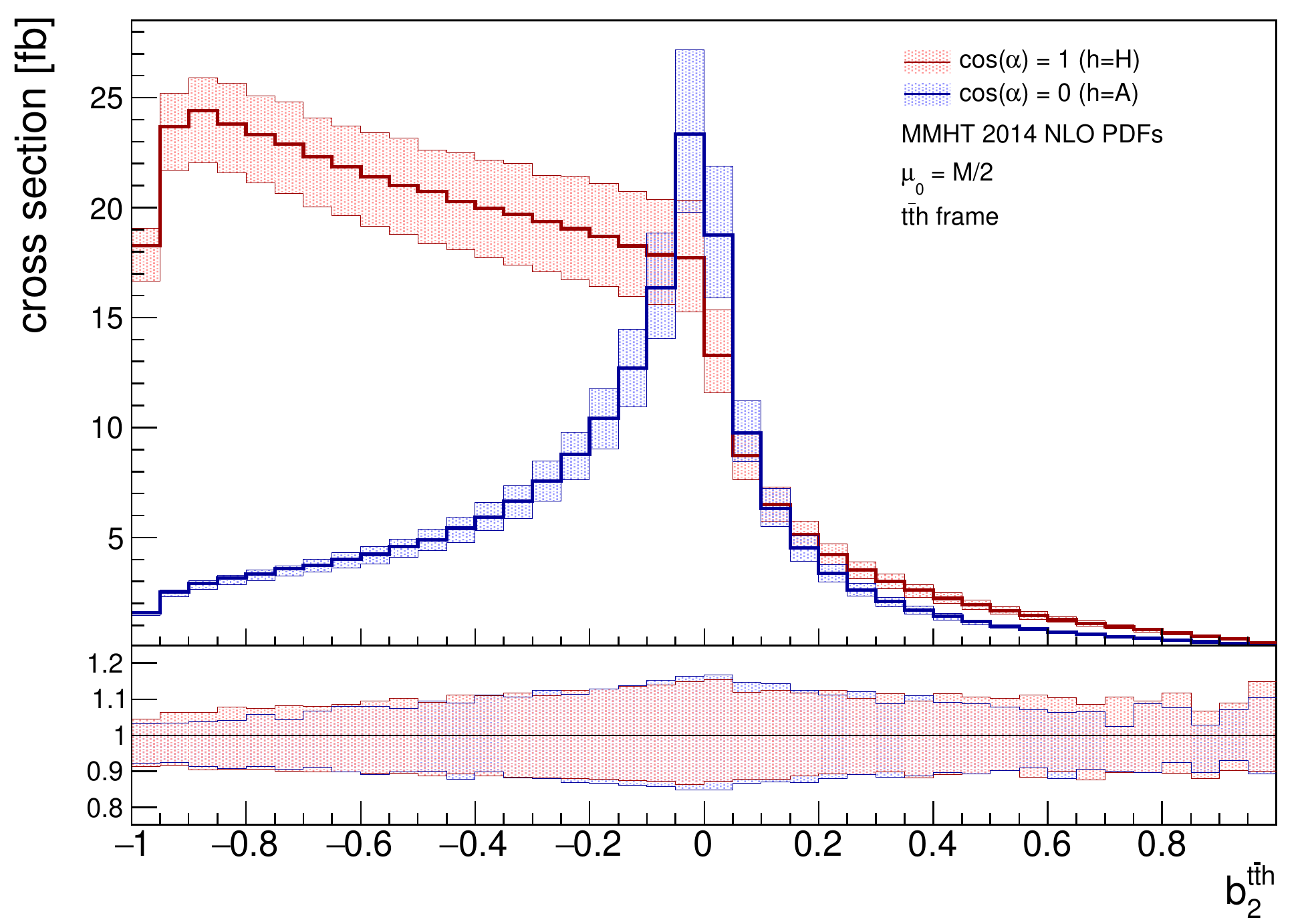,height=4.cm,clip=} & 
\hspace*{-1mm}\epsfig{file=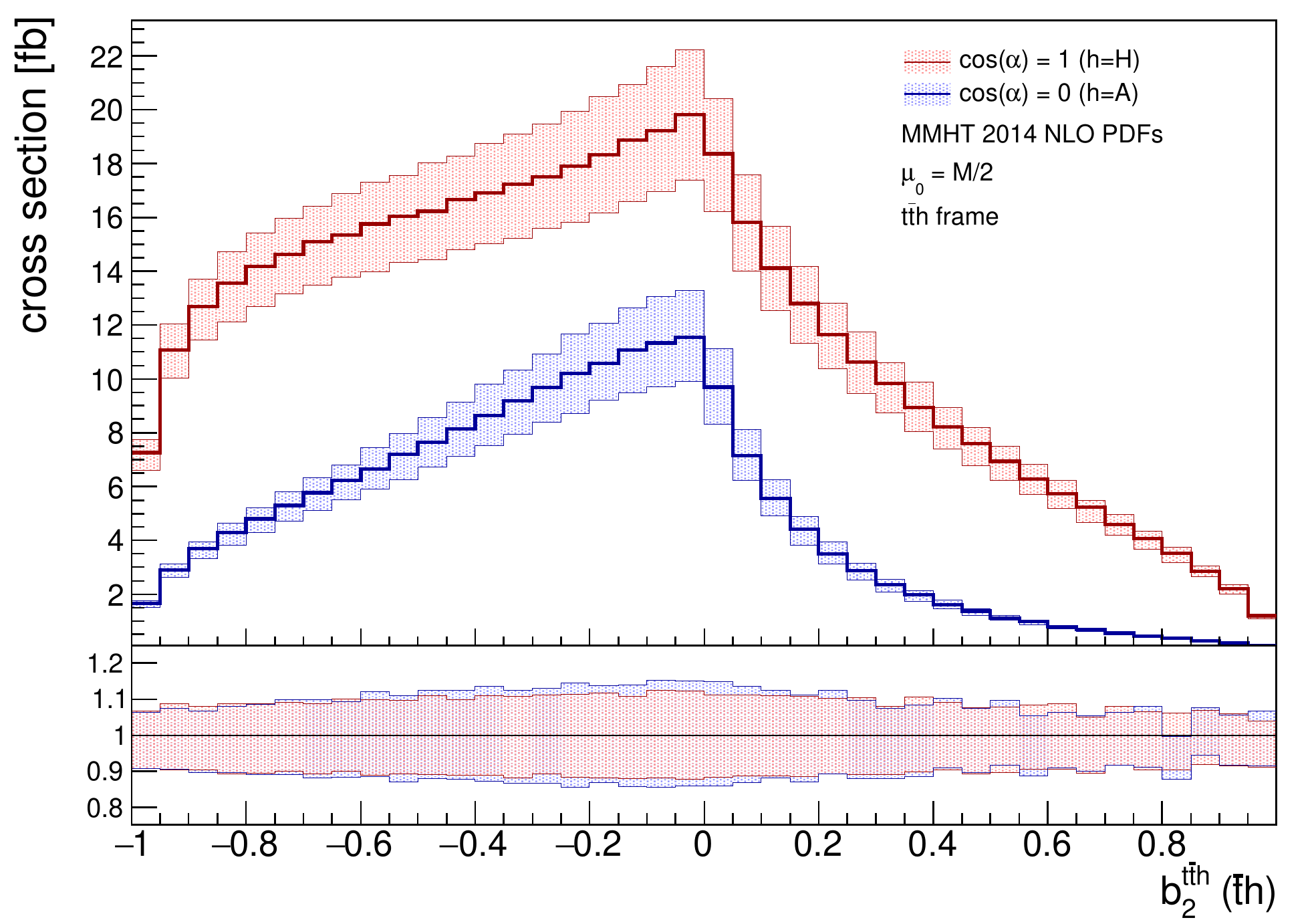,height=4.cm,clip=} &
\hspace*{-1mm}\epsfig{file=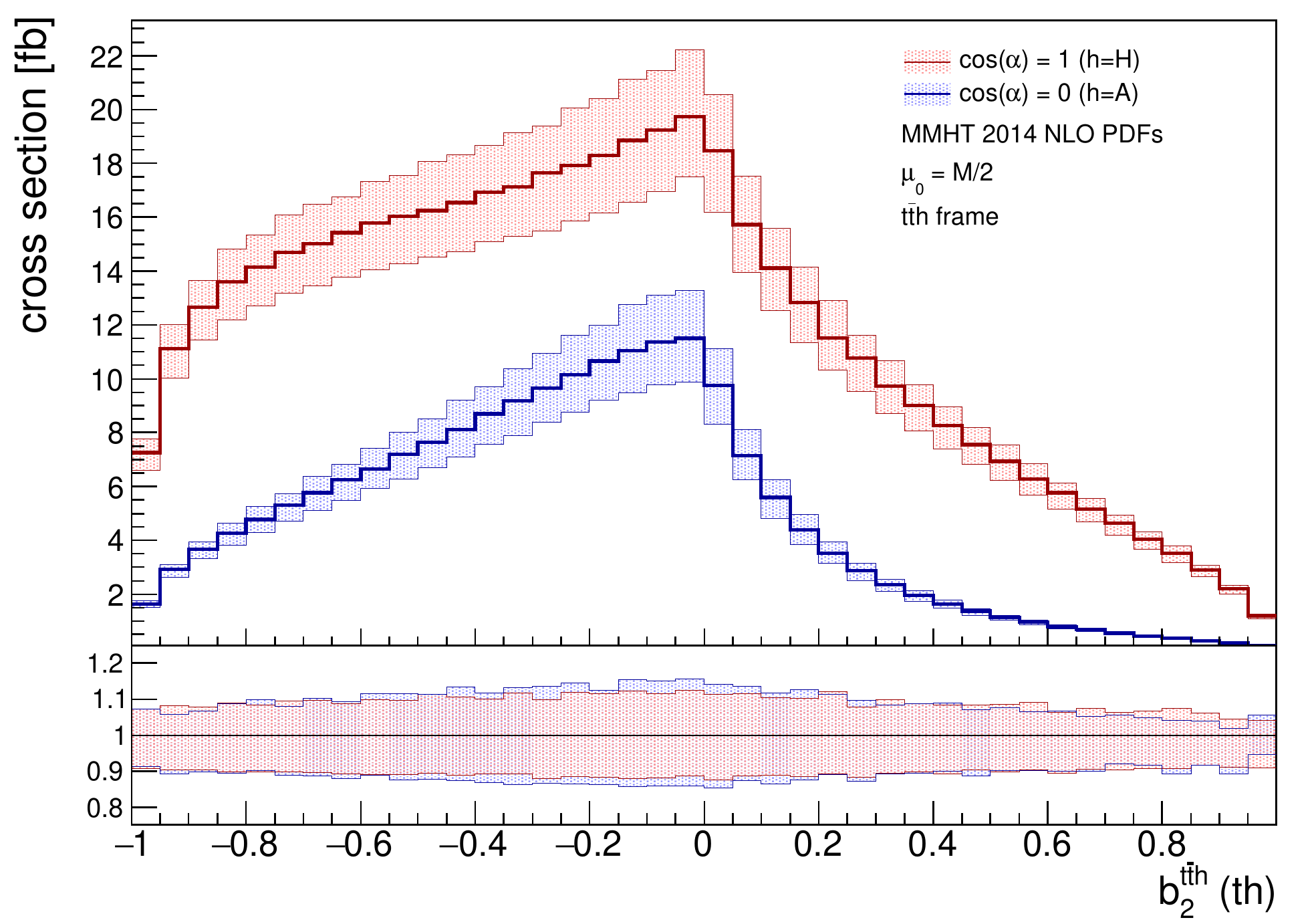,height=4.cm,clip=}  \\
\hspace*{-1mm}\epsfig{file=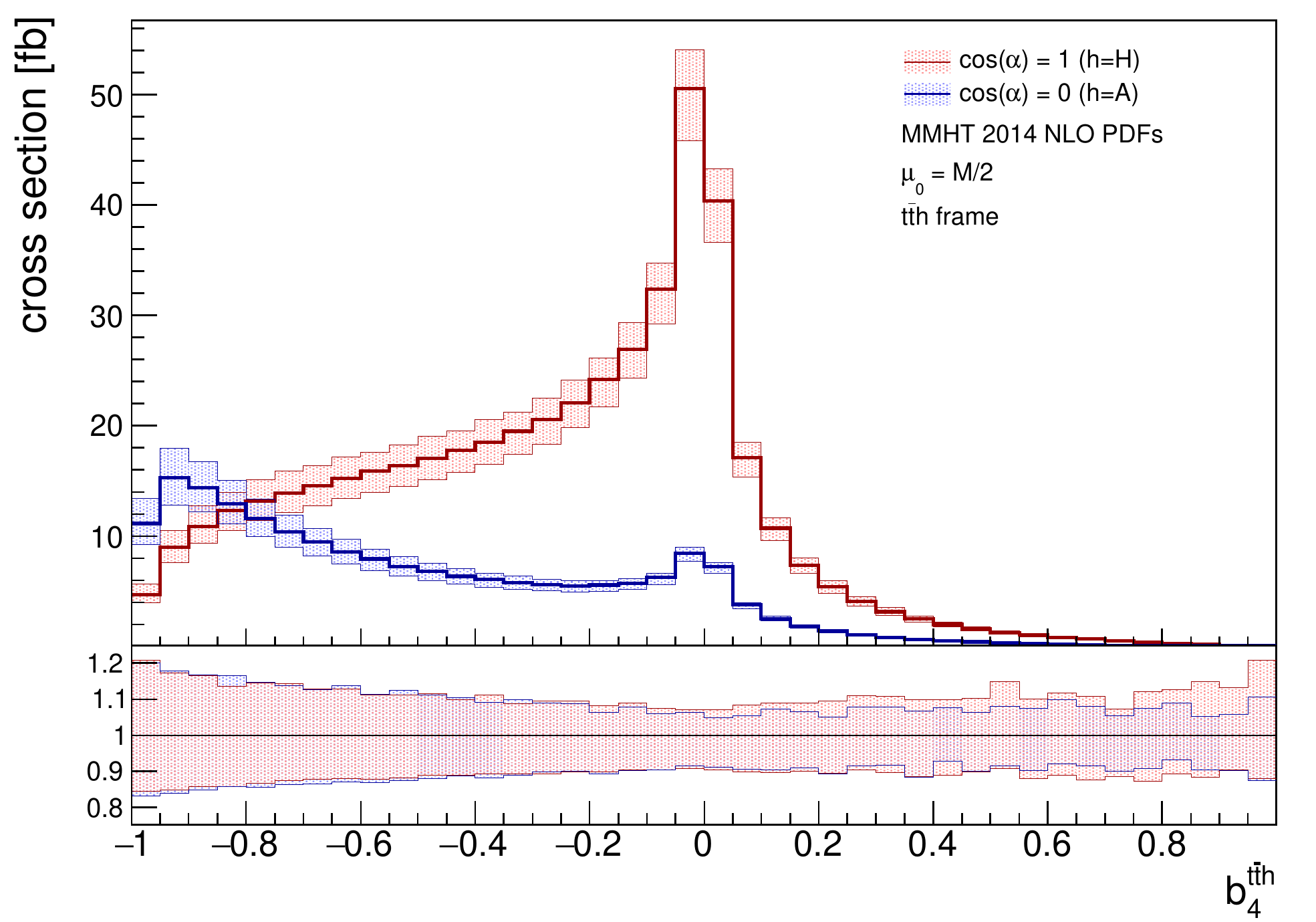,height=4.cm,clip=} & 
\hspace*{-1mm}\epsfig{file=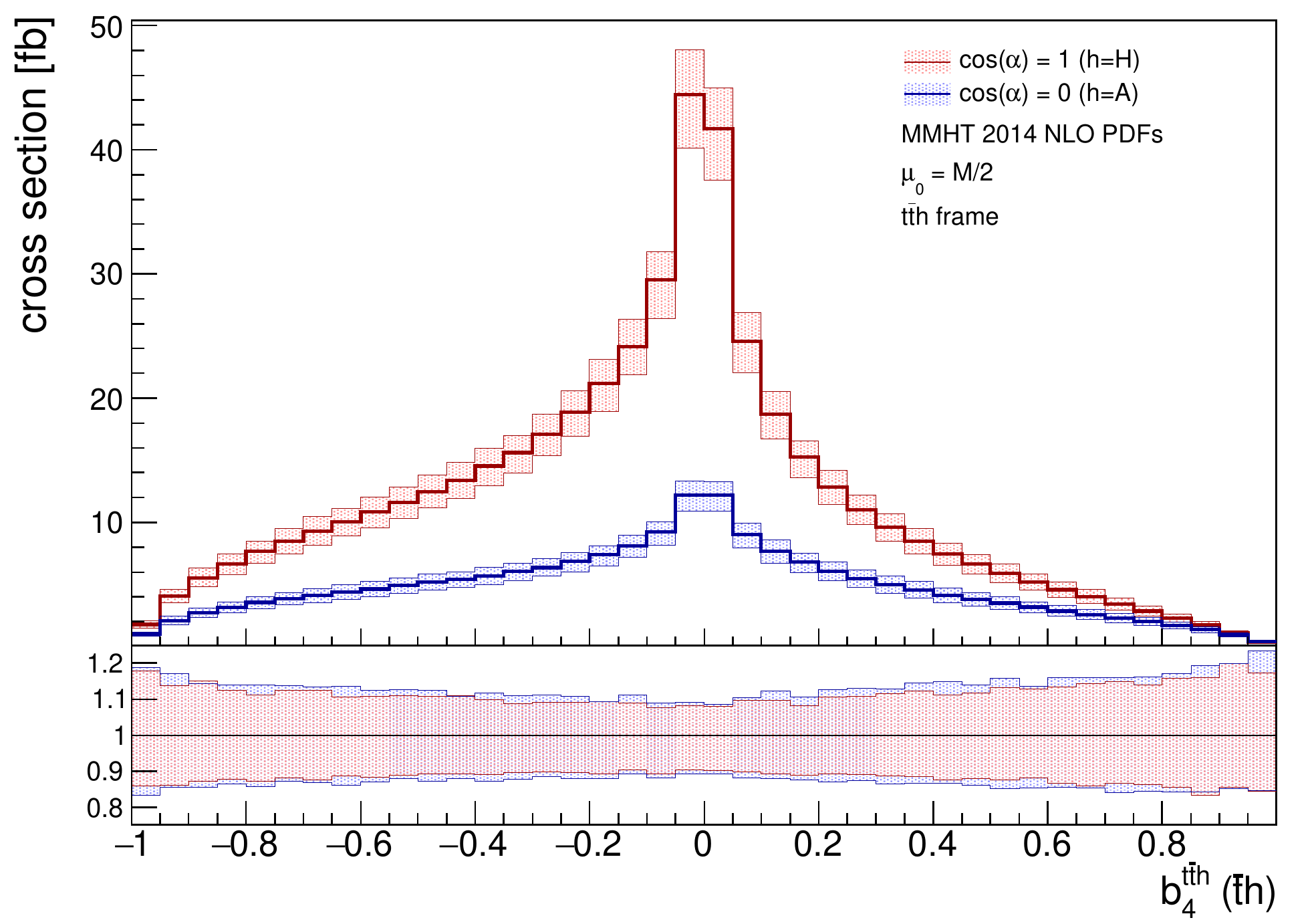,height=4.cm,clip=} &
\hspace*{-1mm}\epsfig{file=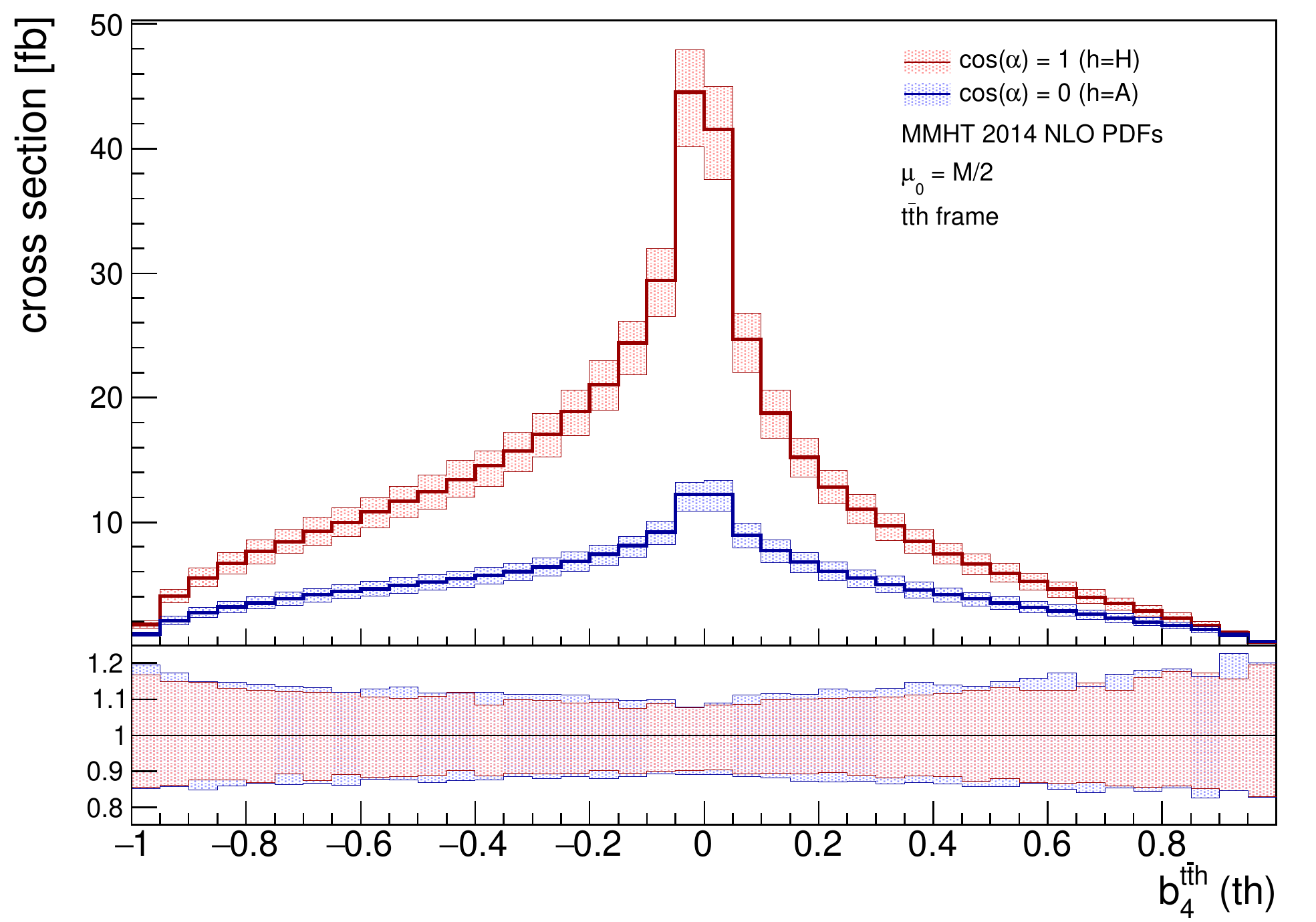,height=4.cm,clip=}
\end{tabular}
\caption{Parton-level $b^{t\bar{t}h}_2$ and $b^{t\bar{t}h}_4$ distributions in the $t\bar{t}h$ center-of-mass system. The $b^{t\bar{t}h}_2$ distributions for  $t\bar{t}$, $\bar{t}h$ and $th$ are shown in the upper left, middle and right panels, respectively. The $b^{t\bar{t}h}_4$ distributions for $t\bar{t}$, $\bar{t}h$ and $th$ are shown in the bottom left, middle and right panels, respectively. For completeness, distributions for both $th$ and $\bar th$ are included, although they are equivalent.}
\label{fig:partonlevelBs}
\end{center}
\end{figure*}

\begin{figure*}
\begin{center}
\begin{tabular}{ccc}
\hspace*{-1mm}\epsfig{file=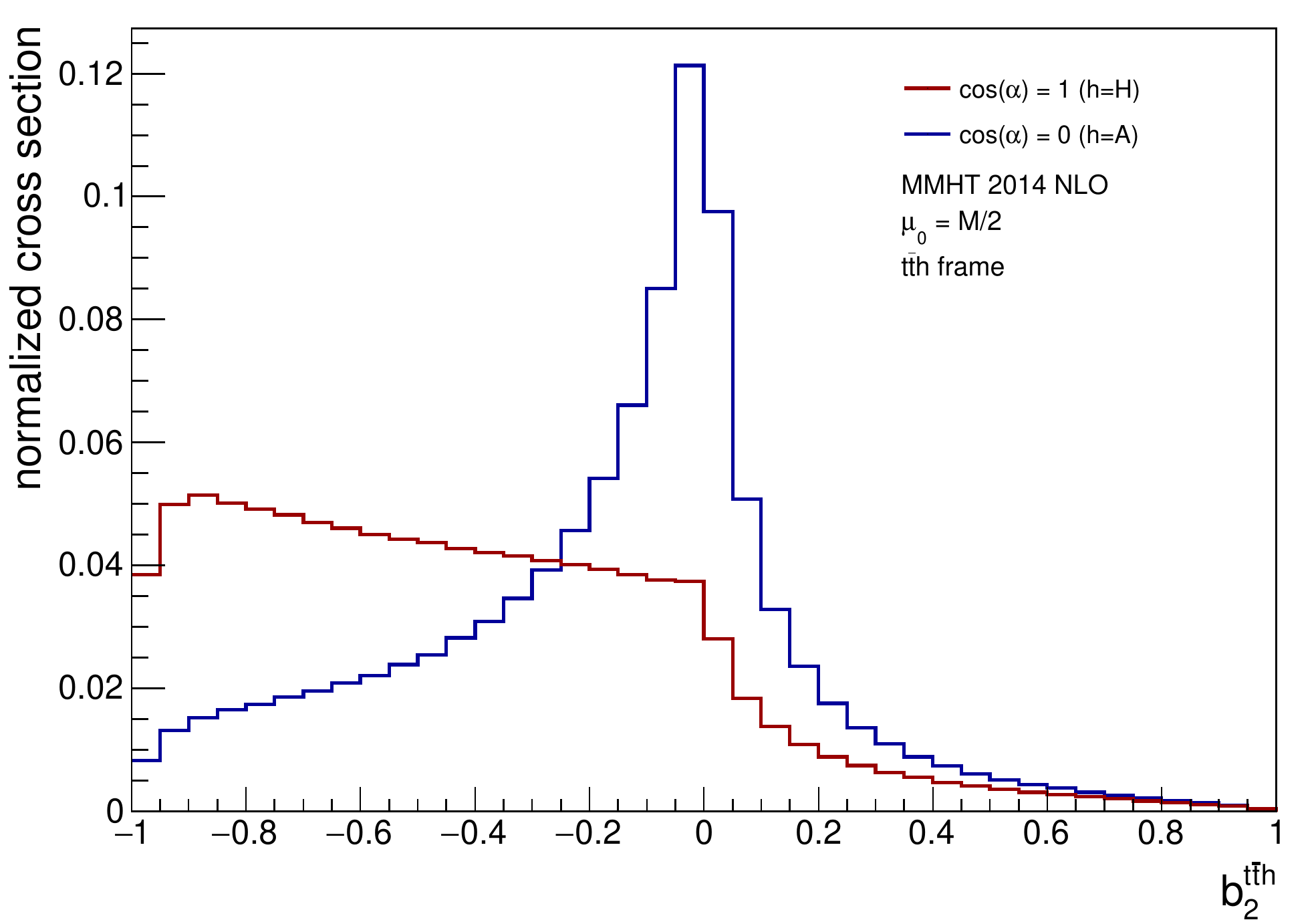,height=4.cm,clip=} & 
\hspace*{-1mm}\epsfig{file=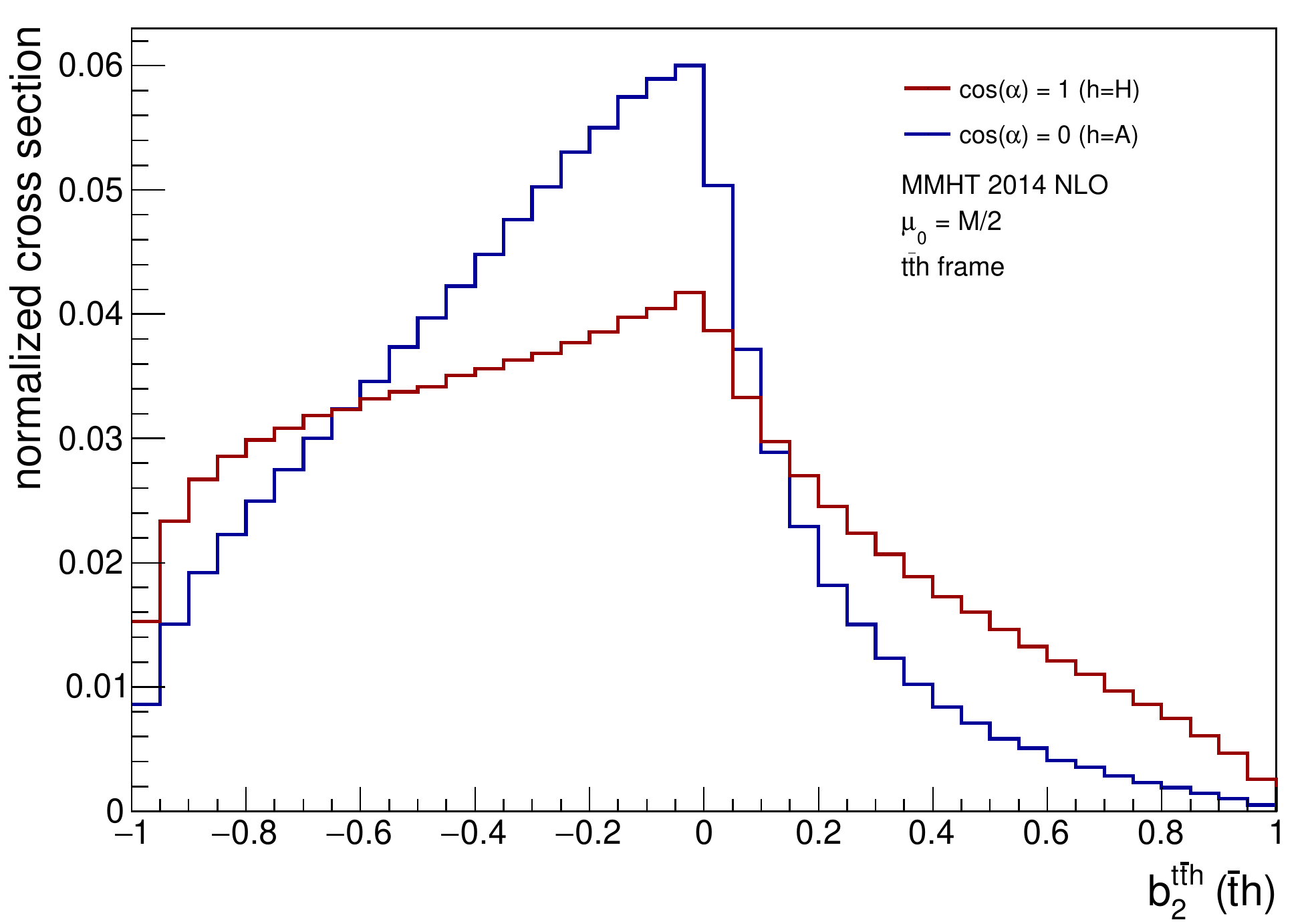,height=4.cm,clip=} &
\hspace*{-1mm}\epsfig{file=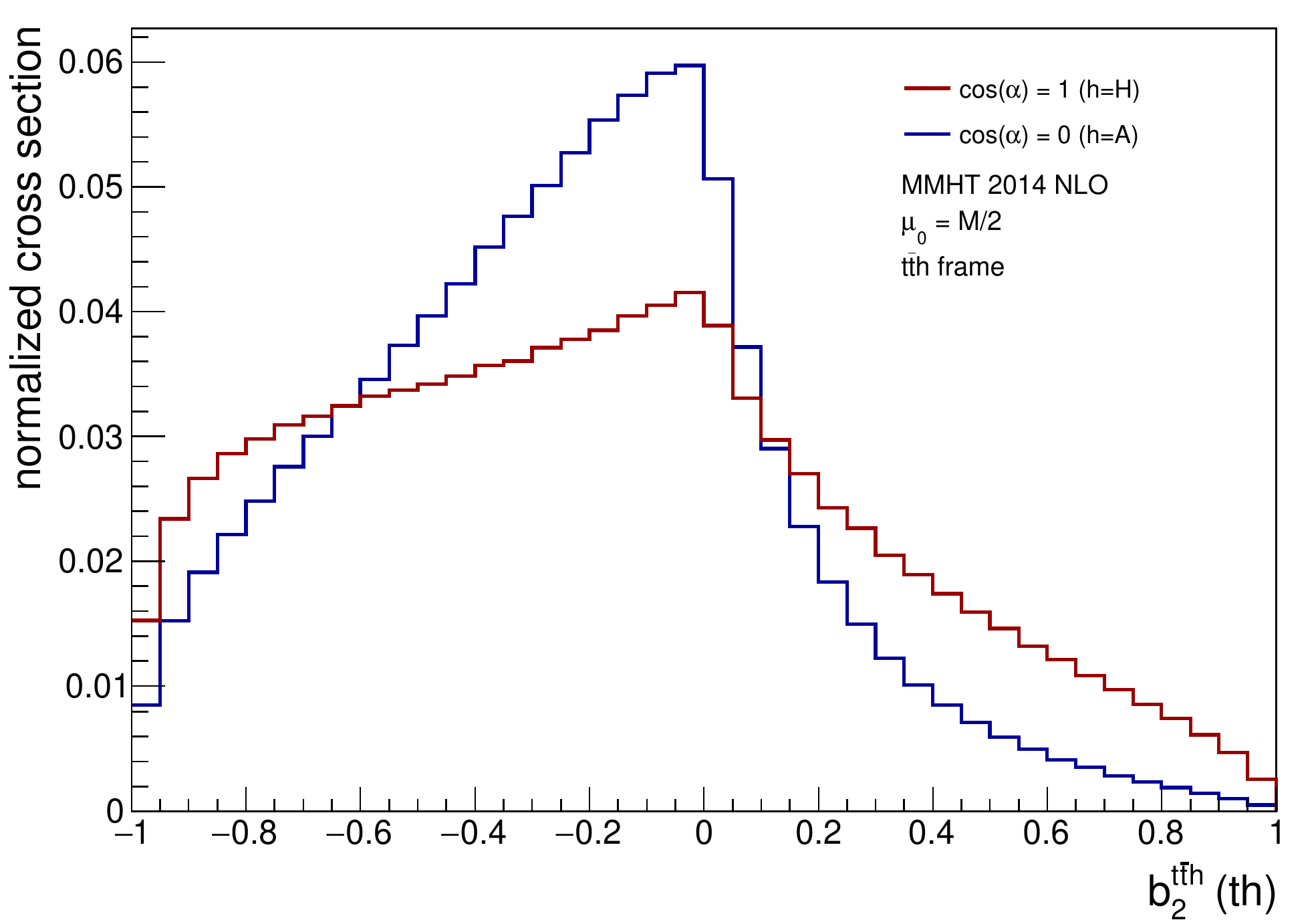,height=4.cm,clip=}  \\
\hspace*{-1mm}\epsfig{file=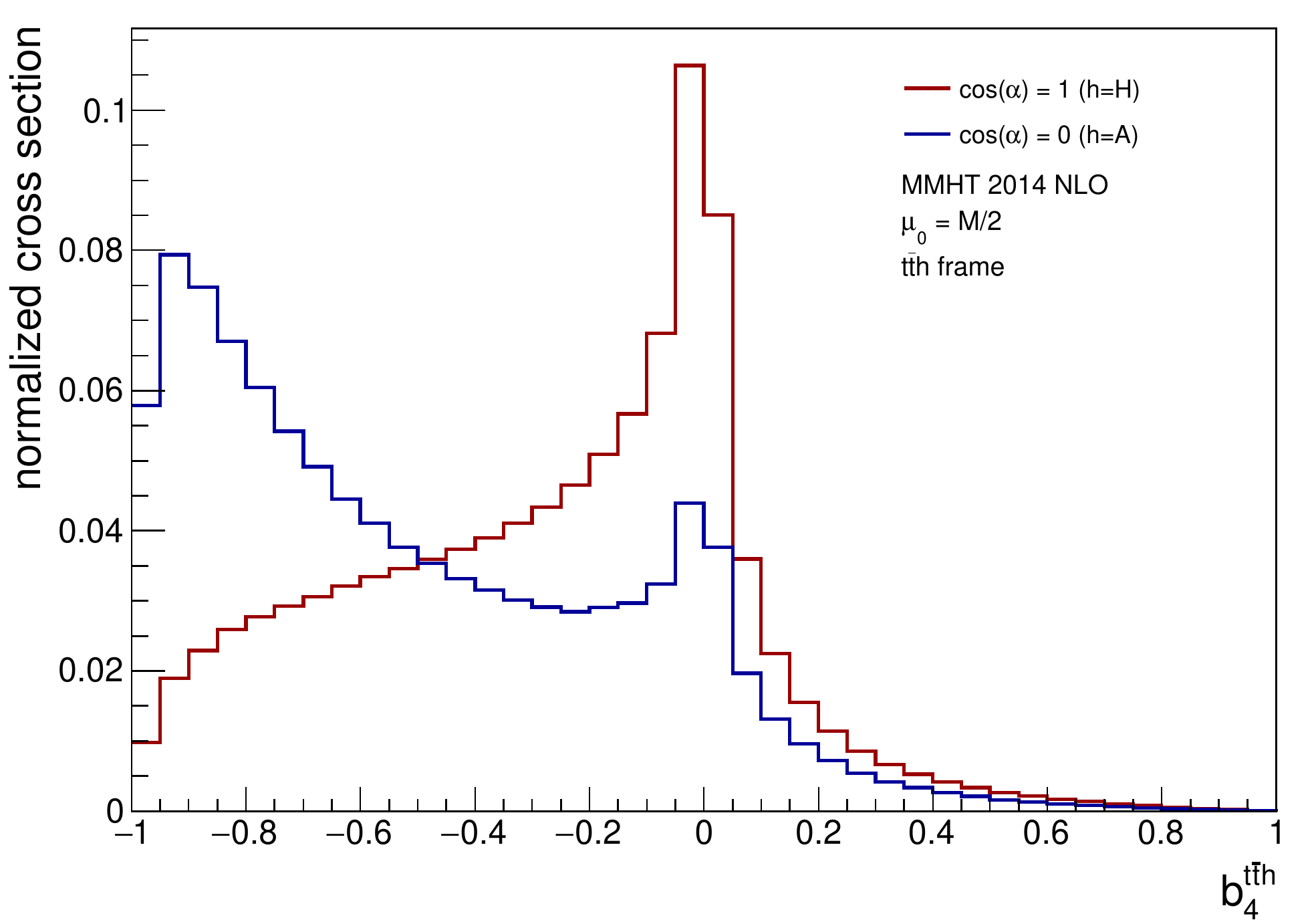,height=4.cm,clip=} & 
\hspace*{-1mm}\epsfig{file=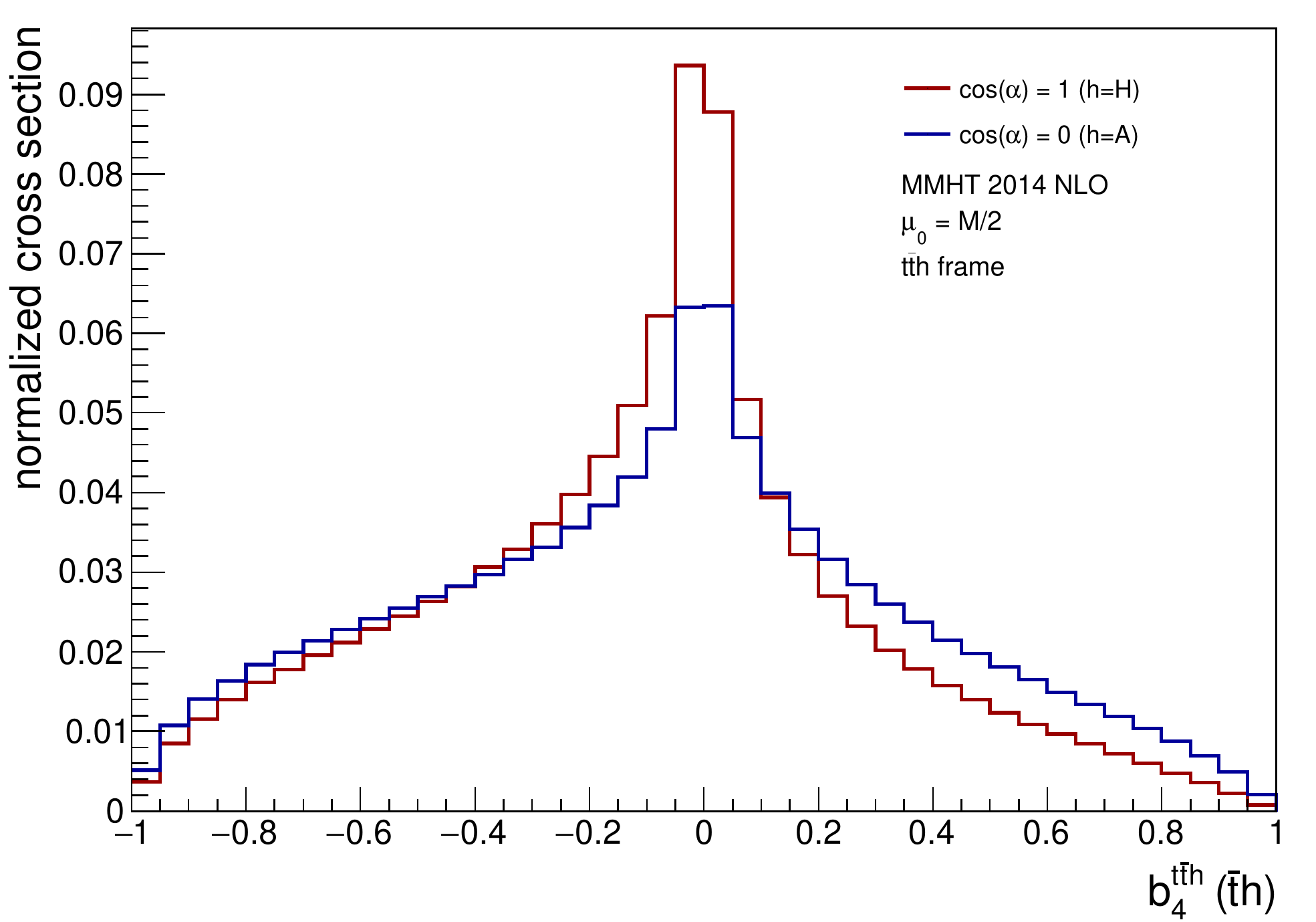,height=4.cm,clip=} &
\hspace*{-1mm}\epsfig{file=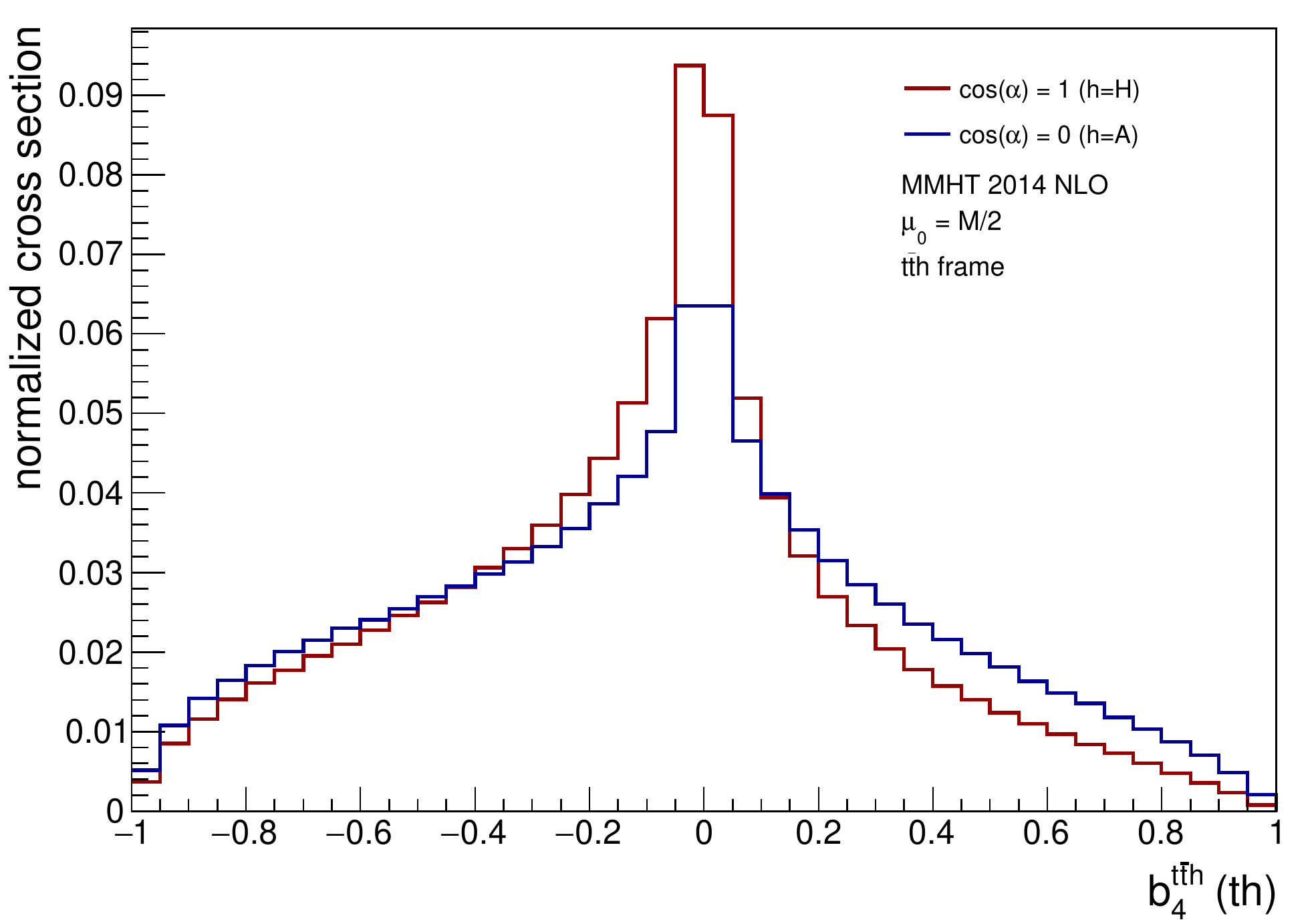,height=4.cm,clip=}
\end{tabular}
\caption{Normalized parton-level $b^{t\bar{t}h}_2$ and $b^{t\bar{t}h}_4$ distributions in the $t\bar{t}h$ center-of-mass system. The $b^{t\bar{t}h}_2$ distributions for  $t\bar{t}$, $\bar{t}h$ and $th$ are shown in the upper left, middle and right panels, respectively. The $b^{t\bar{t}h}_4$ distributions for $t\bar{t}$, $\bar{t}h$ and $th$ are shown in the bottom left, middle and right panels, respectively. For completeness, distributions for both $th$ and $\bar th$ are included, although they are equivalent.}
\label{fig:normpartonlevelBs}
\end{center}
\end{figure*}

Figure~\ref{fig:partonlevelBs} shows the differential distributions with respect to $b^{t \bar{t} h}_2(i,j)$ and $b^{t \bar{t} h}_4(i,j)$ for the three possible choices of $i,j$. As in the previous figures, the red bands correspond to the pure scalar case while the blue bands correspond to the pure pseudoscalar case. The bands' width indicates the scale uncertainty. The inset below each plot shows the width of the scale uncertainty band in each bin as a fraction of the central value in the bin. A better impression of the discriminating powers of these observables is obtained by looking at Figure~\ref{fig:normpartonlevelBs}, which shows the normalized distributions. By looking at that figure one can see that the shapes of the $b^{t \bar{t} h}_2(t,\bar{t})$ and $b^{t \bar{t} h}_4(t,\bar{t})$ distributions  (first column in the figure) are very different in the scalar (red line) and pseudoscalar (blue line) cases. On the contrary, the $b^{t \bar{t} h}_2(t,h)$ distributions and especially the $b^{t \bar{t} h}_4(t,h)$ distributions look quite similar in the case of a pure scalar and a pure pseudoscalar coupling. 

\begin{figure*}
\begin{center}
\begin{tabular}{cc}
\hspace*{-6mm} \epsfig{file=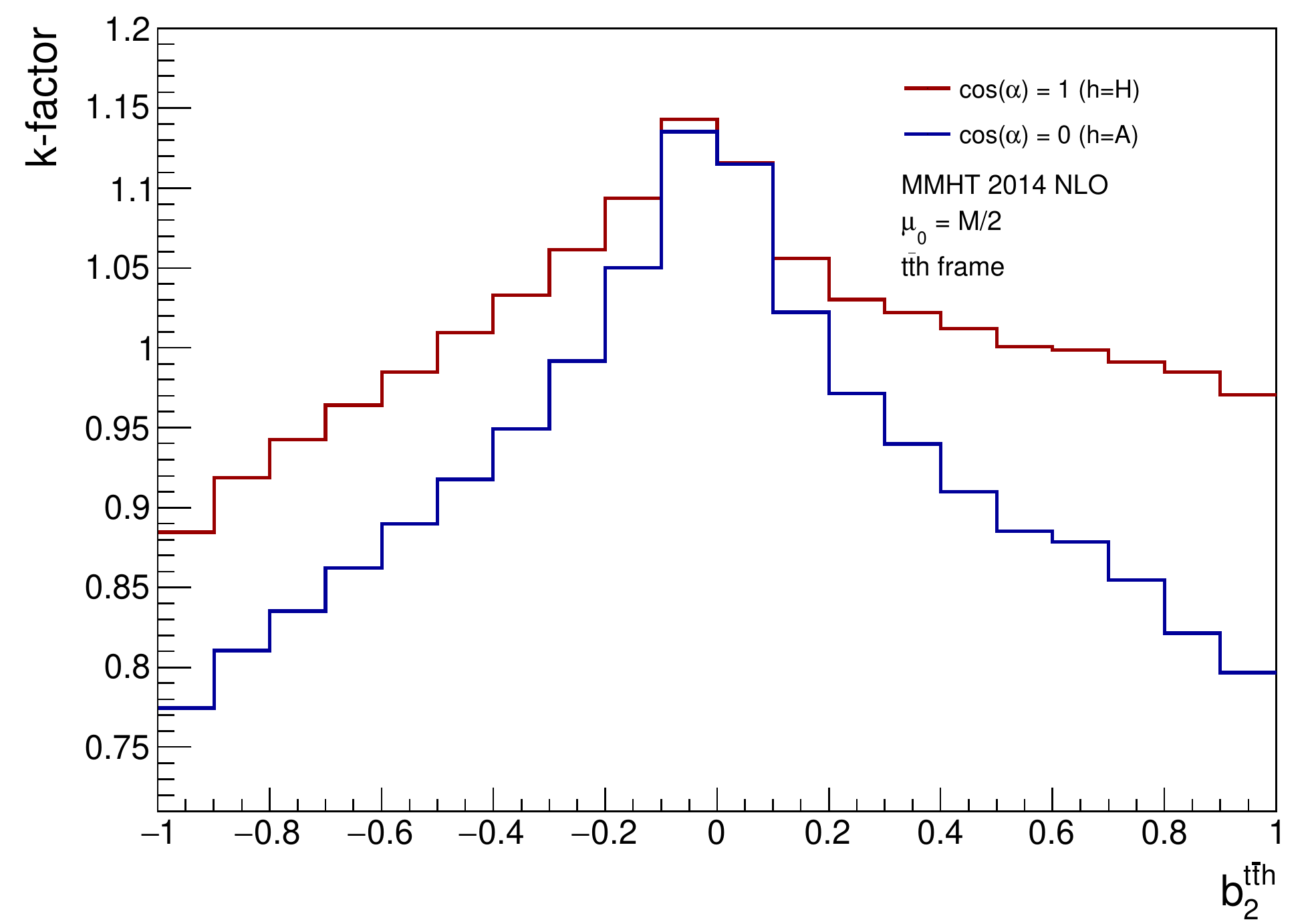,height=4.2cm,clip=} & 
                           \epsfig{file=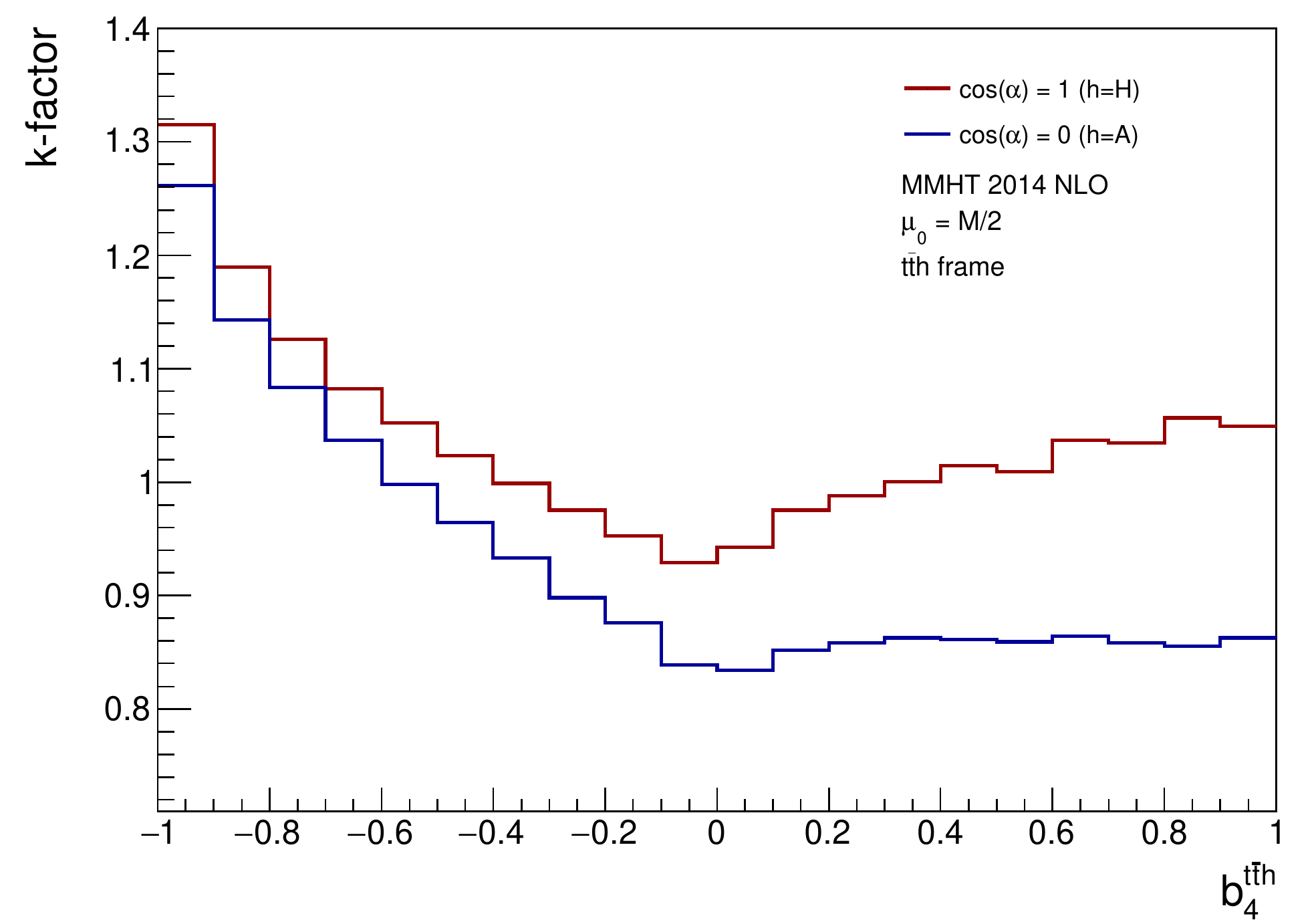,height=4.2cm,clip=} \\
\hspace*{-6mm} \epsfig{file=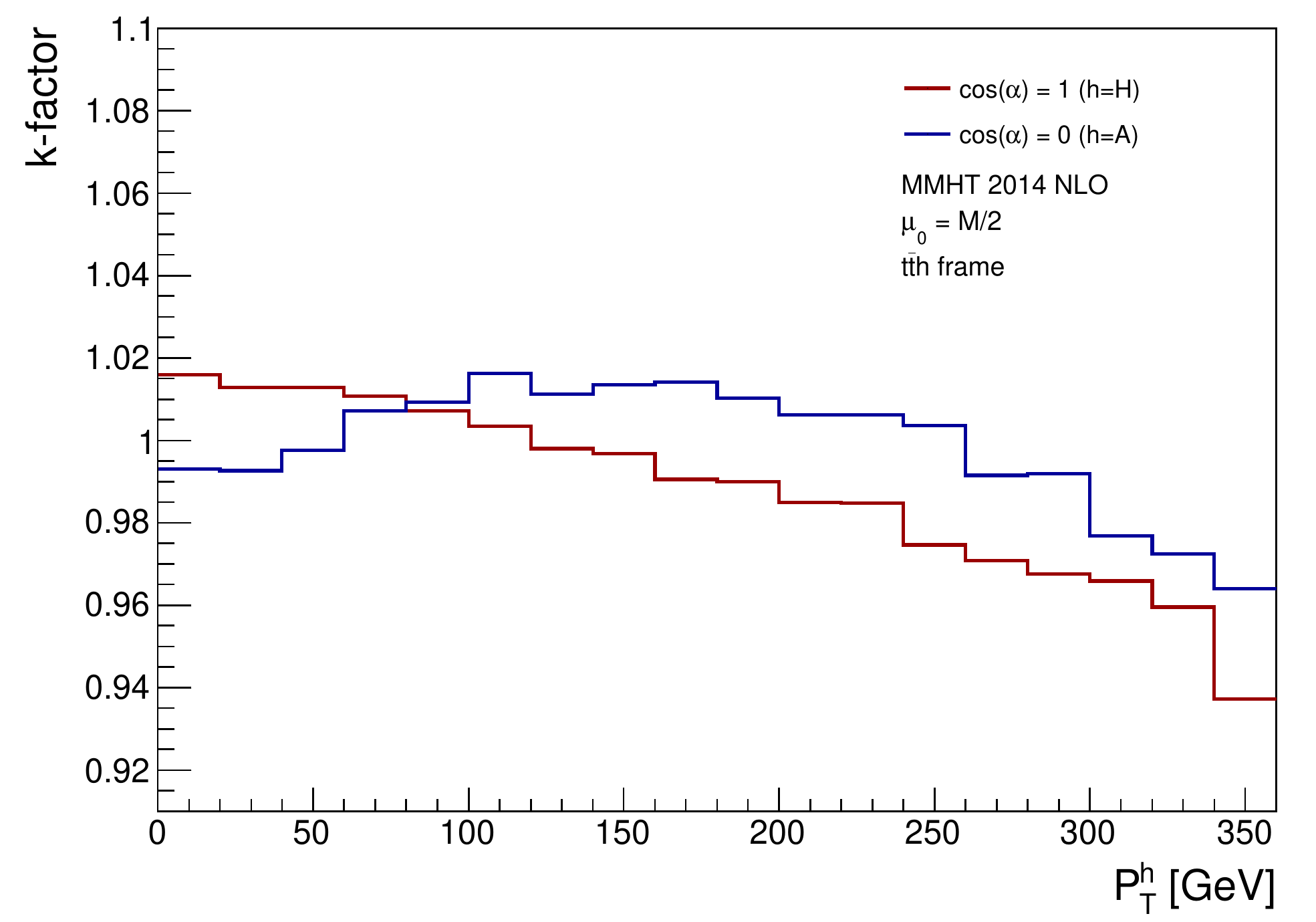,height=4.2cm,clip=} &
                           \epsfig{file=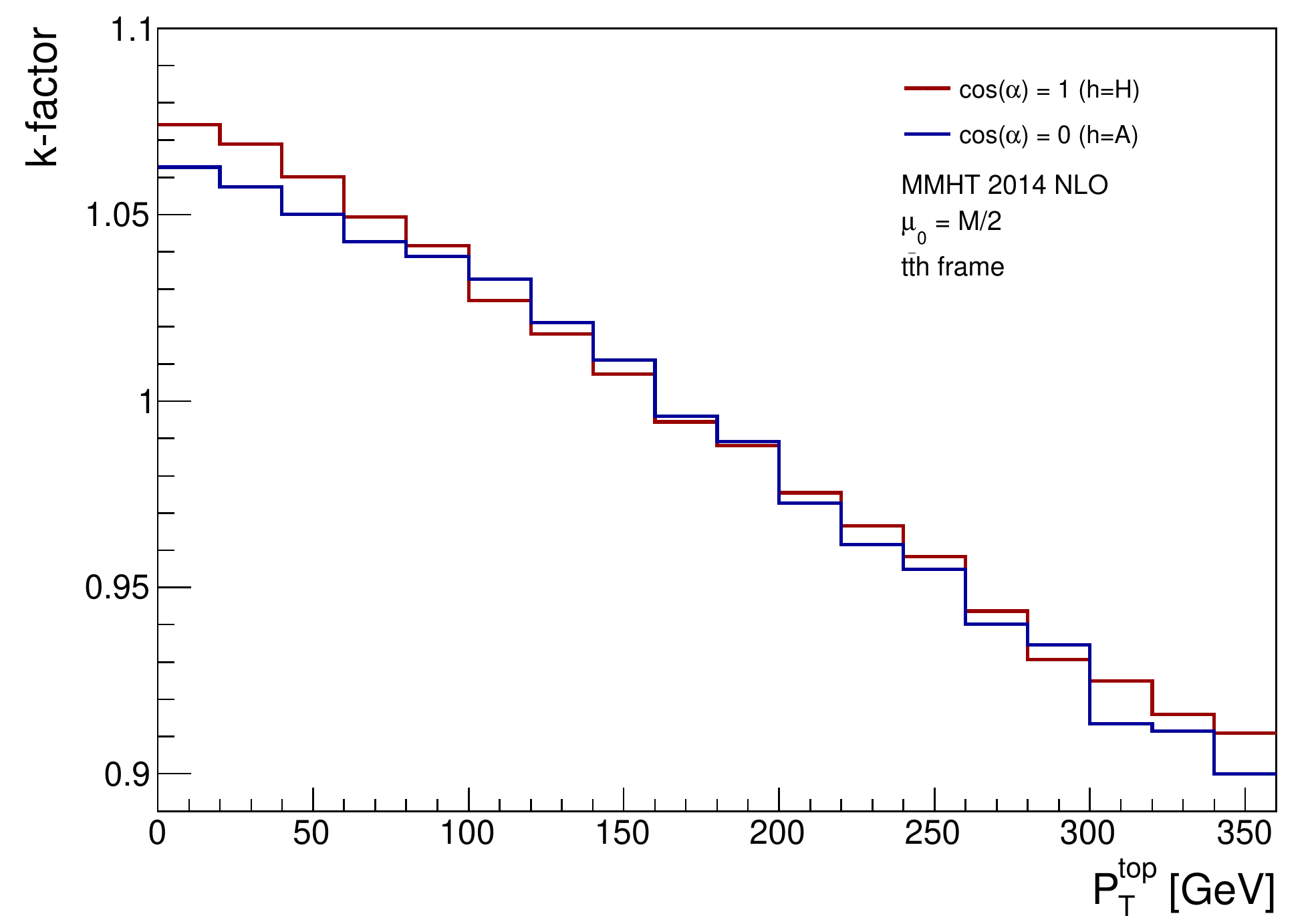,height=4.2cm,clip=} \\
\end{tabular}
\caption{$k$-factor distributions in the $t\bar{t}h$ center-of-mass system. The $b^{t\bar{t}h}_2(t,\bar t)$ and $b^{t\bar{t}h}_4(t, \bar t)$ distributions are shown in the upper left and right panels, respectively. The $h$ boson and top-quark transverse momentum distributions are shown in the bottom left and right panels, respectively.}
\label{fig:kfactors}
\end{center}
\end{figure*}

In this context, it is interesting to study the impact of the NLO QCD corrections on the shape of various distributions in the $t\bar{t}h$ center-of-mass frame. Figure~\ref{fig:kfactors} shows the bin-by-bin ratio of the NLO and LO normalized differential distributions of the variables $b^{t \bar{t} h}_2(t, \bar{t}), b^{t \bar{t} h}_4(t,\bar{t}), p_T^h$ and $p_T^t$. As usual, red lines refer to the pure scalar case, while blue lines refer to to pure pseudoscalar case. If NLO QCD corrections would not distort at all the shape of the normalized distributions, Figure~\ref{fig:kfactors} would show flat horizontal lines at $k=1$.

By observing the figure one can notice that NLO QCD corrections have a considerable impact on the shape of the $b^{t \bar{t} h}_2(t,\bar{t})$ and $b^{t \bar{t} h}_4(t,\bar{t})$ distributions, and a milder impact on the shape of the  $p_T^h$ and $p_T^t$ distributions. (It should be stressed that the scale on the k-factor axis is different in the four panels in Figure~\ref{fig:kfactors}.) Moreover, for the $b^{t \bar{t} h}_2(t,\bar{t})$ and $b^{t \bar{t} h}_4(t,\bar{t})$ distributions in particular, the impact of the NLO QCD corrections on the shape of the distribution is different for the scalar and pseudoscalar cases.

For all of the absolute distributions considered in this section, the distribution for an arbitrary value of the angle $\alpha$
can be obtained by combining the distributions for $\alpha = 0$ (scalar case) and for $\alpha = \pi/2$ (pseudoscalar case) in each bin as indicated in Eq.~(\ref{eq:totcsalpha}) for the case of the total cross section. Normalized distributions and $k$ factor distributions for arbitrary $\alpha$ can then be obtained starting from the non-normalized distributions and total cross section for the chosen value of $\alpha$.

This preliminary study of the differential distributions for on-shell top-antitop pair and $h$ boson leads to conclude that the reconstruction of the massive particle momenta in the $t \bar{t} h$ frame can give a significant contribution in identifying a possible pseudoscalar component in the top-quark Yukawa coupling.

\section{Parton-Level Distributions with showering \label{sec:showering}}

In this section several observables, and in particular $b^f_2$  and $b^f_4$ defined in Eqs.~(\ref{eq:b2},\ref{eq:b4}),  are re-analyzed for event samples after parton shower.
In order to produce the distributions presented in this section, samples for $t \bar{t} H$, $t \bar{t} A$ and $t \bar t b \bar b$ production were generated with {\scshape MadGraph5\Q{_}aMC@NLO}~\cite{Alwall:2014hca}. For the $t \bar{t} H$ and $t \bar{t} A$ signals, the \texttt{HC\Q{_}NLO\Q{_}X0} model~\cite{Artoisenet:2013puc} was used. The samples have NLO accuracy in QCD and were generated by employing NNPDF2.3 PDFs~\cite{Ball:2012cx,Ball:2014uwa}. The input mass parameters used are the same ones that were employed in the parton-level calculations described in Section~\ref{sec:partonleveldist}. Dynamical factorization and renormalization scales, set equal to the sum of the transverse masses of all final state particles and partons, were used. The distributions were obtained by using the NLO four-momenta of top quarks, $h$ boson and $b$ quarks (for $t\bar{t}b\bar{b}$), before decay but after parton-shower effects,
\emph{i.e.} using the four-momenta information of the last corresponding particle found in the event history.

\begin{figure*}
\begin{center}
\epsfig{file=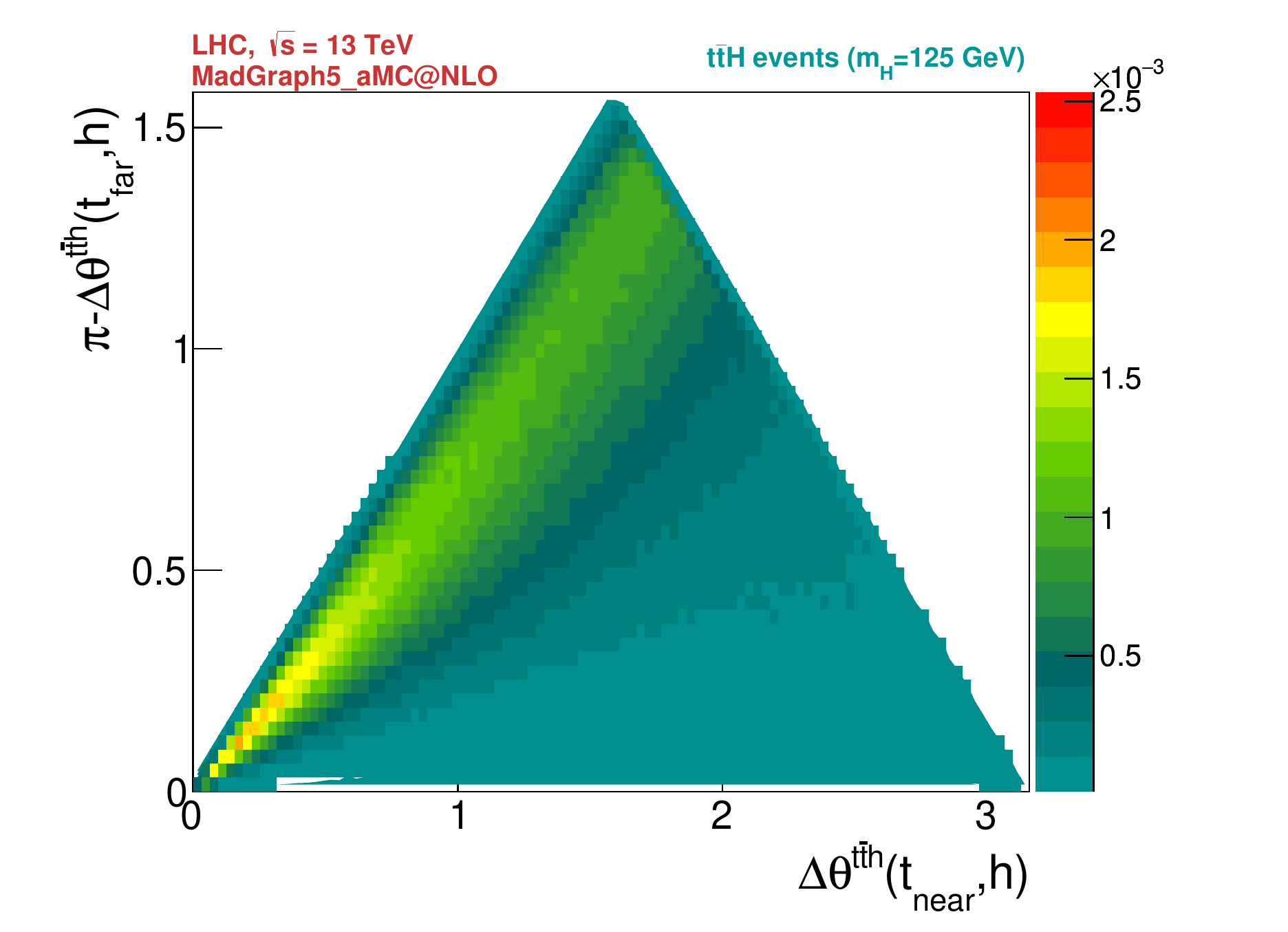,height=5.4cm,clip=} 
\hspace*{-8mm}\epsfig{file=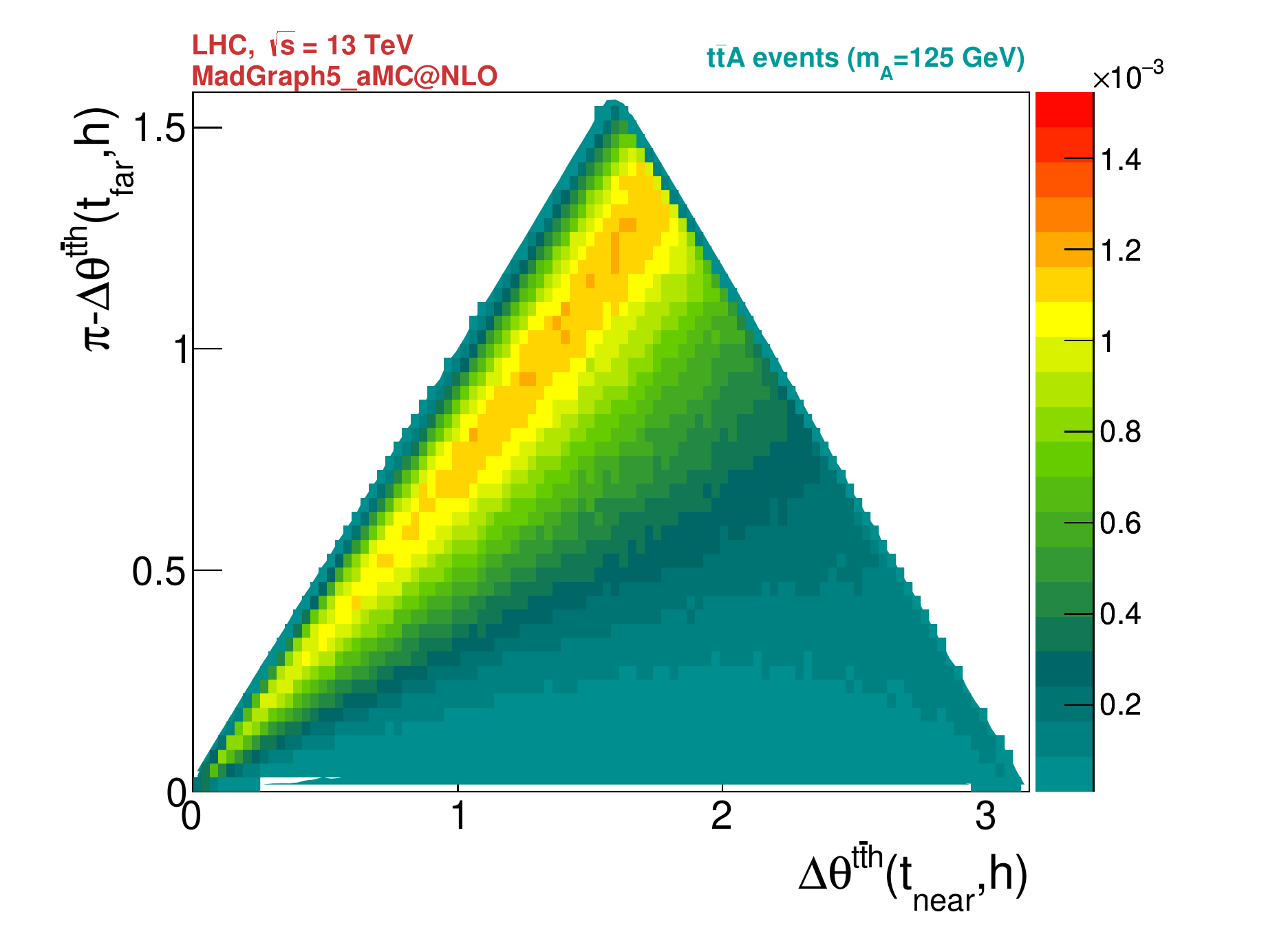,height=5.4cm,clip=} \\
\epsfig{file=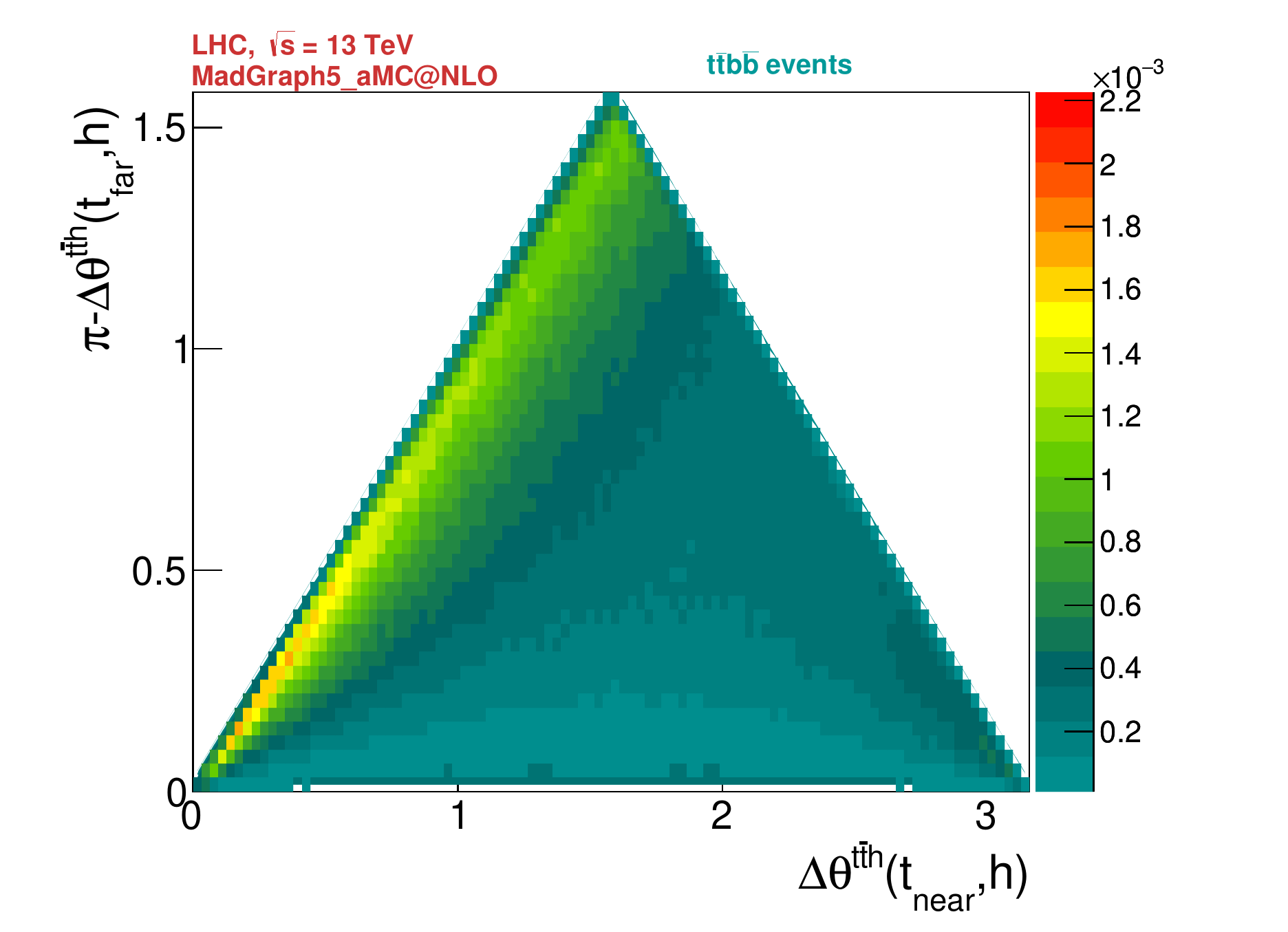,height=5.4cm,clip=}  \\[-2mm]
\vspace*{-0.2cm}
\caption{Normalized two-dimensional distributions at NLO including shower effects: ($x$-axis) the angle between the $h$ boson and the closest top quark ($t$ or $\bar{t}$), plotted against ($y$-axis) the angle between the $h$ boson and the farthest top quark ($\bar{t}$ or $t$), in the $t\bar{t}h$ center-of-mass frame. The pure scalar SM $t\bar{t}H$ distribution (top left), the pure pseudoscalar signal $t\bar{t}A$ (top right) and the dominant $t\bar{t}b\bar{b}$ background (bottom) are shown. In the latter, the role of $h$ is played by the $b\bar b$ system. }
\label{fig:TrianglePlots}
\end{center}
\end{figure*}

As mentioned in the introduction, angular distributions show clear evidence of kinematic differences between the scalar and pseudoscalar type of signals. Moreover, these signals also show significant kinematic differences with respect to the dominant $t \bar t b\bar b$ background distributions.
An example of two-dimensional distributions can be found in  Figure~\ref{fig:TrianglePlots}. In the first two panels of the figure, the $x$-axis corresponds to the angle between the $h$ boson ($h  = H$  in the top left panel, $h = A$ in the top right panel) and the top quark ($t$ or $\bar{t}$) closest to it, evaluated in the $t\bar{t}h$ center-of-mass frame. The $y$-axis corresponds to the angle supplementary to the angle between the $h$ boson and the farthest top quark ($\bar{t}$ or $t$), in the $t\bar{t}h$ center-of-mass system.
In the lower panel in Figure~\ref{fig:TrianglePlots}, which deals with $t \bar{t} b \bar{b}$ events, the role of the $h$ boson is played by the $b \bar{b}$ system, so that the angle on the $x$-axis is the angle between the momentum of the $b \bar{b}$ system and the closest top quark. The colors indicate the normalized number of events in each bin in the $x-y$ plane, evaluated at NLO with parton shower. 
The top left panel in Figure~\ref{fig:TrianglePlots}  shows that, in the case of the pure scalar SM $t\bar{t}H$ production at the LHC, the Higgs boson tends to be produced very close to one of the top quarks and almost back-to-back to the other one. For the pure $t\bar{t}A$ pseudoscalar signal, shown in the top right panel in  Figure~\ref{fig:TrianglePlots},  one can see that $A$ is found to have wider angular distances with respect to both top quarks. The main dominant background $t\bar{t}b\bar{b}$, shown in the lower panel in  Figure~\ref{fig:TrianglePlots}, is such that the angles between the $b\bar{b}$ system and the top quarks are distributed differently with respect to both the $t\bar{t}H$ and $t \bar{t} A$ cases. This shows that the kinematic properties of top quarks and $h$ boson in $t\bar{t}h$ associated production at the LHC are quite different for the scalar ($t\bar{t}H$) signal, the pseudoscalar ($t\bar{t}A$) signal, and for the dominant background $t\bar{t}b\bar{b}$. 
When these angular distributions are studied in the $t\bar{t}h$ center-of-mass frame, the differences between the scalar signal, pseudoscalar signal and background cases emerge clearly.
Moreover, as the spin information survives parton showering, detector simulation, event selection and event reconstruction~\cite{Santos:2015dja,AmorDosSantos:2017ayi,Azevedo:2017qiz}, differential distributions can be used to separate the scalar and pseudoscalar signals. These distributions can also be employed to disentagle the $t \bar{t} h$ signal from the dominant background contributions.

\begin{figure*}
\begin{center}
\vspace*{3cm}
\begin{tabular}{cc}
\hspace*{-5mm} \epsfig{file=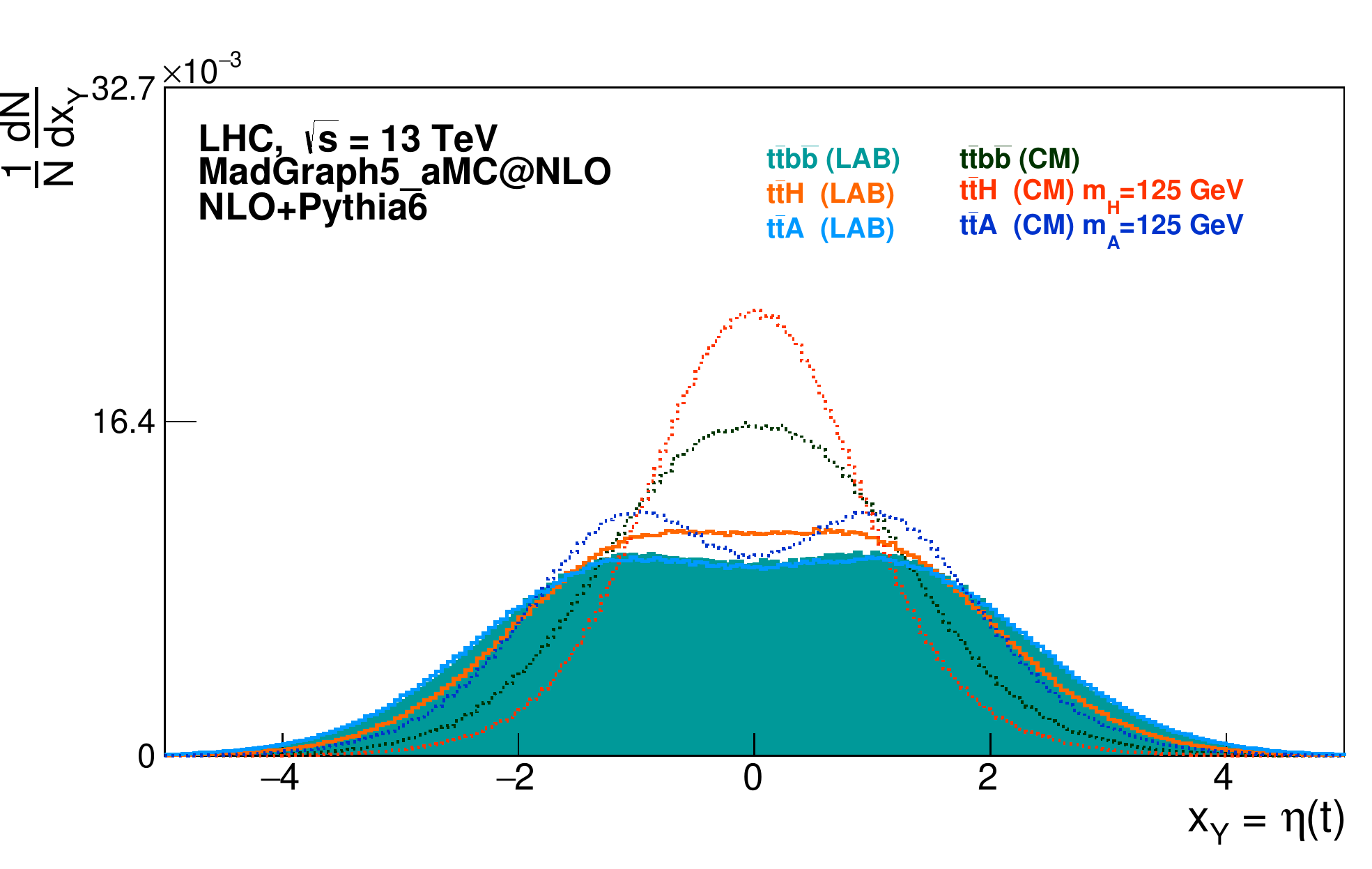,height=5.5cm,clip=} &
                           \epsfig{file=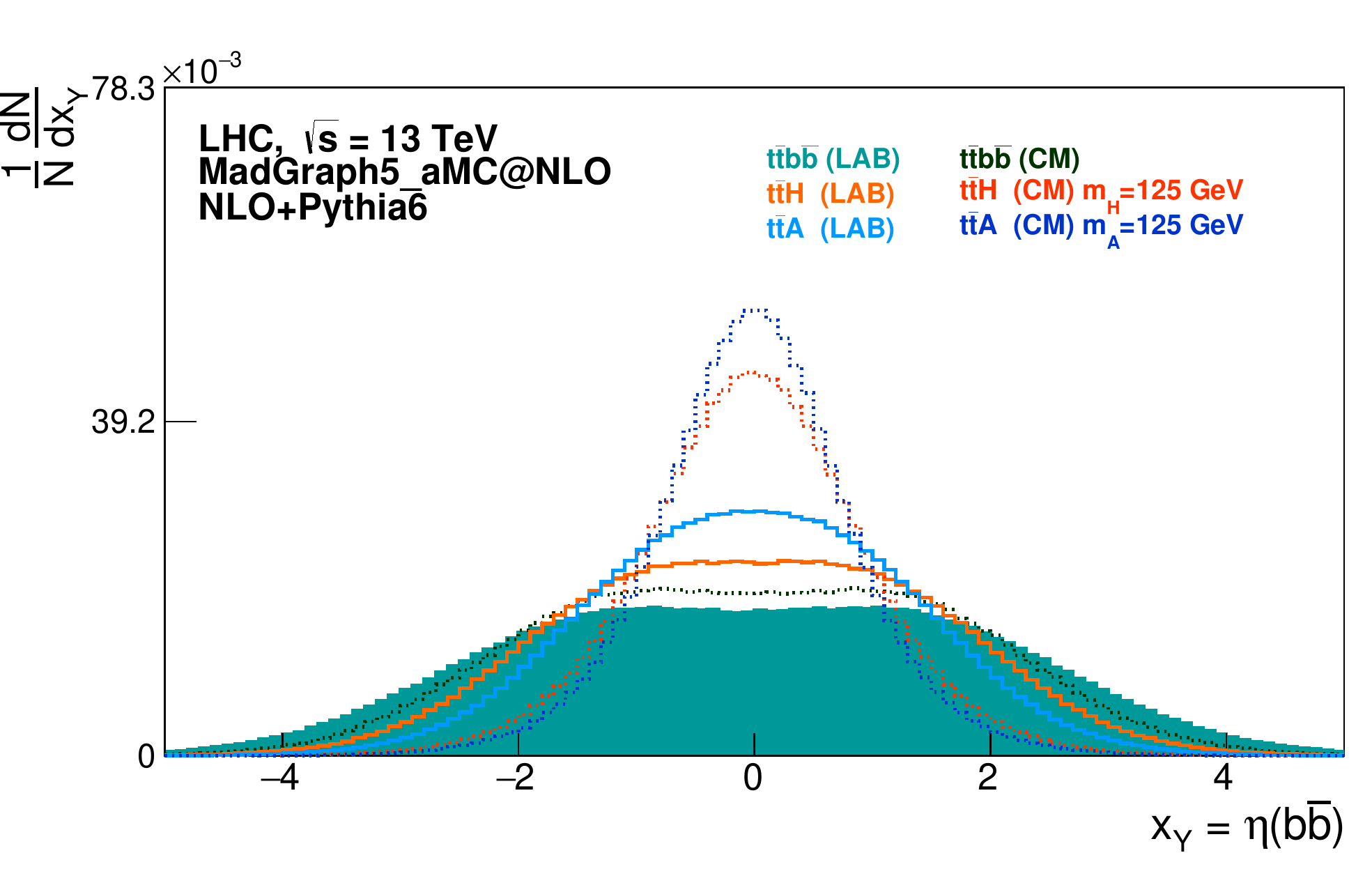,height=5.5cm,clip=} \\[-2mm]
\end{tabular}
\caption{Normalized pseudorapidity ($\eta$) distributions for the top quark (left) and the $b\bar{b}$ system (right), at NLO including shower effects. The distributions measured in the laboratory frame (solid lines) are compared with the ones in the $t\bar{t}h$ center-of-mass system (dotted lines). For comparison, the results of the $t\bar{t}b\bar{b}$ dominant background, the scalar $t\bar{t}H$ and pure pseudoscalar $t\bar{t}A$, are shown.}
\label{fig:PtEtaDist}
\end{center}
\end{figure*}

In Figure~\ref{fig:PtEtaDist}, relevant one-dimensional {normalized} differential distributions are shown. 
The differential distribution with respect to the pseudorapidity $\eta$ of the top quark is shown in the left panel, while the pseudorapidity of the $b\bar{b}$ system is shown in the right panel. The laboratory frame distributions are represented by solid lines, while the distributions in the $t\bar{t}h$ center-of-mass system are represented by dotted lines. For completeness, the distributions for the $t\bar{t}b\bar{b}$ dominant background are also shown together with the pure scalar and pure pseudoscalar signals. 
For the scalar case, the top quark $\eta$ distribution becomes more peaked at the center in the $t\bar{t}H$ center-of-mass frame than in the laboratory frame. For the pseudoscalar case, the top-quark $\eta$ distribution shows a marked minimum at the center in the $t \bar{t} A$ rest frame which is not present in the laboratory frame. {The shapes of the top-quark $\eta$ distributions in the $t\bar{t}h$ center-of-mass frame are in agreement with the NLO calculations for on-shell $t \bar{t} h$  discussed in Section~\ref{sec:partonleveldist}.}
By looking at the right panel in Figure~\ref{fig:PtEtaDist} one can see that the distributions with respect to the pseudorapidity of the $b\bar{b}$ system are more peaked at the center in the $t\bar{t}h$ frame than in the laboratory frame, for both the scalar and the pseudoscalar cases.

\begin{figure*}
\begin{center}
\vspace*{1cm}
\begin{tabular}{cc}
\hspace*{-5mm} \epsfig{file=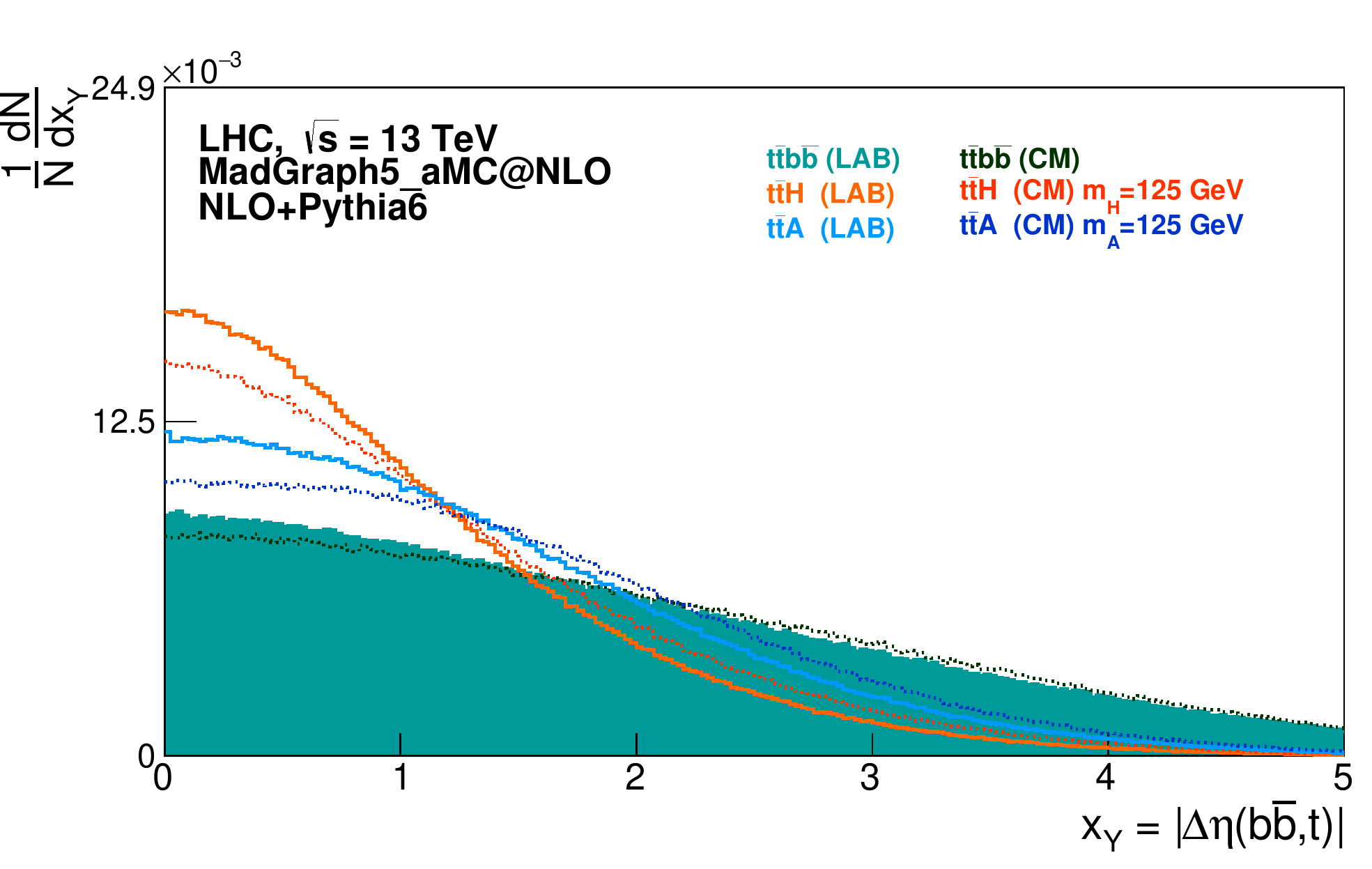,height=5.5cm,clip=} & 
\epsfig{file=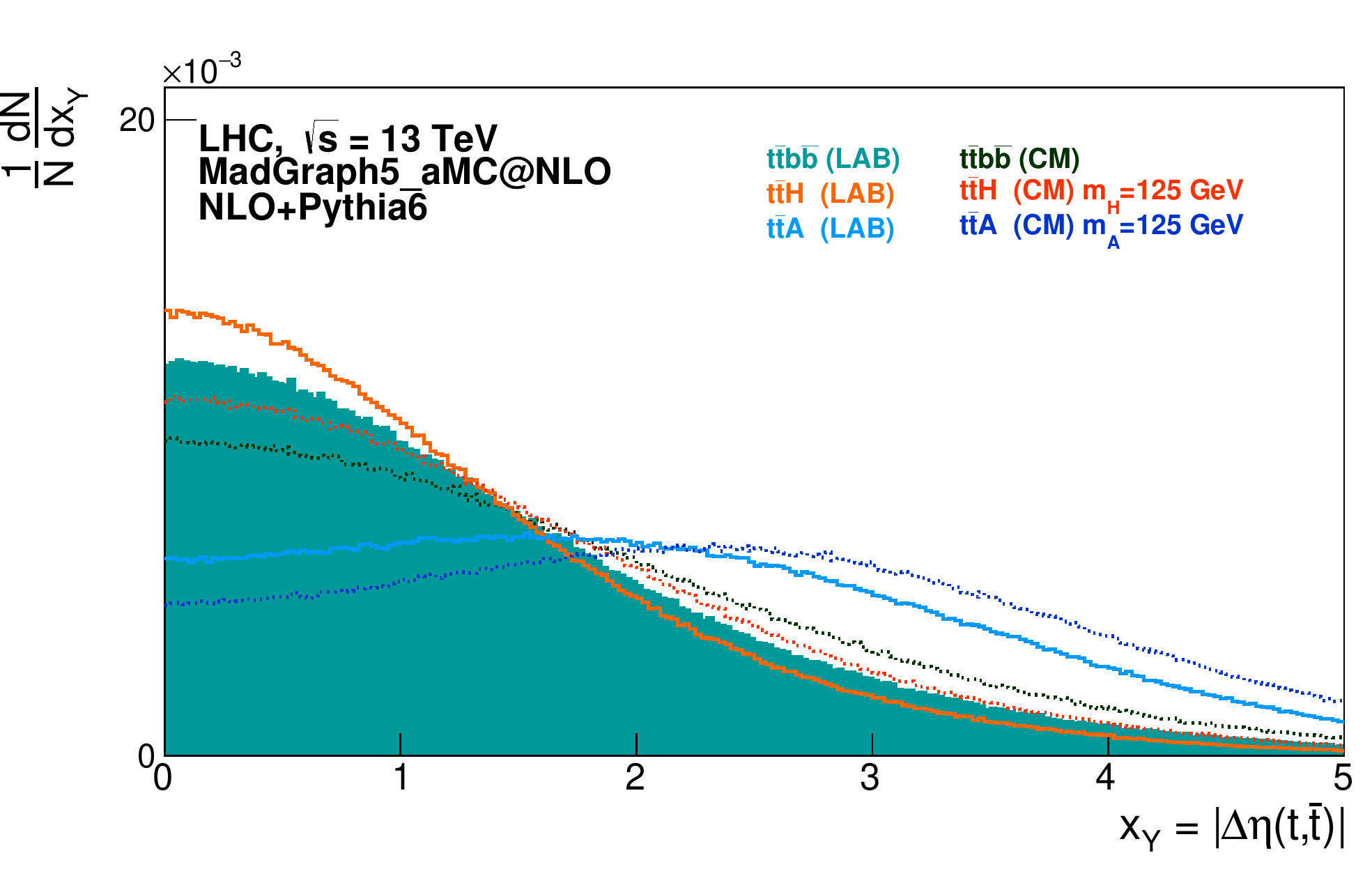,height=5.5cm,clip=} \\[-2mm] 
\end{tabular}
\caption{Comparison of normalized $|\Delta\eta|$ distributions, in the laboratory system (solid lines) and in the $t\bar{t}h$ center-of-mass frame (dotted lines), at NLO including shower effects. The $|\Delta\eta|$ between the $b\bar{b}$ system and the top quark (left) and between top and anti-top quarks (right), are shown. The results of the $t\bar{t}b\bar{b}$ dominant background, the scalar $t\bar{t}H$ and pure pseudoscalar $t\bar{t}A$, are shown for comparison.}
\label{fig:DeltaETA}
\end{center}
\end{figure*}

The left panel of Figure~\ref{fig:DeltaETA} shows the differential distribution with respect to the difference in pseudorapidity, $\Delta \eta$, between the top quark and the $b\bar{b}$ system. The right panel of Figure~\ref{fig:DeltaETA} shows the differential distribution of the difference in pseudorapidity between the top and antitop quarks. In both cases, the signal and background distributions become more populated for higher values of $\Delta \eta$ in the $t\bar{t}h$ center-of-mass system when compared with the laboratory rest frame, while the shape of the distributions remains similar. 

\begin{figure*}
\begin{center}
\vspace*{3cm}
\begin{tabular}{cc}
\hspace*{-5mm} \epsfig{file=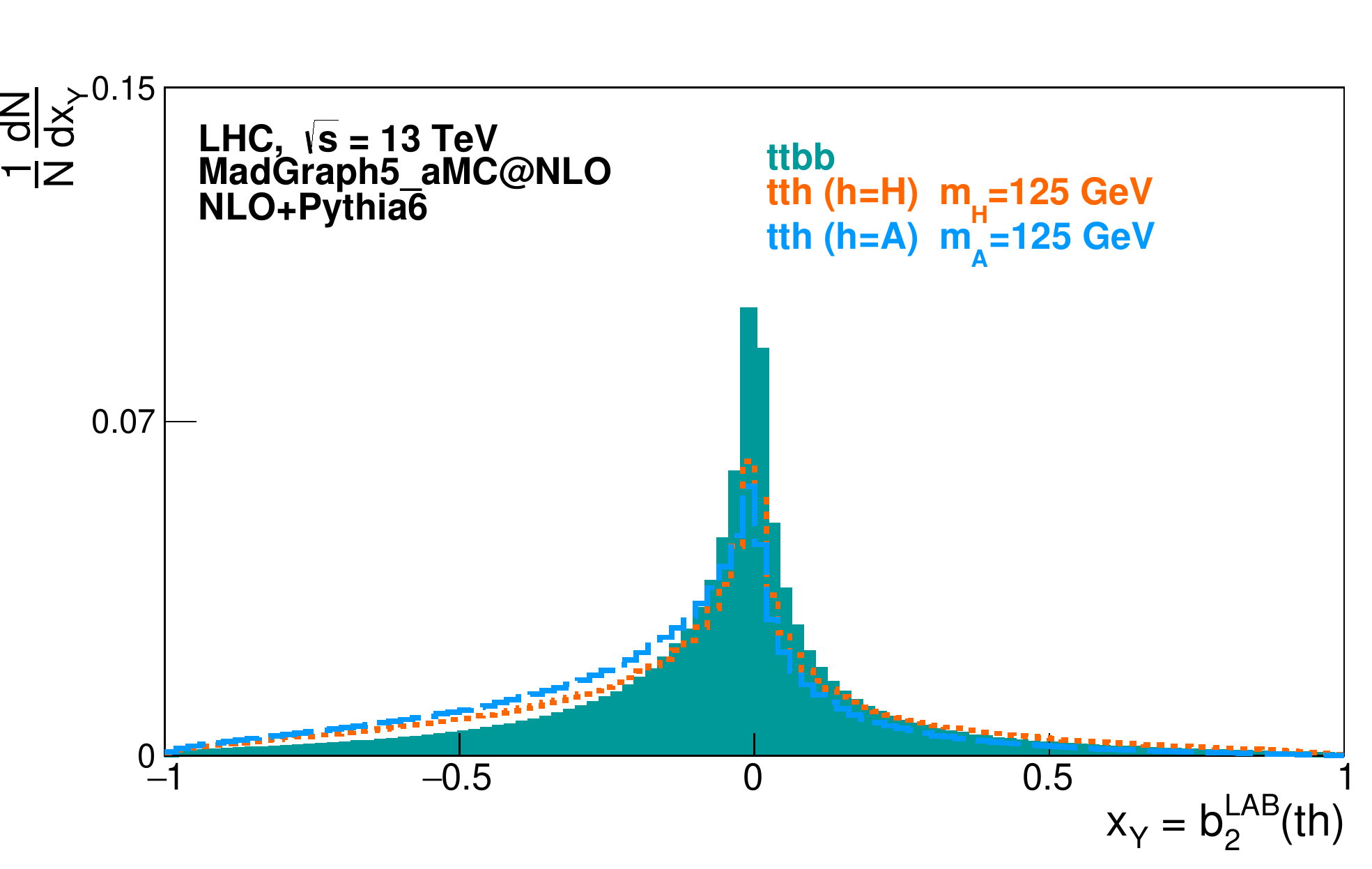  ,height=5.5cm,clip=} & \epsfig{file=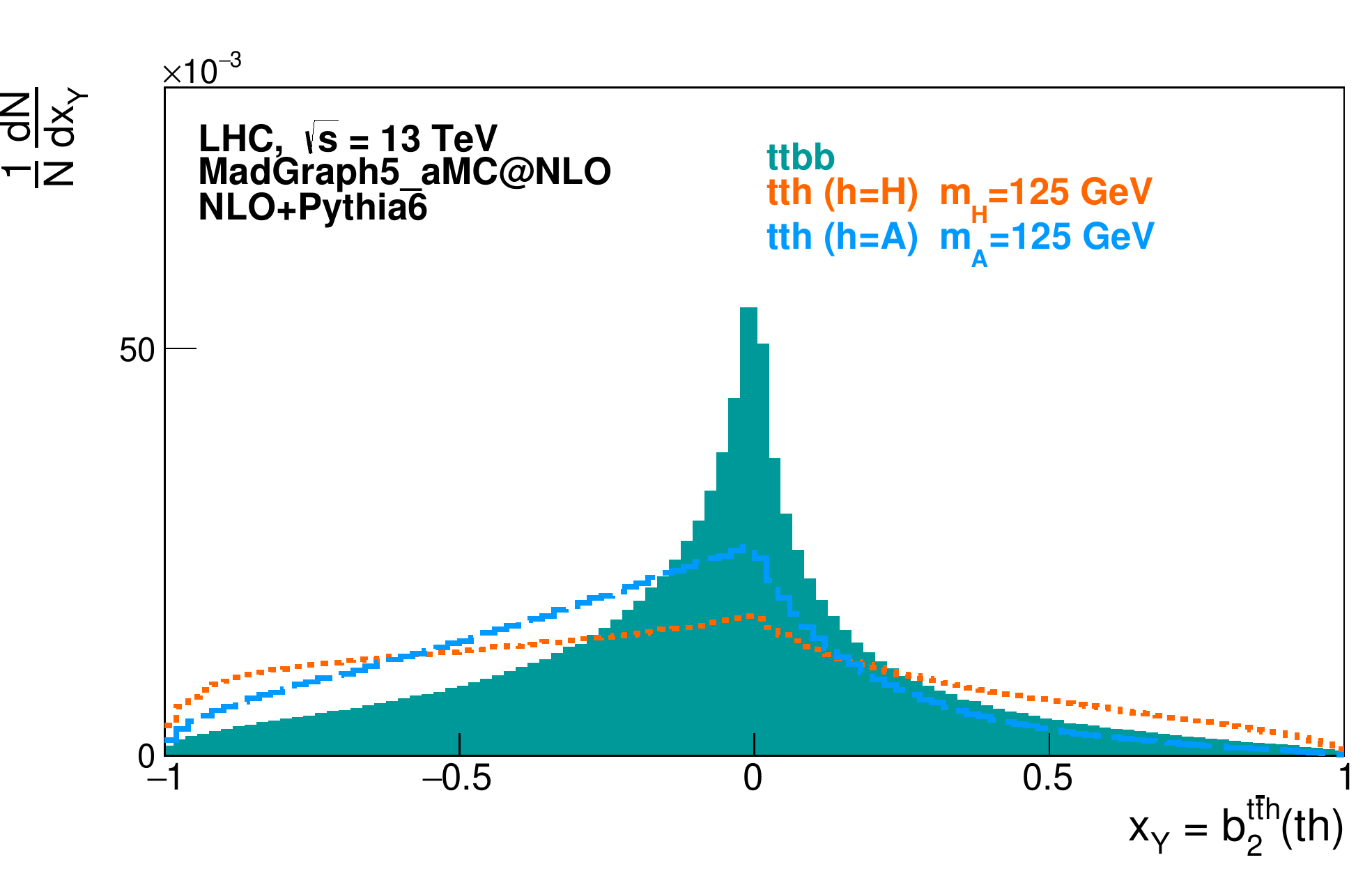  ,height=5.5cm,clip=} \\[-1mm] 
\hspace*{-5mm} \epsfig{file=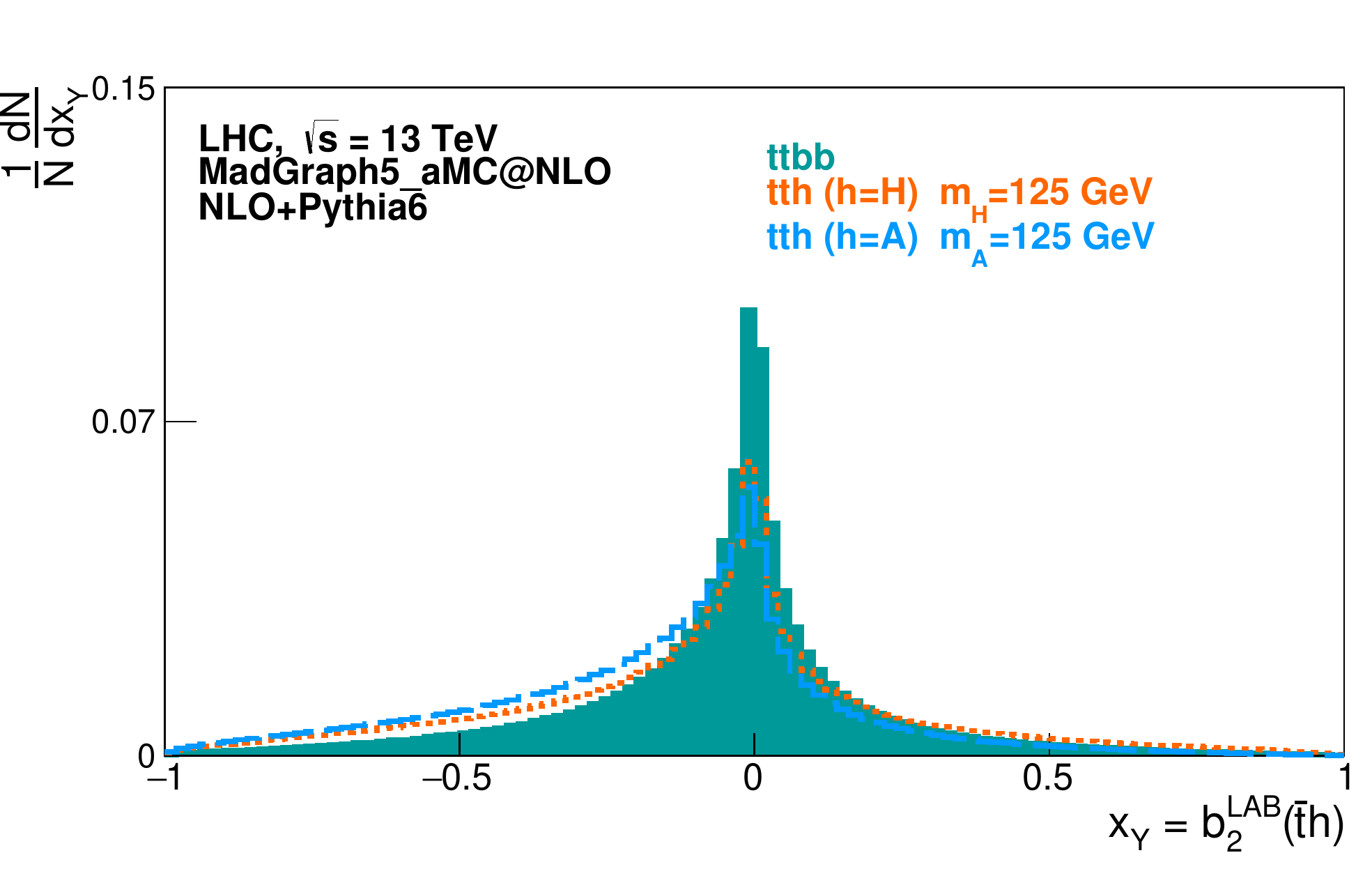 ,height=5.5cm,clip=} & \epsfig{file=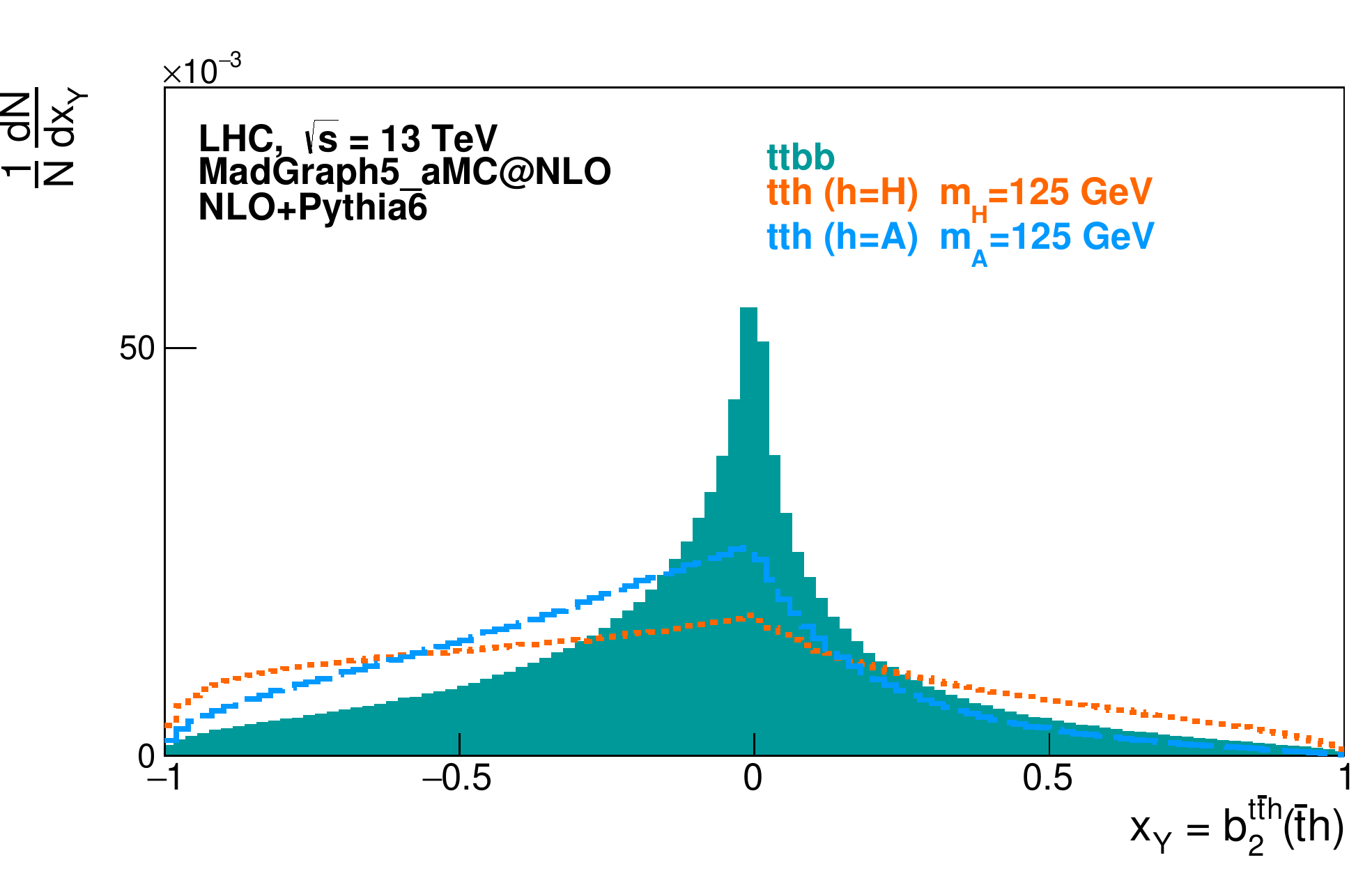,height=5.5cm,clip=} \\[-1mm]
\hspace*{-5mm} \epsfig{file=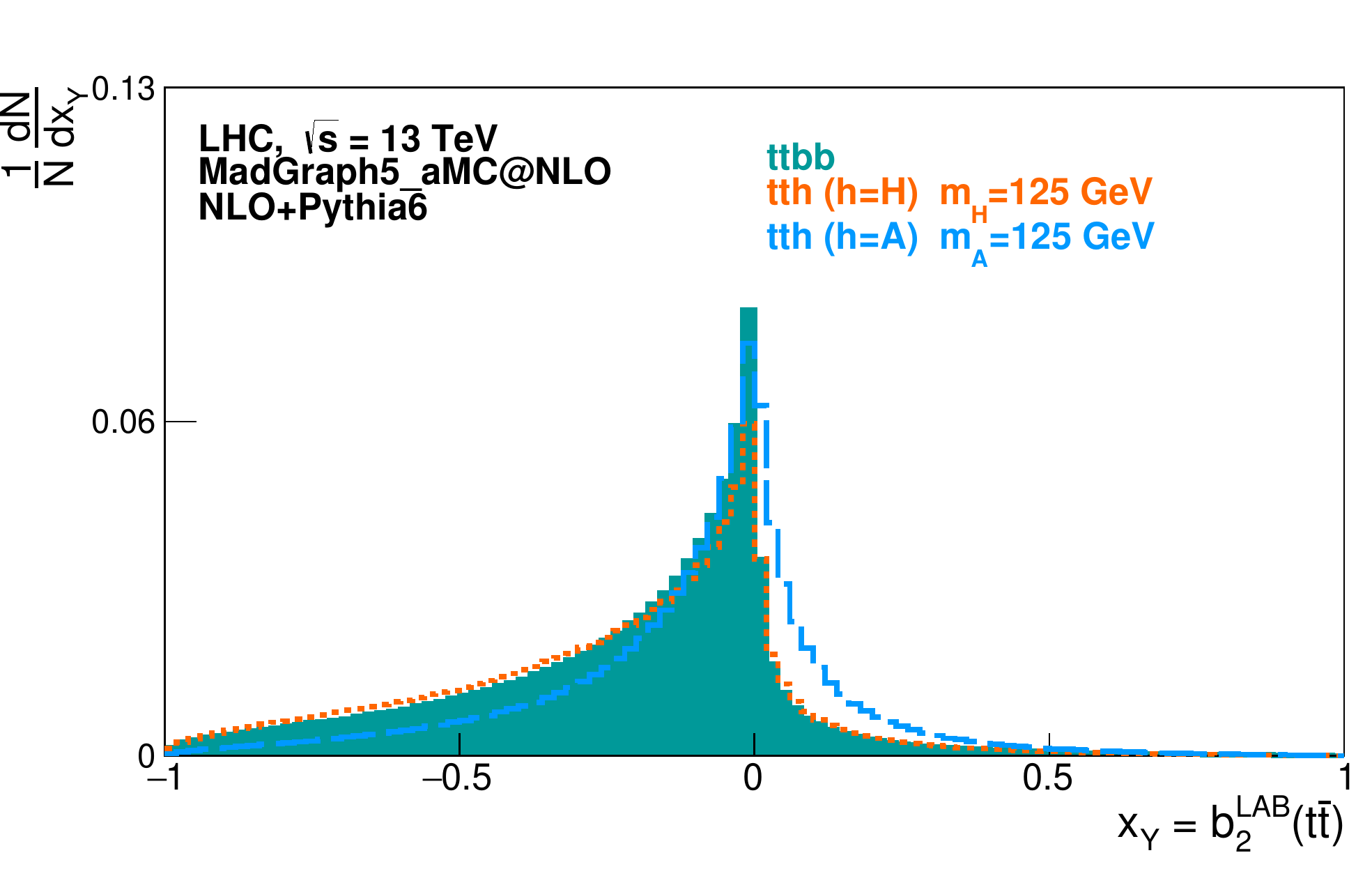 ,height=5.5cm,clip=} & \epsfig{file=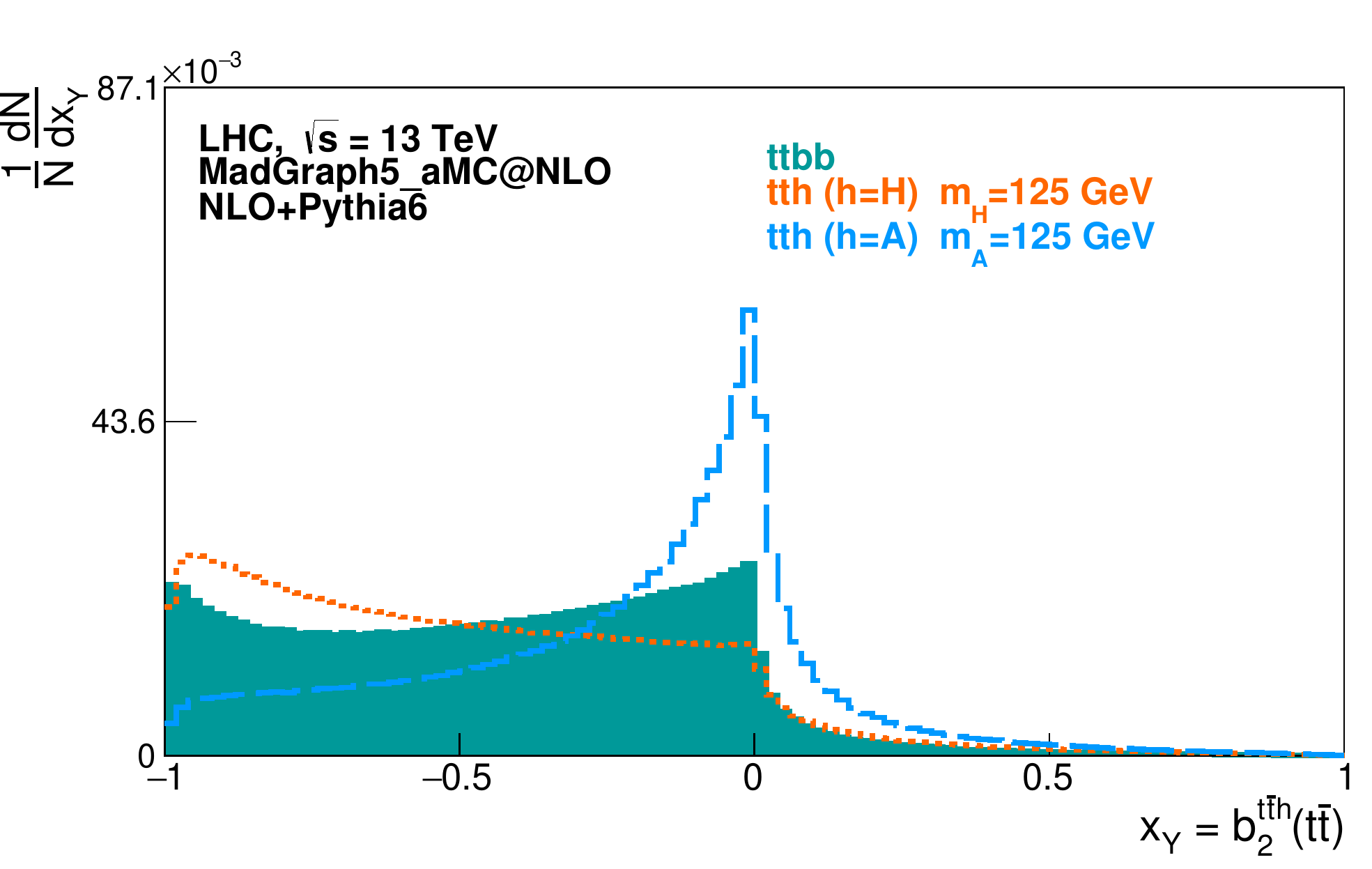  ,height=5.5cm,clip=} \\[-1mm]
\end{tabular}
\caption{Normalized $b^f_2$ distributions for the $th$ (top), $\bar{t}h$ (middle) and $t\bar{t}$ (bottom), evaluated in the laboratory (left) and $t\bar{t}h$ center-of-mass (right) frames, at NLO including shower effects. Distributions for both $th$ and $\bar th$ are included for completeness, although they are equivalent.
The results of the $t\bar{t}b\bar{b}$ dominant background (shaded area), the scalar $t\bar{t}H$ (dashed) and pure pseudoscalar $t\bar{t}A$ (dotted), are shown. In the case of the $t\bar{t}b\bar{b}$ background, the $h$ boson is replaced by the pair of $b$-quarks not coming from top-quark decays.}
\label{fig:b2}
\end{center}
\end{figure*}

\begin{figure*}
\begin{center}
\vspace*{3cm}
\begin{tabular}{cc}
\hspace*{-5mm} \epsfig{file=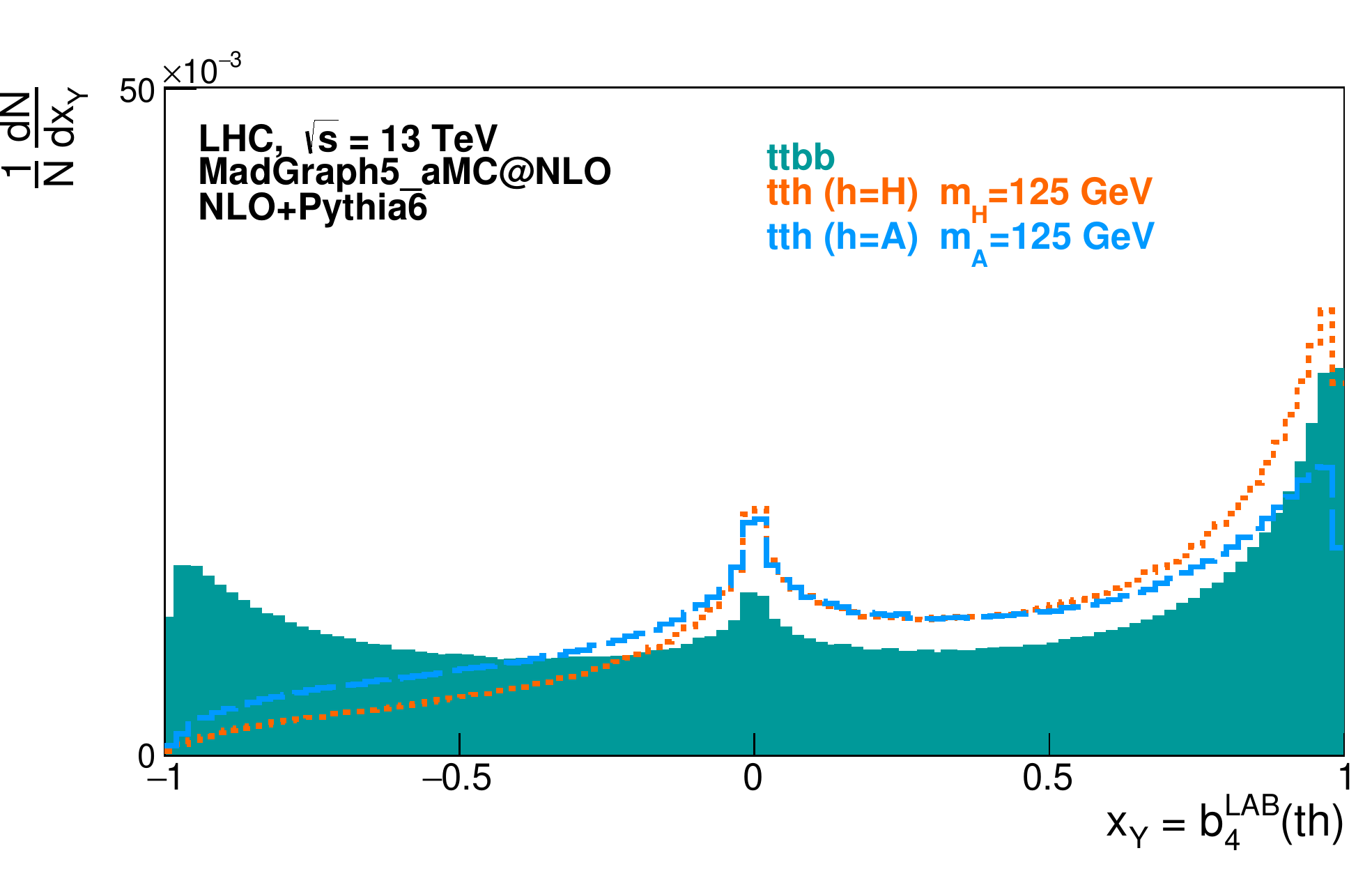  ,height=5.5cm,clip=} & \epsfig{file=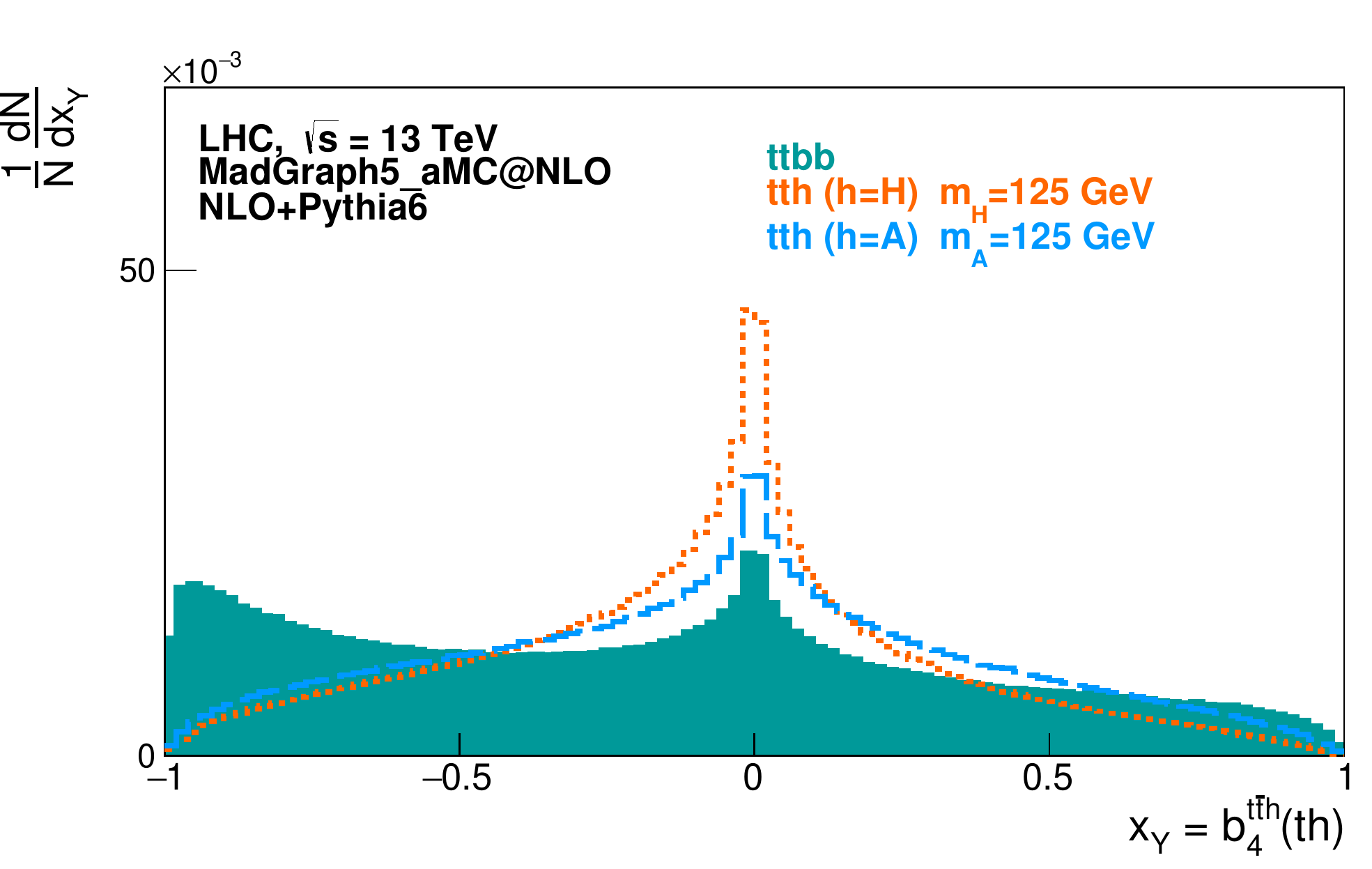  ,height=5.5cm,clip=} \\[-1mm] 
\hspace*{-5mm} \epsfig{file=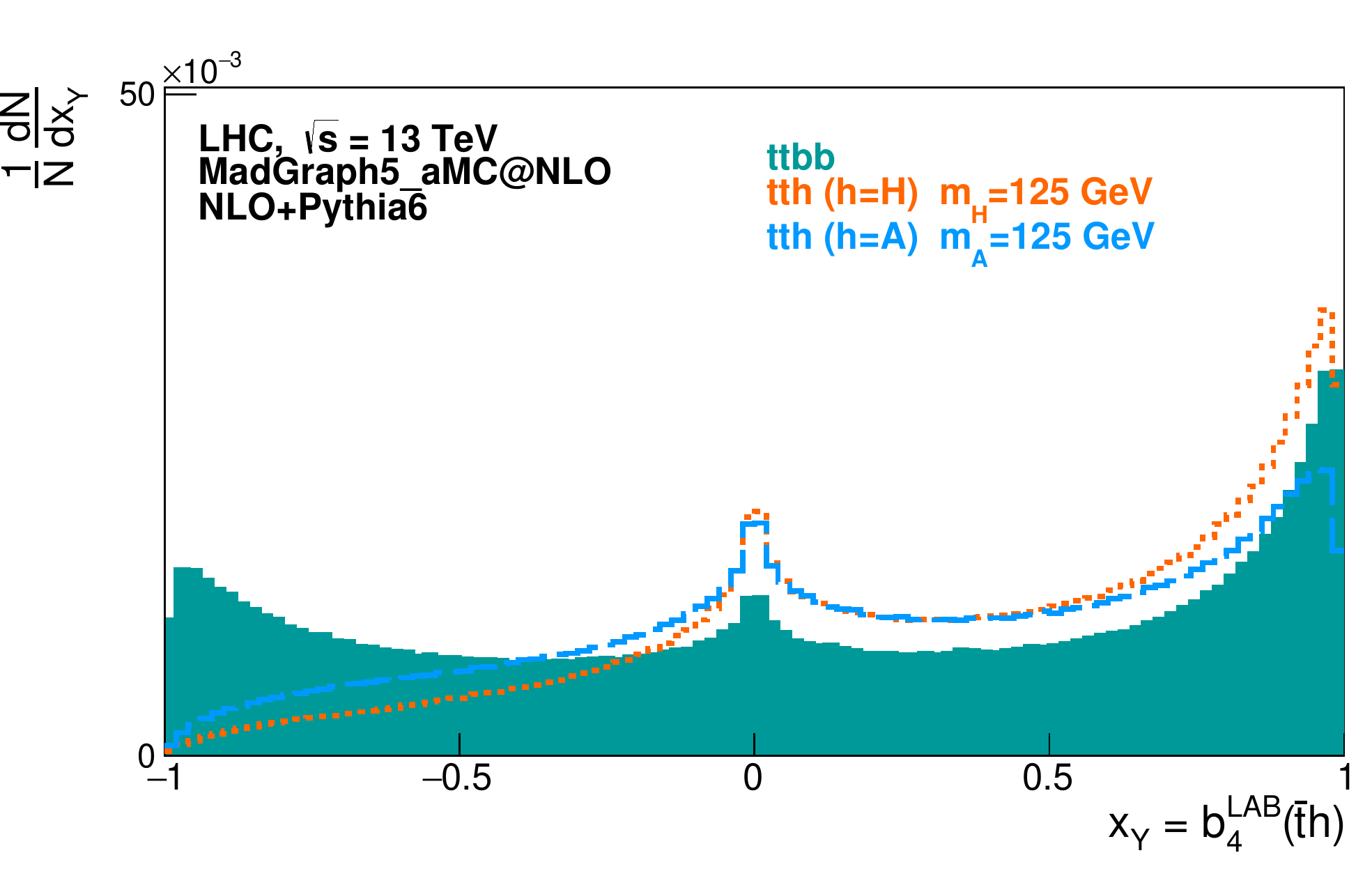,height=5.5cm,clip=} & \epsfig{file=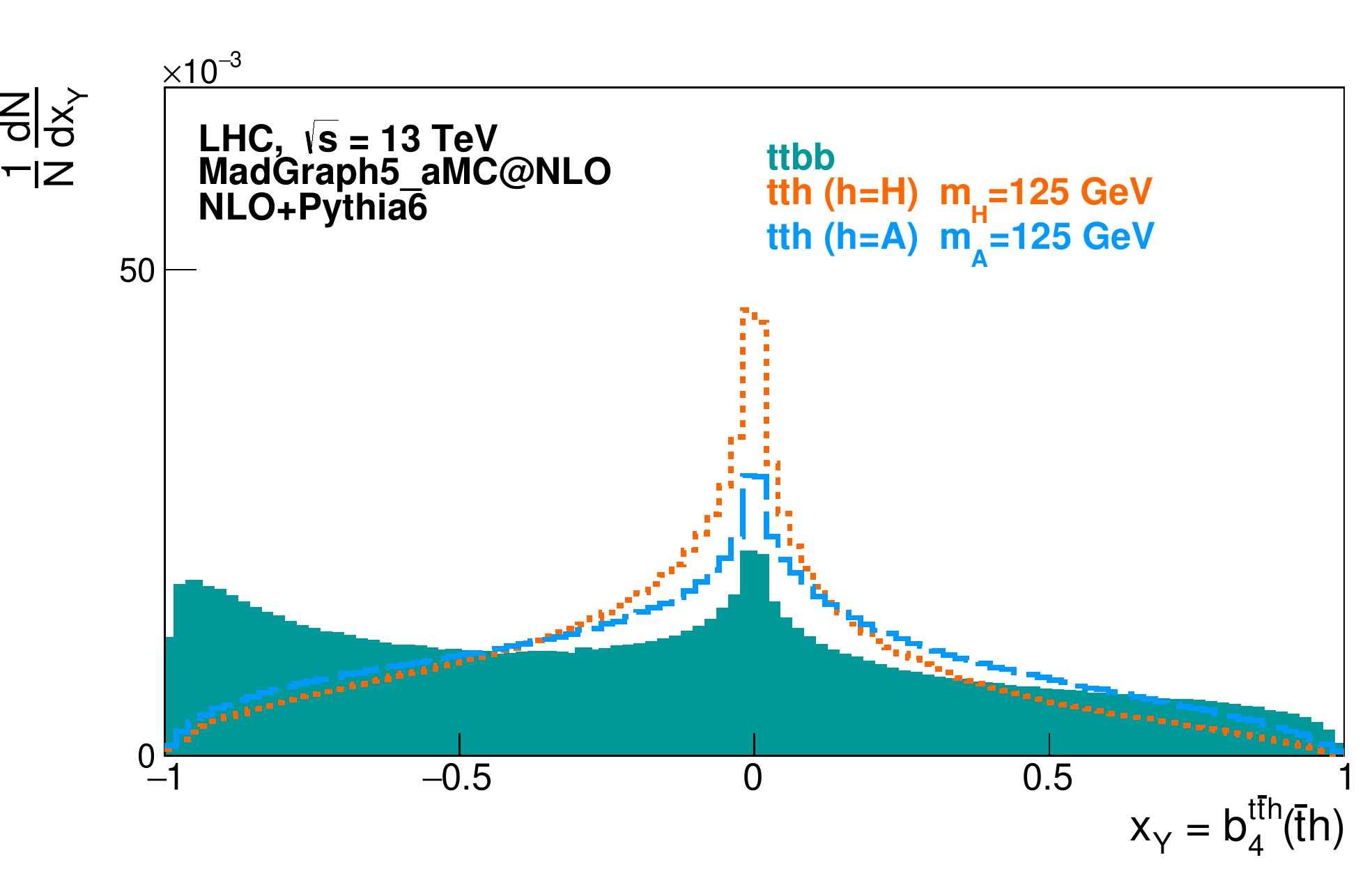,height=5.5cm,clip=} \\[-1mm]
\hspace*{-5mm} \epsfig{file=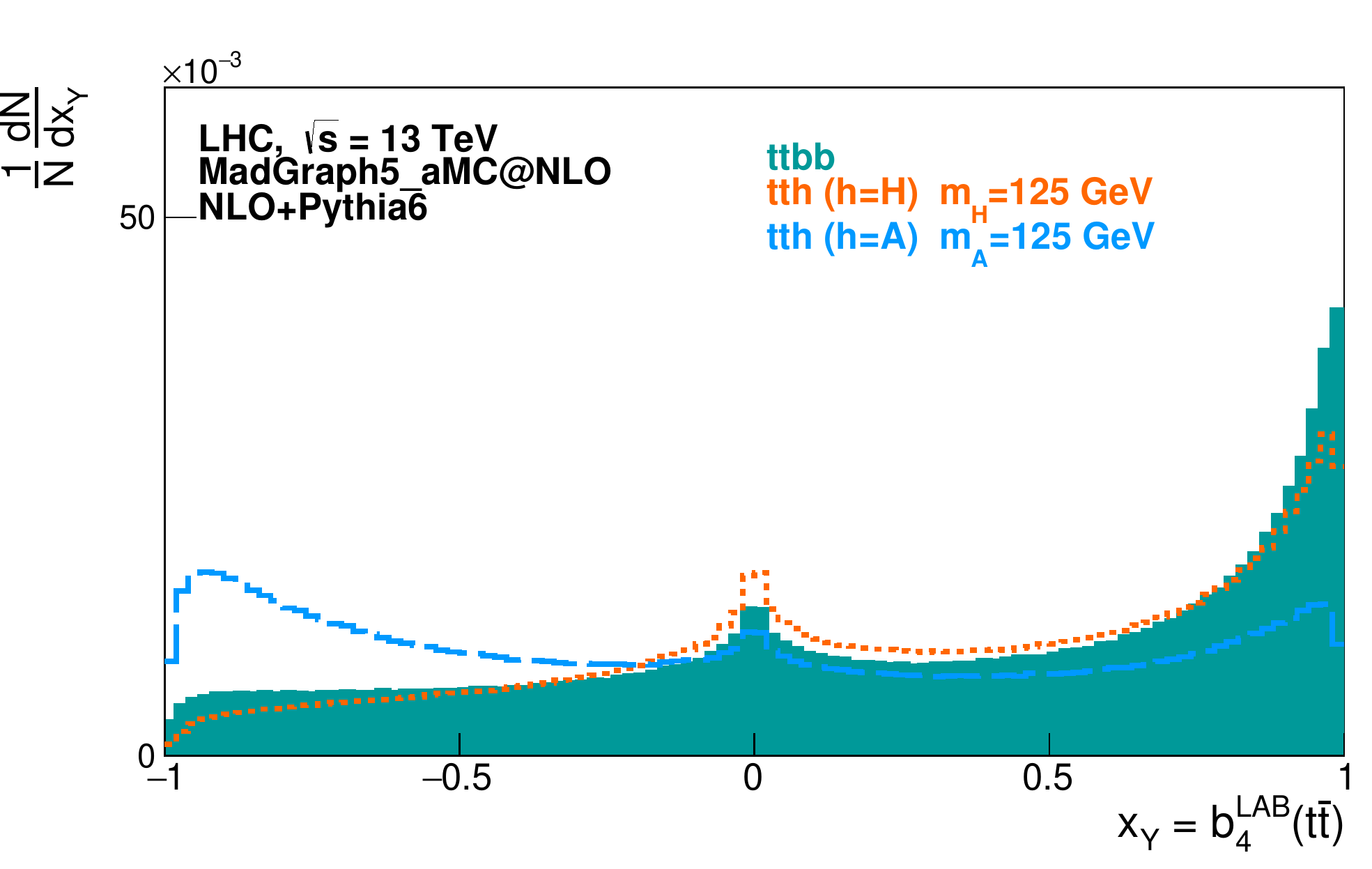 ,height=5.5cm,clip=} & \epsfig{file=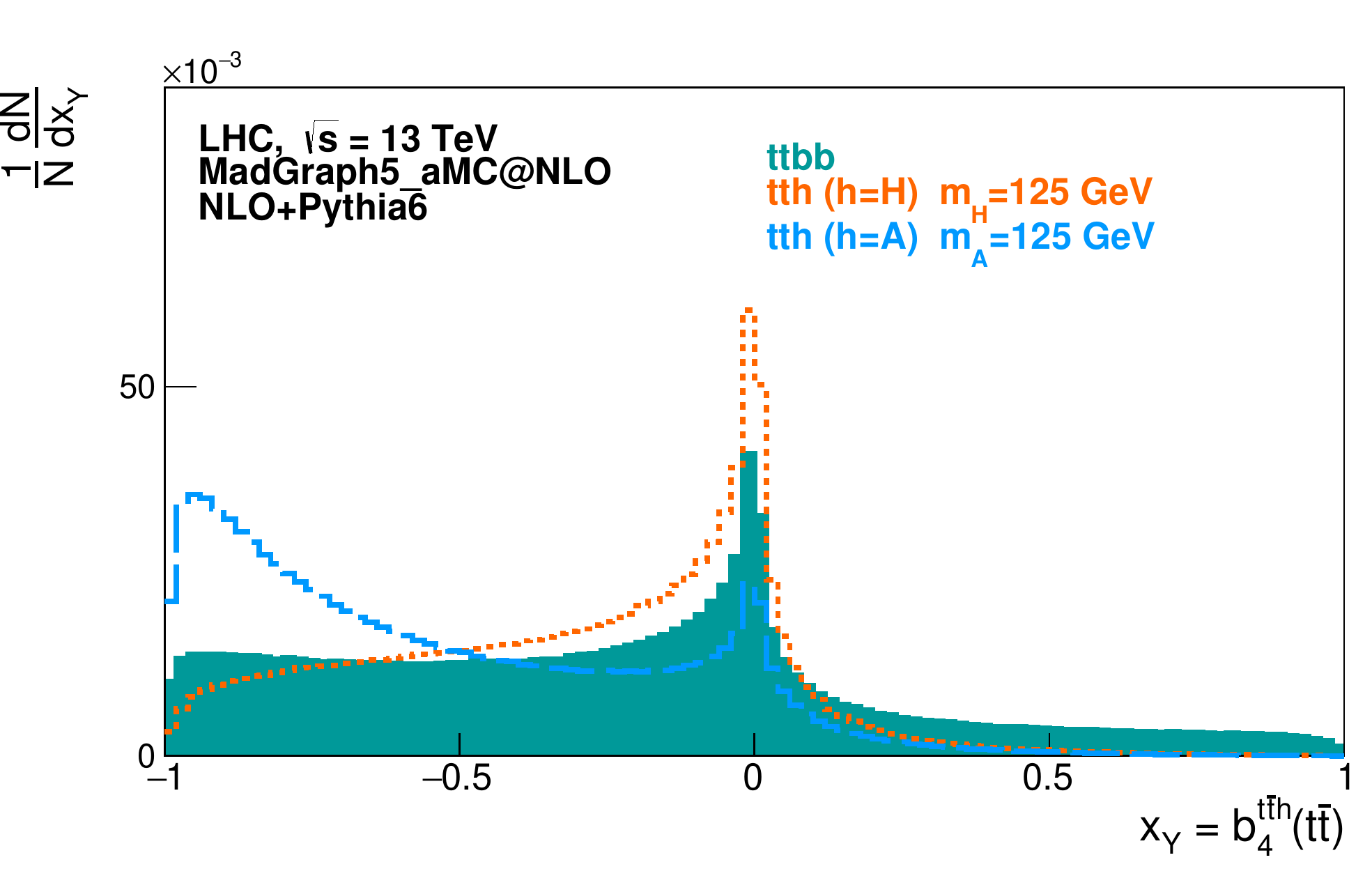  ,height=5.5cm,clip=} \\[-1mm]
\end{tabular}
\caption{Normalized $b^f_4$ distributions for the $th$ (top), $\bar{t}h$ (middle) and $t\bar{t}$ (bottom), evaluated in the laboratory (left) and $t\bar{t}h$ center-of-mass (right) frames, at NLO including shower effects. Distributions for both $th$ and $\bar th$ are included for completeness, although they are equivalent.
The results of the $t\bar{t}b\bar{b}$ dominant background (shaded area), the scalar $t\bar{t}H$ (dashed) and pure pseudoscalar $t\bar{t}A$ (dotted), are represented for completeness. In the case of the $t\bar{t}b\bar{b}$ background, the $h$ boson is replaced by the pair of $b$-quarks not coming from top-quark decays.}
\label{fig:b4}
\end{center}
\end{figure*}

Figure~\ref{fig:b2} and Figure~\ref{fig:b4} show the $b^f_2$ and $b^f_4$ distributions at NLO with parton shower effects and without any selection cuts, respectively, in the laboratory (left panel) and $t\bar{t}h$ (right panel) systems.

The distributions evaluated in the $t\bar{t}h$ center-of-mass system show an increased discriminating power with respect to the ones evaluated in the laboratory frame. This is particularly true for the $b^f_2$ variable, and the effects are more visible in the case of $b_2^{t \bar{t} h} (t,\bar{t})$, than in the cases of $b_2^{t \bar{t} h} (t,h)$ and $b_2^{t \bar{t} h} (\bar{t},h)$. 

Again, the shapes of the $b^{t \bar{t} h}_2$ and $b^{t \bar{t} h}_4$ distributions  shown in Figures~\ref{fig:b2} and \ref{fig:b4} are in agreement with the shape of the corresponding distributions for on-shell $t\bar{t}h$ production, discussed in Section~\ref{sec:partonleveldist}.

\section{Case Study: Dileptonic decays \label{sec:dileptonic}}
In order to evaluate the impact of the observables defined in the $t\bar{t}h$ rest frame, a $t\bar{t}h$ dileptonic analysis was implemented~\cite{Santos:2015dja,AmorDosSantos:2017ayi,Azevedo:2017qiz}, where event generation, simulation and kinematic reconstruction were performed for the conditions of LHC Run 2 proton-proton collisions ($\sqrt{s}= 13$~TeV). As the event analysis was discussed in detail in~\cite{Santos:2015dja,AmorDosSantos:2017ayi,Azevedo:2017qiz}, only a brief reference to its main features is included here.

In addition to the $t \bar{t} H$, $t \bar{t} A$ and $t \bar t b \bar b$ samples presented in the previous sections, backgrounds from $t\bar{t}+jets$ (with up to 3 additional non-$b$ jets), $t\bar{t}V+jets$ (where $V=\{Z,W^{\pm}\}$ and $jets$ include up to 1 additional jet), single top quark production ($t$-channel, $s$-channel and $Wt$ with up to 1 additional jet), diboson ($WW,WZ,ZZ+jets$ with up to 3 additional jets), $W+jets$ and $Z+jets$ (with up to 4 additional jets), and $Wb\bar{b}+jets$ and $Zb\bar{b}+jets$ (with up to 2 additional jets), were generated at LO accuracy in QCD with {\scshape MadGraph5\Q{_}aMC@NLO}, using NN23LO1 PDFs~\cite{Ball:2012cx,Ball:2014uwa}.
{\scshape MadSpin}~\cite{Artoisenet:2012st} was used to decay the top quarks and heavy bosons ($A$, $H$, $W^\pm$, $Z$). Top quarks were decayed through the leptonic decay channel $t(\bar{t})\rightarrow W^+ b (W^-\bar{b})\rightarrow\ell^+\nu b(\ell^-\bar{\nu}\bar b)$, and the $H$ or $A$ boson was decayed through the $b\bar{b}$ channel. {\scshape Pythia6}~\cite{Sjostrand:2006za} was used for parton shower and hadronization. The matching between the generator and the parton shower was carried out by using the MLM scheme~\cite{Alwall:2007fs} for LO events and MC@NLO~\cite{Frixione:2002ik} for NLO events. {\scshape Delphes}~\cite{deFavereau:2013fsa} was used for a fast simulation of a general-purpose collider experiment, using the default ATLAS parameter card. The analysis of the generated and simulated events was performed with \mbox{\scshape MadAnalysis 5}~\cite{Conte:2012fm} in the expert mode~\cite{Conte:2014zja}. Full kinematic event reconstruction was applied, by assuming that the total missing energy originates from the undetected neutrinos. Further details on event generation, simulation and kinematic reconstruction can be found in~\cite{Santos:2015dja,AmorDosSantos:2017ayi,Azevedo:2017qiz}.

\begin{figure*}
\begin{center}
\begin{tabular}{ccc}
\hspace*{-3mm}\epsfig{file=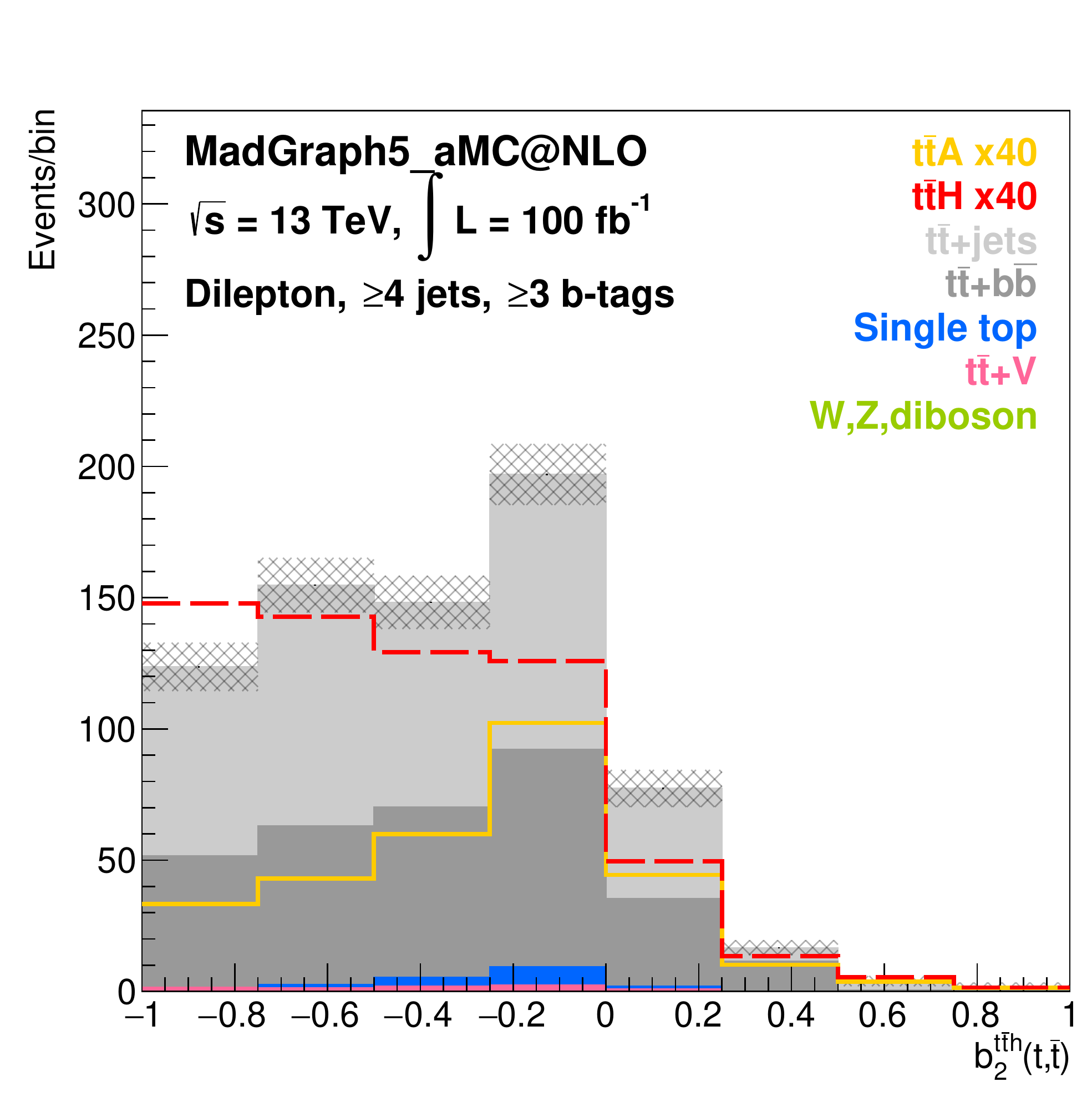,height=8cm,clip=} & \quad & 
\hspace*{-3mm}\epsfig{file=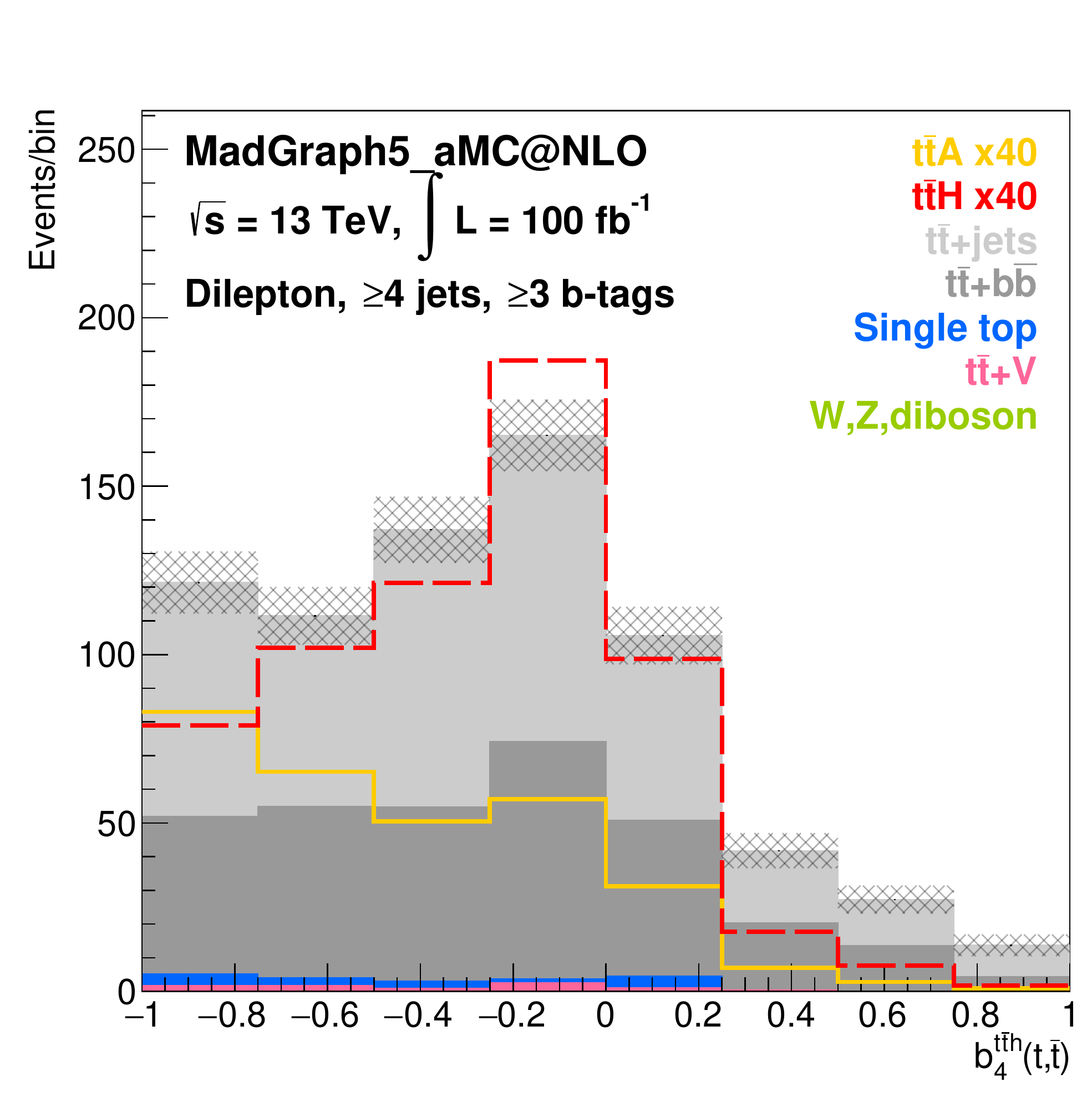,height=8cm,clip=} \\[-2mm]
\end{tabular}
\caption{Distributions of $b^{t \bar{t} h}_2$ (left) and $b^{t \bar{t} h}_4$ (right), for the reconstructed $t\bar{t}$ pair, in the center-of-mass frame of the reconstructed $t\bar{t}h$ system, after event selection and full kinematic reconstruction. The scalar ($t\bar{t}H$) and pseudoscalar ($t\bar{t}A$) signals are scaled by a factor 40 for visibility.}
\label{b2_b4_ttb_reco}
\end{center}
\end{figure*}

\begin{figure*}
\begin{center}
\begin{tabular}{ccc}
\hspace*{-3mm}\epsfig{file=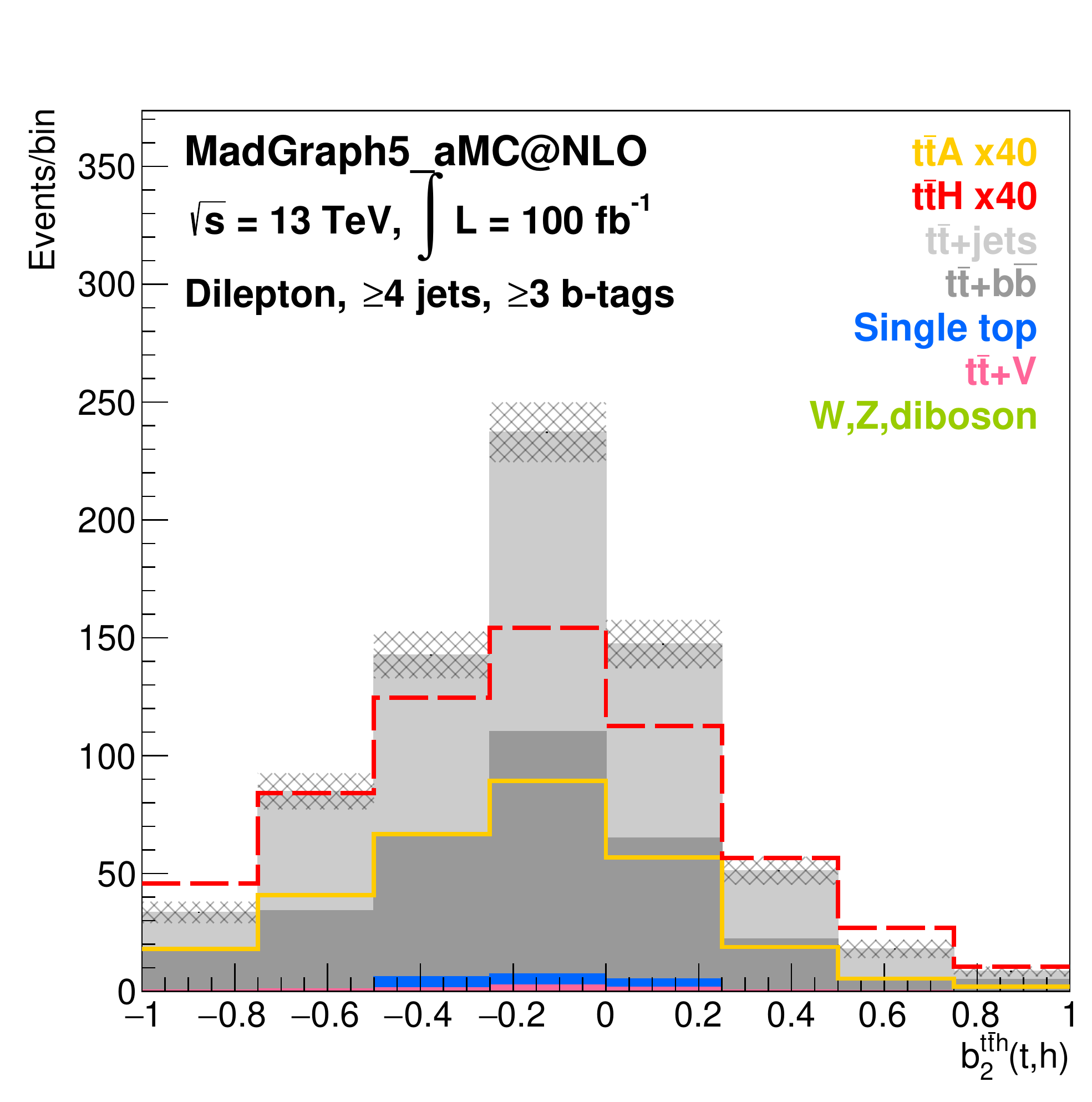,height=8cm,clip=} & \quad & 
\hspace*{-3mm}\epsfig{file=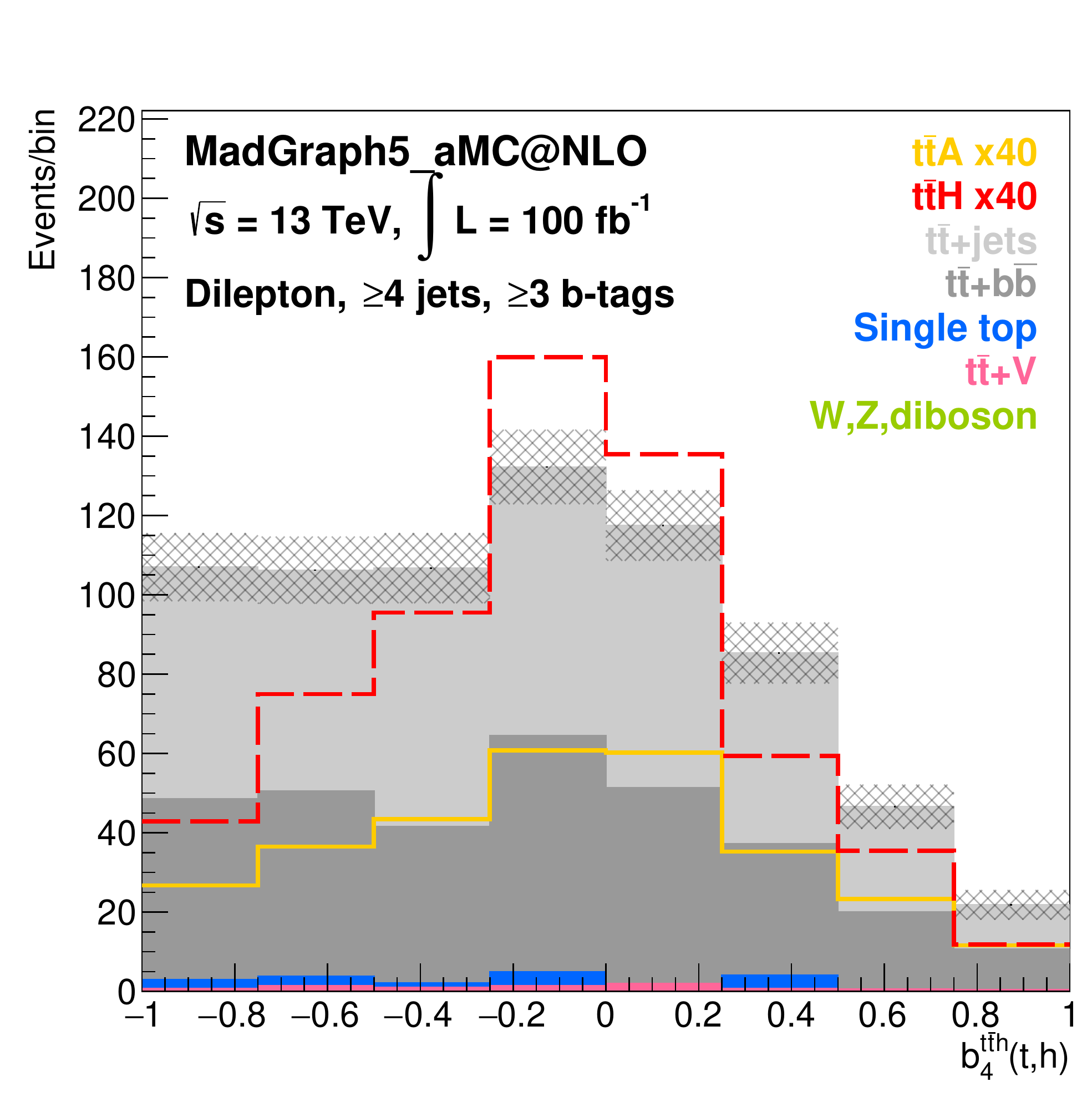,height=8cm,clip=} \\[-2mm]
\end{tabular}
\caption{Distributions of $b^{t \bar{t} h}_2$ (left) and $b^{t \bar{t} h}_4$ (right), for the reconstructed $th$ pair, in the center-of-mass frame of the reconstructed $t\bar{t}h$ system, after event selection and full kinematic reconstruction. The scalar ($t\bar{t}H$) and pseudoscalar ($t\bar{t}A$) signals are scaled by a factor 40 for visibility.}
\label{b2_b4_th_reco}
\end{center}
\end{figure*}

Following the event selection and full kinematic reconstruction, the distributions of different \CP-sensitive observables were obtained, for the scalar and pseudoscalar signals, as well as for the SM backgrounds. The selection targets $t\bar{t}h$ dileptonic final states, in events with at least four jets, of which at least three are identified as coming from the hadronization of $b$-quarks ($b$-tagged). Figure~\ref{b2_b4_ttb_reco}(\ref{b2_b4_th_reco}) show the distributions of $b^{t \bar{t} h}_2$, on the left, and $b^{t \bar{t} h}_4$, on the right, for the reconstructed $t\bar{t}$($th$) pair, in the center-of-mass frame of the reconstructed $t\bar{t}h$ system. The number of events is scaled to an integrated luminosity of 100~fb$^{-1}$ at the LHC. The signal distributions are further scaled by a factor 40 for better visibility. Although resolution effects from detector simulation smear out the kinematic properties of the events when compared to the parton-level distributions, it is still possible to see distinct shape differences between the signals and between signal and the SM backgrounds for $b^{t \bar{t} h}_2$ and $b^{t \bar{t} h}_4$.

\section{CL results for \CP-odd exclusion} \label{sec:CL}

In this section, CLs on the exclusion of a pure pseudoscalar scenario are calculated. The binned distributions of different \CP-sensitive observables presented in the previous sections were used to this effect. From each one of these distributions, which include the contribution from all the expected SM backgrounds, 100,000 pseudo-experiments were generated for two cases, \emph{i.e.} the pure scalar and the pure pseudoscalar. These pseudo-experiments were built bin-by-bin according to a Poisson distribution, using the expected number of events in each individual bin as the mean value.
For each pseudo-experiment, the probability of obtaining such a pseudo-experiment was computed, under the scalar and pseudoscalar hypotheses. A likelihood ratio, defined as the ratio between the pseudoscalar and the scalar probabilities, was used as the test-statistics to compute the CL to which the pure \CP-odd scenario can be excluded, assuming the pure SM scalar scenario. 
The expected exclusion CL was calculated as a function of the integrated luminosity, in the range 100-3,000~fb$^{-1}$. The CL was computed per observable and per luminosity point.

\begin{figure*}
\begin{center}
\begin{tabular}{ccc}
\hspace*{-5mm}\epsfig{file=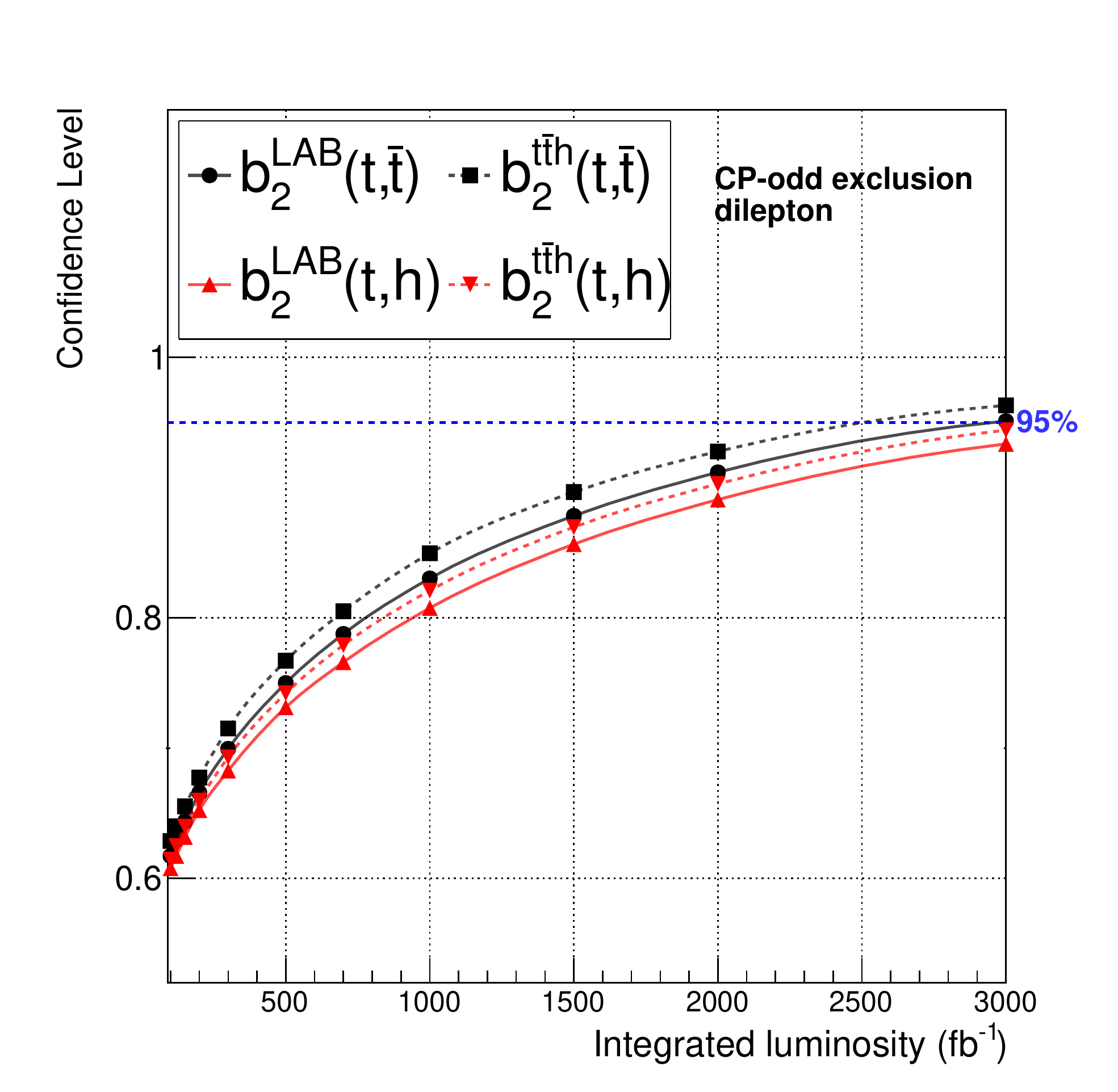,height=8.0cm,clip=} & \quad & 
\hspace*{-5mm}\epsfig{file=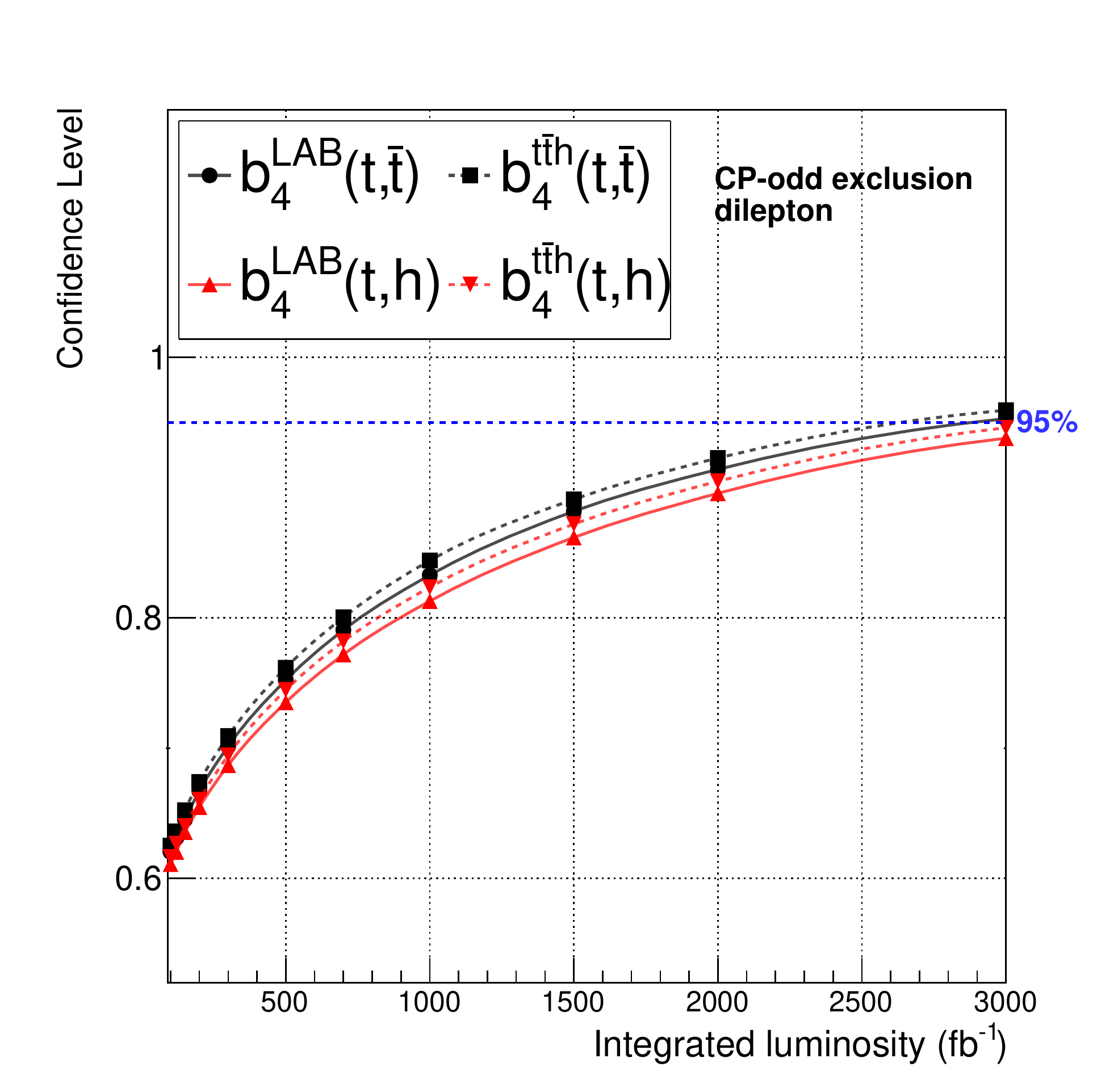,height=8.0cm,clip=} \\[-2mm]
\end{tabular}
\caption{Expected CLs, assuming the SM, for exclusion of the pure \CP-odd scenario, as a function of the integrated luminosity, using the $t\bar{t}h$ ($h\rightarrow b\bar{b}$) dileptonic analysis only. A likelihood ratio computed from the binned distributions of the $b^f_2$ (left) and $b^f_4$ (right) discriminant observables were used as test statistic, evaluated both in the lab and $t\bar{t}H$ frames. Only statistical uncertainties were considered.}
\label{fig:CLdilep}
\end{center}
\end{figure*}

\begin{figure*}
\begin{center}
\begin{tabular}{ccc}
\hspace*{-5mm}\epsfig{file=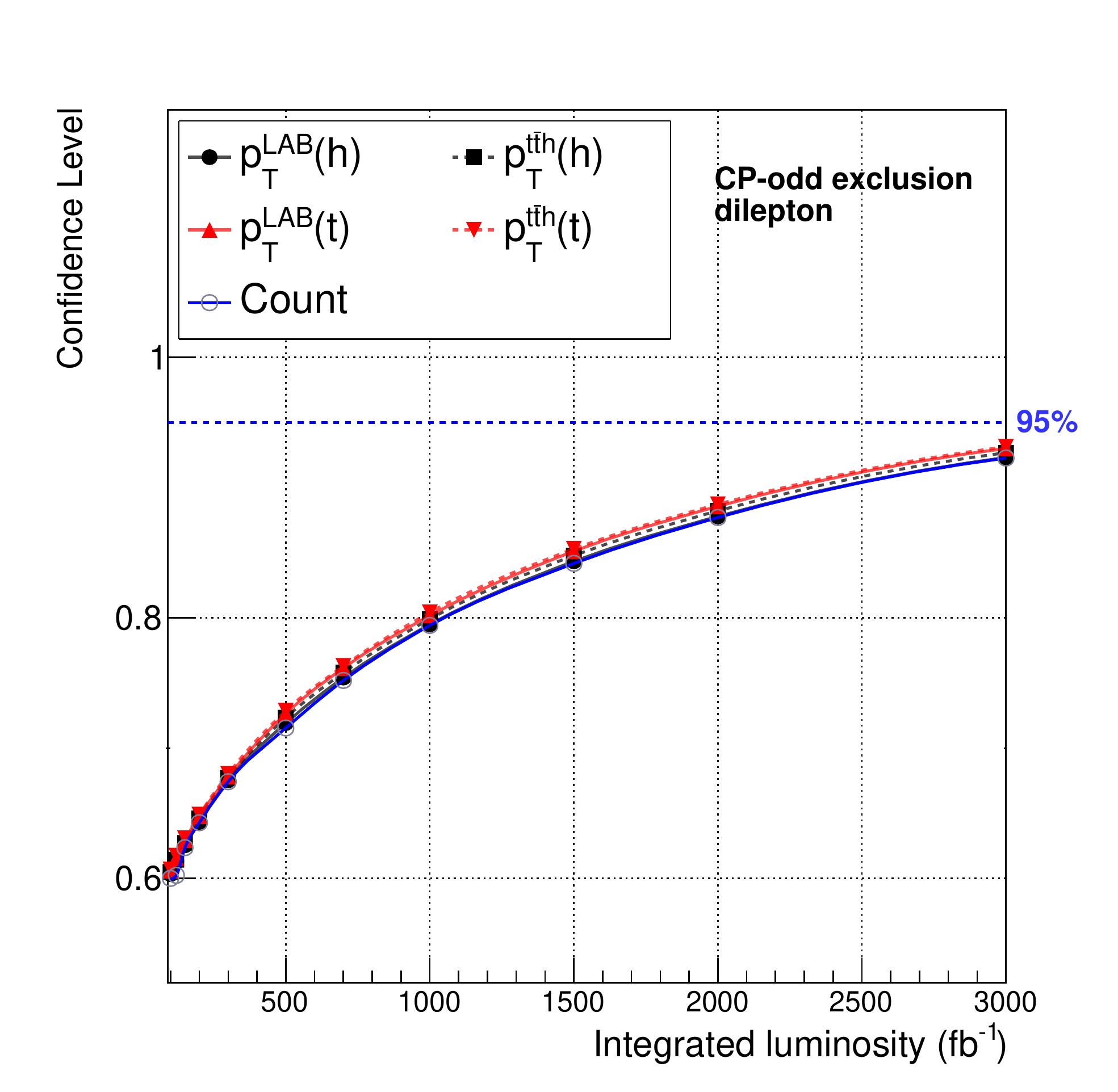,height=8.0cm,clip=} & \quad & 
\hspace*{-5mm}\epsfig{file=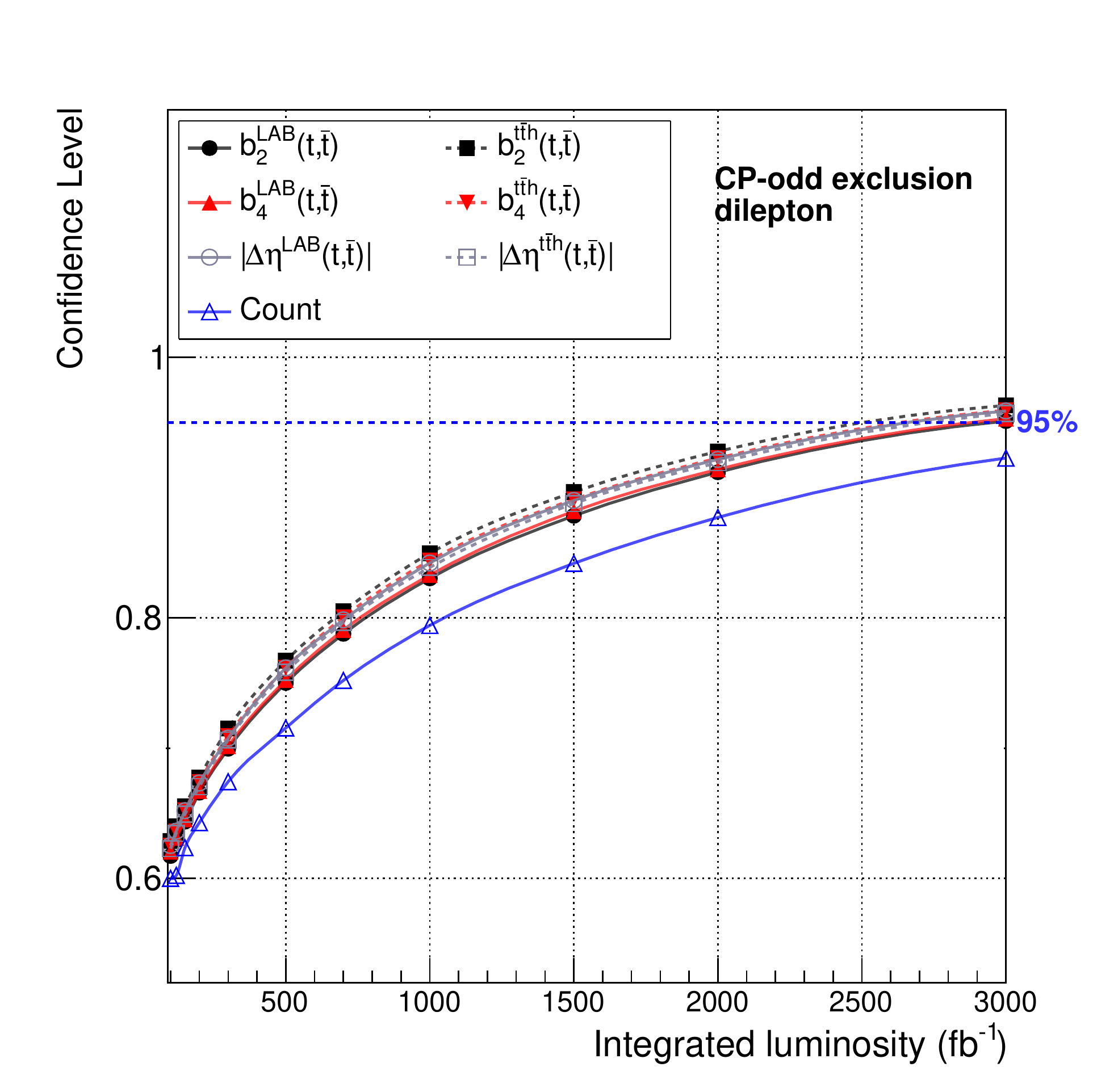,height=8.0cm,clip=} \\[-2mm]
\end{tabular}
\caption{Expected CLs obtained with a test statistic derived from the top quark and Higgs boson $p_T$ distributions (left) and from a set of the most sensitive observables (right), assuming the SM, for exclusion of the pure \CP-odd scenario, as a function of the integrated luminosity, using the $t\bar{t}h$ ($h\rightarrow b\bar{b}$) dileptonic analysis alone. Only statistical uncertainties were considered.}
\label{fig:CLdilep_ttH}
\end{center}
\end{figure*}

Figures~\ref{fig:CLdilep} and~\ref{fig:CLdilep_ttH} show the expected CLs, assuming the SM, to exclude the pure \CP-odd scenario as a function of the integrated luminosity, for different sets of observables. The results are presented using the dileptonic analysis alone, and only statistical uncertainties are considered. Figure~\ref{fig:CLdilep} (left) shows the expected CLs using the $t\bar{t}h$ center-of-mass observables $b_2^{t \bar{t} h} (t,\bar{t})$ and $b_2^{t \bar{t} h} (t,h)$, compared with the ones measured in the laboratory frame. In Figure~\ref{fig:CLdilep} (right) the corresponding $b^{t \bar{t} h}_4$ distributions are shown. One can already see an improvement at this point when the observables are evaluated in the $t\bar{t}h$ center-of-mass frame. For instance, $b^f_2$ requires roughly 250~fb$^{-1}$ less luminosity to achieve the 90\% exclusion CL, when evaluated in the $t\bar{t}h$ center-of-mass frame {than when evaluated in the laboratory frame}.
In Figure~\ref{fig:CLdilep_ttH} (left), the CLs obtained with the top quark and $h$ boson $p_T$ distributions are shown, as a function of the integrated luminosity. One sees that the $p_T$ distributions provide no more discriminant power than counting experiments, such as the measurement of the total cross section. It should be stressed that the present analysis targets events in which the $b$ quarks from the decay of $h$ result in two resolved small-radius jets in the detector. For $p_T(h)\gtrsim200$~GeV, the fraction of events in which this is not the case becomes significant, and boosted analysis techniques may help to improve sensitivity in this region~\cite{Buckley:2015vsa}. In Figure~\ref{fig:CLdilep_ttH} (right), a comparison between
{the CLs obtained with observables sensitive to the \CP properties of the Yukawa coupling is shown, together with the expected exclusion CLs for a counting experiment}. The angular distributions can indeed improve the expected exclusion obtained from a cross section measurement alone. At this point, it is appropriate to mention that these results may be significantly improved by taking into account additional $t\bar{t}h$ final states~\cite{Cepeda:2019klc}. For instance, when considering observables measured in the laboratory frame, the luminosity required to achieve a given level of sensitivity in the single-lepton final state for $t\bar{t}h$($h\rightarrow b\bar{b}$) production, is expected to be roughly five times smaller than the one required for the dileptonic channel alone, see for example Figure~132 of Section~7.7.1 in~\cite{Cepeda:2019klc}. A combination of the single-lepton and dileptonic channels can visibly improve the result even more, providing a powerful and direct test of the nature of the top-quark Yukawa coupling.

\section{Conclusions \label{sec:conclusions}}
In this paper, several observables defined in the $t{\bar t} h$ center-of-mass frame
are proposed. These observables are sensitive to the nature of the top-quark Yukawa coupling, in particular to a possible BSM pseudoscalar component of the coupling. The observables were defined in terms of simple angular distributions, as well as transverse and longitudinal projections of the $t$ quark, $\bar{t}$ quark, and $h$ boson momenta with respect to the beam axis. The differential distributions of these observables in the $t{\bar t} h$ center-of-mass frame are calculated up to NLO in QCD, with and without the effects of parton showering. In both cases, the scalar and pseudoscalar distributions have different shapes. Consequently,  they can be employed as a powerful tool to explore the \CP nature of the coupling.

A dileptonic analysis is implemented to study the impact of the aforementioned observables in a realistic physics scenario, using the fast DELPHES detector simulation and a full kinematic reconstruction of the $t{\bar t} h$ final state. As a result, CLs are presented for the exclusion of the specific \CP-odd scenario as a function of the integrated luminosity. It is shown that by evaluating observables in the $t{\bar t} h$ center-of-mass frame one can significantly reduce the required total integrated luminosity for a given CL, when compared with variables evaluated in the laboratory frame. For example, the $b^f_2$ observable requires approximately 250~fb$^{-1}$ less luminosity to achieve the 90\% exclusion CL, when evaluated in the $t\bar{t}h$ center-of-mass frame. The combination of several $t\bar{t}h$ decay channels should further reduce the luminosity required to directly probe the \CP structure of the top-quark Yukawa coupling by at least a factor five, when compared to a dileptonic analysis alone.

\section*{Acknowledgements}
This work was supported by Funda\c{c}\~ao para a Ci\^encia e Tecnologia, FCT (projects CERN/FIS-PAR/0034/2017, contract SFRH/BSAB/139747/2018 Ref CRM:0061260 and scholarship PD/BD/128231/2016). The in-house Monte Carlo code which we developed and employed to evaluate the (differential) cross sections presented in this paper was run on the computer cluster of the Center for Theoretical Physics at the Physics Department of New York City College of Technology. This work was supported by the PSC-CUNY Awards 61085-00 49, 61151-00 49 and 62187-00 50.


\newpage


\newpage

\newpage

\newpage

\newpage

\end{document}